\documentclass{article}
\usepackage{graphicx}
\def\ligne#1{\hbox to \hsize{#1}}
\def\PlacerEn#1 #2 #3 {\rlap{\kern#1\raise#2\hbox{#3}}}
\newtheorem{thm}{Theorem}
\newtheorem{fig}{Figure}
\newtheorem{algo}{Algorithm}

\def\leurre{\noindent\leftskip0pt\small\baselineskip 10pt}
\def\grostrait{\ligne{\vrule height 1pt depth 1pt width \hsize}}
\def\demitrait{\ligne{\vrule height 0.5pt depth 0.5pt width \hsize}}

\setbox221=\hbox{\includegraphics{cercle_L20.eps}}

\def\encercle#1#2{\hbox{\raise-5pt\copy221\hskip#2#1}}

\title{
Cellular Automata and Discrete Geometry}
\author{Isabelle Debled-Rennesson$^1$, Maurice Margenstern$^2$
}

\begin{document}

\maketitle

%\begin{center}
\ligne{\hfill
$^1$ \vtop{\leftskip 0pt\parindent 0pt\hsize=180pt
ADAGIo team, LORIA, UMR 7503, Nancy University,\\
Campus Scientifique $-$ BP 239\\
54506 Vand\oe uvre-l\`es-Nancy Cedex, France\\
Email: debled@loria.fr
}
\hskip 20pt
$^2$ \vtop{\leftskip 0pt\parindent 0pt\hsize=180pt
LITA, EA 3097, Universit\'e Paul Verlaine $-$ Metz,\\
        and LORIA, CNRS, Campus du Saulcy,\\
 57045 Metz Cedex, France\\
Email: margens@univ-metz.fr
}
\hfill}
%\end{center}

\begin{abstract}
   In this paper, we look at the possibility to implement the algorithm
to construct a discrete line devised by the first author in cellular automata.
It turns out that such an implementation is feasible.
\end{abstract}

\def\cqfd{\hbox{\kern 2pt\vrule height 6pt depth 2pt width 8pt\kern 1pt}}
\def\Hii{$I\!\!H^2$}
\def\Hiii{$I\!\!H^3$}
\def\Hiv{$I\!\!H^4$}
\def\norm{\hbox{$\vert\vert$}}
\def\grossone{\hbox{\encercle{\bf 1}{5pt}$\,\,$}}
\section{{\Large Introduction}}

   In Section~\ref{geom}, we remind the basic features of discrete geometry
and, in particular, the construction of a line in this framework.
In Section~\ref{CA}, we remind of the basic principles of cellular automata.
In Section~\ref{scenario}, after reminding of the algorithm to construct a
discrete line devised by the first author, see~\cite{DEB95}, we explain the 
guidelines which allow us to implement this algorithm
into cellular automata in the plane. In Section~\ref{rules}, we explain how
to transform the scenario of Section~\ref{scenario} into rules which are exhaustively
given in the Appendix. Also, in Section~\ref{rules} we give a sketchy account of the
computer programme devised to construct the rules and to check their correctness.
In section~\ref{conclusion}, we briefly mention how to go on in the line open by
the paper. 

\section{{\Large Discrete Geometry}}
\label{geom}

In this section, we briefly recall some results of \cite{REV91} and
\cite{DEB95} that we shall need.
%\begin{defn}
A {\bf discrete line}\cite{REV91}, named ${\mathcal
D}(a,b,\mu,\omega)$, is the set of integer points $(x,y)$ verifying the
inequalities
$\mu \leq ax - by < \mu + \omega$ where $a,b,\mu,\omega$ are integers.
$\frac{a}{b}$ with $b\not = 0$ and
gcd($a$,$b$)$=1$ is the slope of the discrete line,
$\mu$ is named lower bound and $\omega$ arithmetical thickness.
%\end{defn}
Among the discrete lines we shall distinguish, according to their topology
\cite{REV91}~:
\begin{itemize}
\item [$-$] the {\bf naive lines} which are 8-connected and for which
the thickness
$\omega$ verifies $\omega = max(|a|,|b|)$,
\item [$-$] the {\bf$*$-connected lines} for which the thickness
$\omega$ verifies $max(|a|,|b|) < \omega < |a| + |b|$,
\item [$-$] the discrete lines said {\bf standard} where $\omega = |a| +
|b|$, this thickness is the smallest one for which the discrete line
is 4-connected,
\item [$-$] the {\bf thick lines} where $\omega > |a| +
|b|$, they are 4-connected.
\end{itemize}

\vskip 7pt
\vtop{
\ligne{\hfill
%\centering
\includegraphics[height=3.5cm]{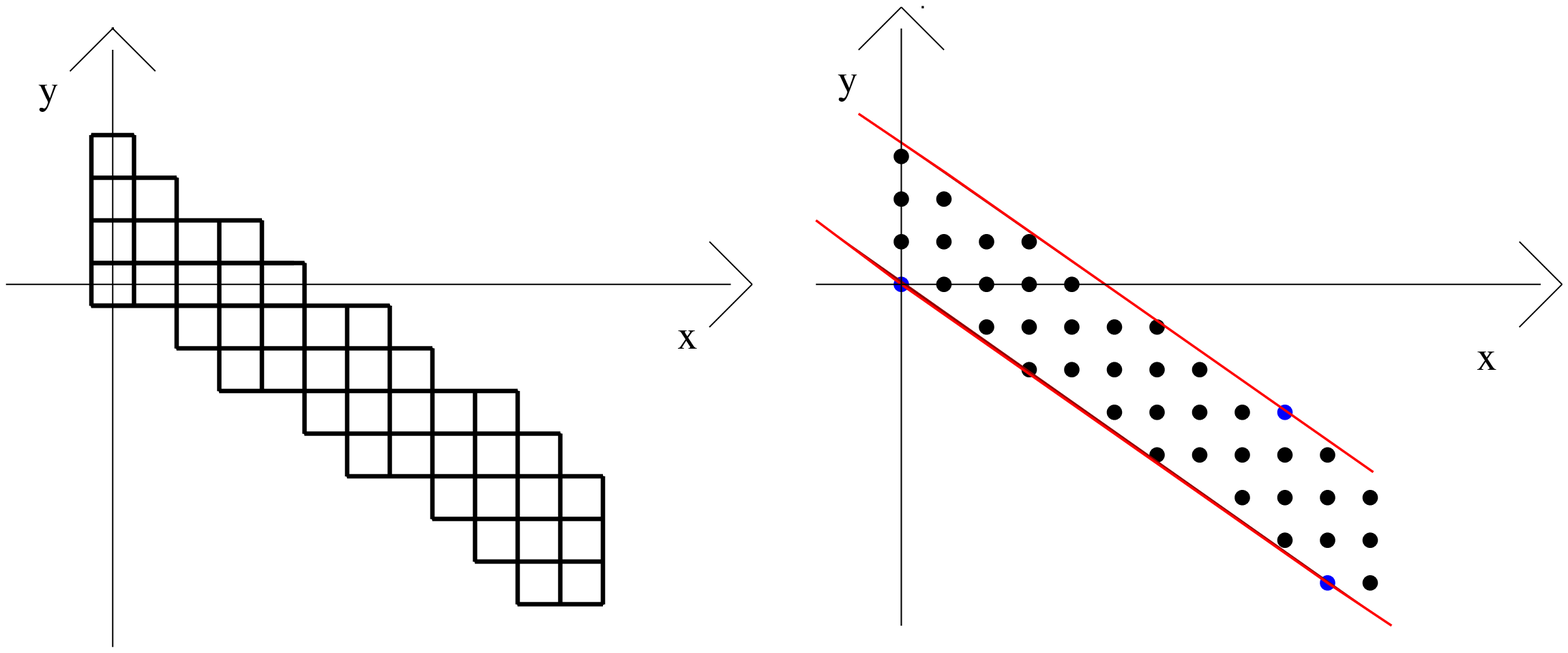}
\hfill}
%\begin{figure}
\begin{fig}
\leurre
%\caption[]{
On the left hand side a representation by pixels (each
integer point is represented by
a square centered at the point) of a segment of the thick line
${\mathcal D}(7,-10,0,34)$ whose equation is
$0 \leq 7x + 10y < 34$, for $x \in [0,10]$, on the right hand side the
points of this line are represented
by disks to get a better visualisation of the leaning lines.%}
%\end{figure}
\end{fig}
}
\vskip 5pt
%\begin{defn}
%interval $[0,l]$, we respectively name $L_F$ and $L_L$ the first and the
%last lower leaning point of this segment, and $U_F$ and $U_L$ the first and
%the last upper leaning point of the segment.

\vtop{
\begin{algo}
\label{algo1}
\leurre
The algorithm for constructing the discrete line \hbox{$0\leq ax-by < b$}
with $0\leq a \leq b$ and $b>0$.
\end{algo}

\def\keyw#1{\hbox{\bf #1}}
\def\iwhile{{\bf while}}
\def\ido{{\bf do}}
\def\iif{{\bf if}}
\def\ithen{{\bf then}}
\def\iendif{{\bf endif}}
\def\iendwhile{{\bf endwhile}}
%\begin {tabbing}
%{\bf The algorithm for constructing the discrete line \hbox{$0\leq
%ax-by < b$}
%with $0\leq a \leq b$ and $b>0$.}\\
%\ \\
\setbox110=\vtop{\leftskip 0pt\parindent 0pt\hsize=200pt
Input: ~$a$, $b$, {\rm characteristics of the discrete line, $n$ number of
points} 
\vskip5pt
{
\obeylines\tt
\leftskip 0pt
\obeyspaces\global\let =\ \parskip=-1pt
$r$ := $0$; $x$ := $0$; $y$ := $0$; $k$ := $1$;
Plotpoint $(x,y)$;
\iwhile{} $k \leq n$ \ido 
   $r$ := $r$ $+$ $a$; 
   $x$ := $x$ $+$ $1$; 
   \iif{} $r \geq b$ 
    \ithen 
       $y$ := $y$ $+$ $1$; 
       $r$ := $r$ $-$ $b$; 
   \iendif;
   Plotpoint $(x,y)$;
   $k$ := $k$ $+$ $1$;
\iendwhile;
\par}
}
\vspace{-13pt}
\grostrait
\ligne{\hfill\box110
\hfill}
\vspace{5pt}
\demitrait
%\end {tabbing}
}
\vskip 10pt

Real straight lines $ax - by = \mu$ et $ax - by = \mu + \omega - 1$
are named the {\bf leaning lines} of the discrete line ${\mathcal
D}(a,b,\mu,\omega)$. An integer point of these lines is named {\bf a
leaning point\/}.
%\end{defn}

The leaning line located above (resp. under) ${\mathcal D}$ in the first
quadrant ($0 \leq a$ and $0 \leq b$) respects
the following equation $ax - by = \mu$ (resp. $ax - by = \mu + \omega -
1$), it is named {\bf {upper leaning line}\/} (resp.
{\bf {lower leaning line}\/}) of ${\mathcal D}$, noted $d_{U}$ (resp.
$d_{L}$).
%If we consider a segment of the line ${\mathcal D}(a,b,\mu,\omega)$ in the
%\begin{defn}
Let $M(x_{M},y_{M})$ be an integer point, the {\bf remainder at the
point $M$} as a function of
${\mathcal D}(a,b,\mu,\omega)$, noted $r(M)$, is defined by:
$$\bf r(M) = ax_{M} - by_{M} $$
%\end{defn}

To simplify the writing, we shall suppose hereafter that {\bf the slope
coefficients verify $0 \leq a \leq b$} which corresponds to the first
octant.

%\begin{algo}
%\label{algo}
%\leurre
%The algorithm for constructing the discrete line \hbox{$0\leq ax-by < b$}
%with $0\leq a \leq b$ and $b>0$.
%\end{algo}

\section{{\Large Cellular Automata}}
\label{CA}

   Devised by Ulam and von Neumann in the late forties, see~\cite{vonNeumann},
cellular automata were studied from various theoretical point of view and
were applied in many different fields as physics, chemistry, biology, economics
and psychology.  Cellular automata are shared by several scientific communities,
mainly physicists, mathematicians and computer scientists. We shall
consider them from the computer science point of view: for us, they are
an algorithmic tool to solve problems. Theoretical computer science
proved the Turing completeness of cellular automata, which means that
they are able to simulate the computation of any Turing machine or,
which is an equivalent formulation, of any partial recursive function
see, for instance \cite{gruska,mazoyer,morita}. They are also considered in
various abstract settings, see~\cite{roka,mmHypCA1,mmHypCA2}.
Accordingly, cellular automata have a great power of simulation, see~\cite{wolframNKS}.
What theoretical computer science tells us is that cellular automata
are more efficient than Turing machines. If the class of traditional cellular 
automata working in polynomial time capture the same algorithms as
the corresponding class of Turing machines and no more, things are different
if we consider specific problems and this matters for us. As an example,
the best algorithms to compute the product of two natural numbers written
in binary has a complexity in $\vert n\vert\log\vert n\vert$, where $\vert n\vert$
is the number of digits in the binary representation of the biggest factor~$n$
in the considered product. With cellular automata, there is a linear algorithm 
in~$\vert n\vert$, see~\cite{atrubin}. While the $\vert n\vert\log\vert n\vert$ 
result involves
non trivial results on Fourier series, the linear algorithm for cellular
automata makes use of a very elementary algorithm: the one which is alike
what children learn at school for multiplying numbers with several digits.
Many interesting aspects of the complexity of cellular automata can be found
in~\cite{gruska,wolfram}.

\subsection{The computation of a cellular automaton}

   Cellular automata consists of a set of {\bf cells}, which is usually
called the {\bf space} of the automaton. The space must be uniform in the sense
that each cell has the same number of neighbours and that the shape of the 
neighbourhood around the cell is the same for all the cells. Each cell is
equipped with a copy of the same finite automaton whose alphabet is called
the set of {\bf states} of the cellular automaton. The transition table
of this automaton defines what we call the {\bf local transition function}
of the cellular automaton. To each neighbourhood of a cell, including the sate
of the cell itself called the {\bf current state} of the cell, the function  
associates a state, called the {\bf next state} of the cell. These names
come from the computation defined for cellular automata as follows.
We have a clock defining a discrete time starting from the {\bf initial} time 
usually called~$0$. At each top of the clock, each cell changes its current
state by taking the new state defined by the local transition function
applied to its neighbourhood. 

   What we have just described is a {\bf deterministic} cellular automaton
as for each neighbourhood, the local transition function defines a single new state.

\subsection{Neighbourhoods}

   The space of the automaton is important. Traditionally, the most studied
cases are the line, identified with~$Z\!\!\!Z$, as an integer can be given
to each cell which is called its {\bf coordinate}, and the Euclidean plane, 
identified with~$Z\!\!\!Z^2$.
The neighbourhood of the cell can be defined in very different ways.
For the line, we shall take what is called the symmetric neighbourhood
of radius~1. This means that the neighbours of the cell with coordinate~$x$,
we shall later say {\bf the cell~$x$}, are the cells~$x$$-$$1$ 
and~$x$$+$$1$. As mentioned above, the neighbourhood of~$x$ thus consists
of~$x$$-$$1$, $x$ and~$x$$+$$1$. 

   In the Euclidean plane, there are traditionally two kinds of neighbourhoods.
If the coordinate of a cell is $(x,y)$, its von Neumann neighbourhood 
consists of the cells $(x,y)$, $(x,y$$+$$1)$, $(x$$-$$1,y)$,
$(x,y$$-$$1)$ and $(x$$+$$1,y)$. This neighbourhood is illustrated by
the left-hand side picture of Figure~\ref{voisinages}. There is another 
neighbourhood which is also much used, for instance in the {\it Game of Life}, 
which is called Moore neighbourhood. Together with the previous neighbours, 
the Moore neighbourhood of~$(x,y)$ also contains the cells $(x$$-$$1,y$$+$$1)$,
$(x$$-$$1,y$$-$$1)$, $(x$$+$$1,y$$-$$1)$ and $(x$$+$$1,y$$+$$1)$.
In Figure~\ref{voisinages}, the neighbourhood is illustrated by the right-hand side
picture.

   Traditionally, alternative names are also given to the neighbours of a 
cell~$(x,y)$ in its von-Neumann neighbourhood: $(x,y$$+$$1)$ is
the {\bf northern} neighbour, $(x$$-$$1,y)$ is the {\bf western} one,
$(x,y$$-$$1)$ is the {\bf southern}  one and $(x$$+$$1,y)$ is the {\bf eastern}
one. These names allow us to not mention the coordinates and we shall use them.
We shall also say that the cell~$(x,y)$ sees $(x,y$$+$$1)$ through its
{\bf northern side}, $(x$$-$$1,y)$ through its {\bf western side},
$(x,y$$-$$1)$ through its {\bf southern side} and $(x$$+$$1,y)$ through
its {\bf eastern side}. Note that these notions are the same as those of
4- and 8-connectedness, see Section~\ref{geom}. More precisely, 4-connectedness
corresponds to von Neumann neighbourhood and 8-connectedness corresponds to Moore
neighbourhood.

\vskip 10pt
\vtop{
\ligne{\hfill\scalebox{1.10}{\includegraphics{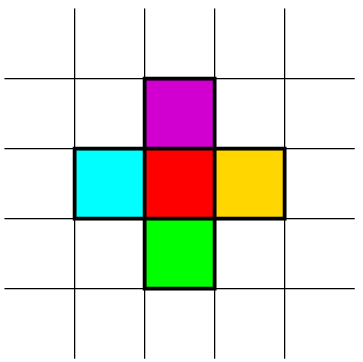}}
\hfill\scalebox{1.10}{\includegraphics{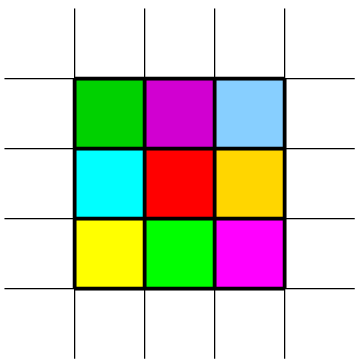}}
\hfill}
\begin{fig}\label{voisinages}
The neighbourhood of a cell in the von Neumann definition.
\end{fig}
}
\vskip 10pt
   We can write the transition function by taking the list of the state
of the neighbours, say the cell, north, west, south and east, which means that
we counter-clockwise turn around the cell, and to such a sequence in this order,
define a state. Such a list of these six states is called a {\bf rule}.
Accordingly, the local transition function can also be represented as a
{\bf table of rules}. We shall adopt this point of view in the rest of the
paper.

   A last but not least notion have to be introduced: the notion of 
{\bf configuration} which is essential in cellular automata.
Formally, it is an application of the space into the set of states of the
automaton. If we apply the local transition function, we define a
new configuration. Going from one configuration to a new one by applying
the rules defines a new function, this time from the set of configurations
into itself which is called the {\bf global function} of the cellular automaton.

   However, we shall not look at the succession of the configurations
in this way, which is the way mathematicians look at them. We shall devise
them one by one, which is a very different point of view.

\subsection{Programming with cellular automata}

   Contrarily to what might suggest the formal definition of cellular automata,
programming a concrete cellular automata never starts by writing the table
of the rules. Programming with cellular automata is a programming through the
data. We have to initially distribute them in an appropriate way and then
look at how we can change this initial configuration to the final one
which represents the solution of our problem for the instance defined by the 
initial configuration.

    This transformation of the initial configuration into the final one
usually involves many steps and except for very small configurations and for
short interval of times, we cannot see all of them in a single glance.
We have to split this path from the initial configuration to the final one
into stages, sometimes into sub-stages and then for these sub-stages,
we can imagine the evolution step by step from the starting point of the
sub-stage to its conclusion.

    We have to see the states of the cellular automaton as colours,
and the changes on the configurations as a kind of painting. But this
painting is moving, it can change parts already painted in one colour into
another one. And in the painting, some part of it can be
interpreted as a {\bf signal} sent from a part of the data to another one in
order to trigger some action. A typical example is the occurrence of a state
somewhere in the data, and we can see that, after a certain time, a
part of the data completely changed their initial colour to another one. 
The writing of the table arrives as almost the last point: when we arrive
to these sub-stages where it is possible to see step by step the transformation
form a configuration to the next one. Usually, in this
step by step transformation, not all cells change their state at the next step
but only a few of them: this allows us to isolate the rules we need for our
table by looking at the neighbourhood of a cell before it changed and the new state
of the state when it changed. In such an approach, if the problem is not very complex,
and for tiny configurations, this can be done by hand. But when it is the case
to check the validity of the rules by applying them to larger configurations,
a computer program is absolutely needed. There are two reasons for that.
First, as our cellular automaton is deterministic, we have to be sure that the 
set of rules does not contain contradictory rules. This means that if two 
rules give different next states, they must also be different at least in one 
of the members of the neighbourhood, the cell itself belonging to the neighbourhood.
Second, when starting from an initial configuration which correctly implements
an instance of our problem, the computation using our table of rules must
lead to a correct implementation of the implementation of the solution.
It is important to indicate here that we assume the initial configuration
to be a correct one: the cellular automaton is devised for them and it does not
check whether the initial configuration is correct or not.

%   We shall see how we deal with these different points in the next section where
%we implement the algorithm described in Section~\ref{geom}.

    In the next section, we give a simplified version of the scenario. We call it
{\it naive} as it clearly separates the various operations which are performed by
the automaton.

\section{{\Large The scenario of the implementation: a naive version}}
\label{scenario}

   From our previous section, we know that our present task is to imagine
a sequence of configurations, from the very initial one to the final one which, in
an informal sense are {\bf key configurations}.

   They are illustrated by Figures~\ref{kconfigpos} and~\ref{etape0neg} for
the computation of the line \hbox{$\mu\leq ax -by < \mu$+$b$}. 
Figure~\ref{kconfigpos} illustrates the case when $\mu\geq 0$ and
Figure~\ref{etape0neg} illustrates the case when $\mu < 0$. The corresponding 
situations are in Sub-section~\ref{scenarpos} and Sub-section~\ref{scenarneg}
respectively.

   In each sub-section, we consider a {\bf cycle} of the computation which
consists in appending a new pixel to the part of the line which is already drawn
by the automaton. Accordingly, the whole work of the automaton is a loop in which
each turn consists in performing such a cycle. 
In these sub-section, the first configuration of a {\bf cycle}
is called the {\bf starting configuration} of the cycle. It is characterized by 
the
position of the data with respect to the part of the line already present. 
In both Sub-section, the data consists in three segments which we call {\bf rows},
the $U$-row, the $V$-row and the $R$-row. Each row consists of cells in the same
state: $U$, $V$ and~$R$ for the $U$-, $V$- and~$R$-row respectively. 
The number of $U$'s and $V$'s is the value of~$a$ and~$b$ respectively in the
equation \hbox{$\mu\leq ax-by < \mu$+$b$}. The number of~$R$'s is the 
the value of the parameter which controls the drawing. At the beginning of the 
cycle, this value is the result~$r$ yielded by the previous cycle. At a 
certain point of the current cycle, the number of~$R$'s will be \hbox{$r+a$}. 
The rest of the cycle will be determined by the comparison of this value 
with \hbox{$\mu+b$}. At last, there is a cell in the state~$W$ which
is the first element of a structure used by the computation. This cell
is placed as both the eastern neighbour of the last written~$X$ and the
northern neighbour of the first element of the $U$-row.

   The three rows are placed one above another in the following order: first,
the $U$-row, below the $V$-row and below again, the $R$-row.
The $V$-row is shifted with respect to
the $U$-row by a number of cells which is the value of~$\mu$: to the right
if $\mu>0$, to the left if $\mu<0$. When $\mu=0$, the $V$-row is aligned with
the others. Also, the position of the $R$-row depends on the sign of~$\mu$, as
well as the number of~$R$'s of which it consists. We shall see that these 
dispositions of the data induces a different working of the automaton at some 
point of the cycle.

\subsection{The case when $\mu$ is non-negative}
\label{scenarpos}

   The starting configuration is given by the first picture of 
Figure~\ref{kconfigpos}.
  
   We notice that the $R$-row is aligned with the $U$-row, but the number of~$R$'s
in the starting configuration is always at least the value of~$\mu$. We also notice 
the presence of a~$W$ to the east of the last~$X$ of the line and to the north of
the first $U$ of the $U$-row. It is the first element of the future $W$-column.

   In this naive representation, the first step of the cycle consists in moving the
data by one step to the east. To this purpose, the automaton creates the $W$-column,
see the second picture of Figure~\ref{kconfigpos}, which erases the first cell
of the~$U$- and the $R$-rows and, when $\mu=0$, the first cell of the $V$-column.
This triggers a process which we shall later describe which pushes the data by one step
to the east. The colours of the rows are changed: $U$ to~$U1$, $V$ to~$V1$
and $R$ to~$R_1$. In this process, the last cell of the $V$-row, at its new place,
is marked as~$V2$. In the $U$- and $R$-rows, the new last element is not marked.

   The next step consists in computing \hbox{$r+a$}. This is obtained by moving a
copy of each cell of the $U$-row and to append this copy to the eastern end of the
$R$-row. This copy is a new~$R$. We shall later describe precisely how this is 
performed. When the computation
is completed, the comparison with \hbox{$\mu+b$} is given by the position
of the last copied~$U$ with respect to~$V2$. This last~$R$ of the $R$-row
can see~$V1$ through its northern side, in which case \hbox{$a$+$r < \mu$+$b$},
or it can see~$V2$, in which case \hbox{$a$+$r = \mu$+$b$} or it can see a blank,
which means that \hbox{$a$+$r > \mu$+$b$}. In both latter cases, we have
to subtract a bloc of $b$~$R$'s from the $R$-row. This is illustrated by 
Figures~\ref{etape10G} to~\ref{etape13G}. Then, in all situations, we have to 
transform the
configuration into the starting one of the next cycle. This is illustrated by the
last pictures of Figures~\ref{kconfigpos} and Figures~\ref{etape14G} 
to~\ref{etape17G}. 

\vskip 10pt
\vtop{
\ligne{\hfill
%\scalebox{0.50}{\includegraphics{etape1_S.eps}}
\scalebox{0.50}{\includegraphics{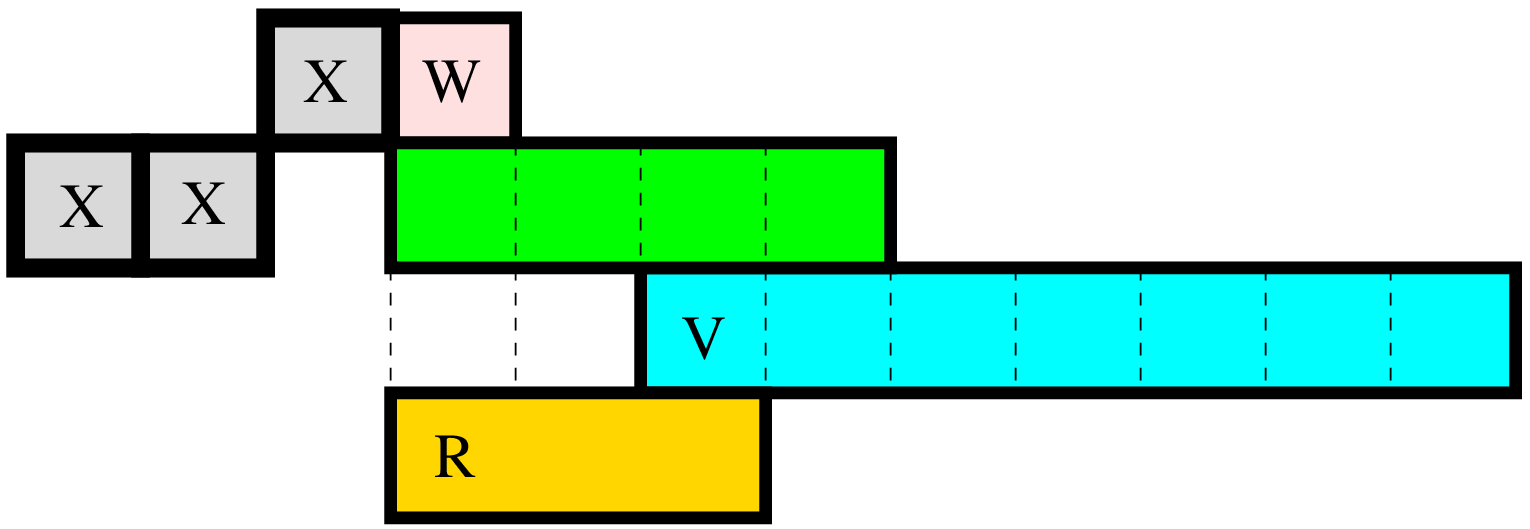}}
\hfill}
\vskip 10pt
\ligne{\hfill
%\scalebox{0.50}{\includegraphics{etape2_S.eps}}
\scalebox{0.50}{\includegraphics{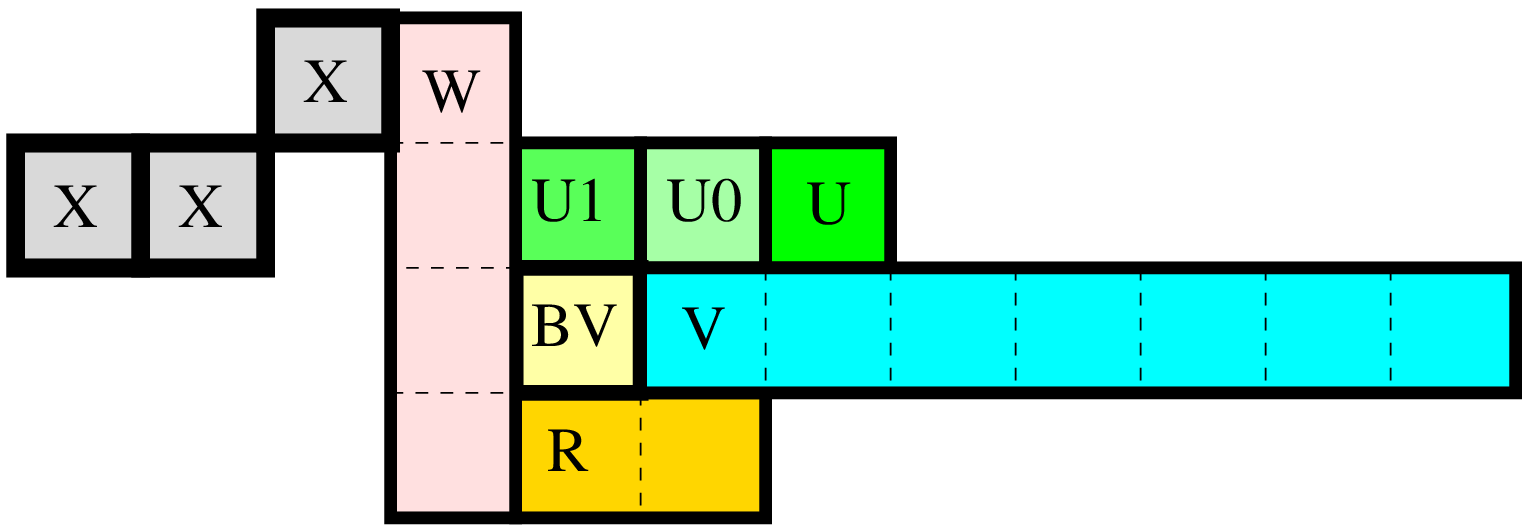}}
\hfill}
\begin{fig}\label{kconfigpos}
\leurre
Two key configurations: the starting one and the configuration when
the $W$-column is completed.Note that the second configuration shows
a typical phenomenon of computation with cellular automata: the possibility
to simultaneously perform transformations which are independent.
\end{fig}
}

   In the following paragraphs, we give the outline of each specific operation
we defined in the above description. A few of them are also used in the case when
$\mu<0$, so that in Sub-section~\ref{scenarneg}, we shall not repeat them.
   
\subsubsection{Shifting the data by one step to the east}
\label{shift}

    As suggested by the second picture of Figure~\ref{kconfigpos}, the first
action performed by the automaton is to construct the {\bf $W$-column}.
As in the case of the $U$-, $V$- and $R$-rows, it consists of a vertical block
of cells in the state after which the column is called. Note that even when
the content of the cell is not~$W$ we shall still say that it is a cell of 
the $W$-column.

    This structure deletes the first cell of the $U$- and $R$-rows, also of the
$V$-row when $\mu=0$. This is to materialize a part of the path that has to be 
followed by the copies of the cells of the $U$-row. Each time $W$~erases $U$,
$V$ or~$B$, $R$ or~$B$, it triggers the process of shifting the corresponding
row by one step to the east. 

   When the process is completed, we obtain the configuration illustrated
by Figure~\ref{etape3}. Later, we look how the process goes on each row
in a detailed way.

\vskip 10pt
\vtop{
\ligne{\hfill
\scalebox{0.50}{\includegraphics{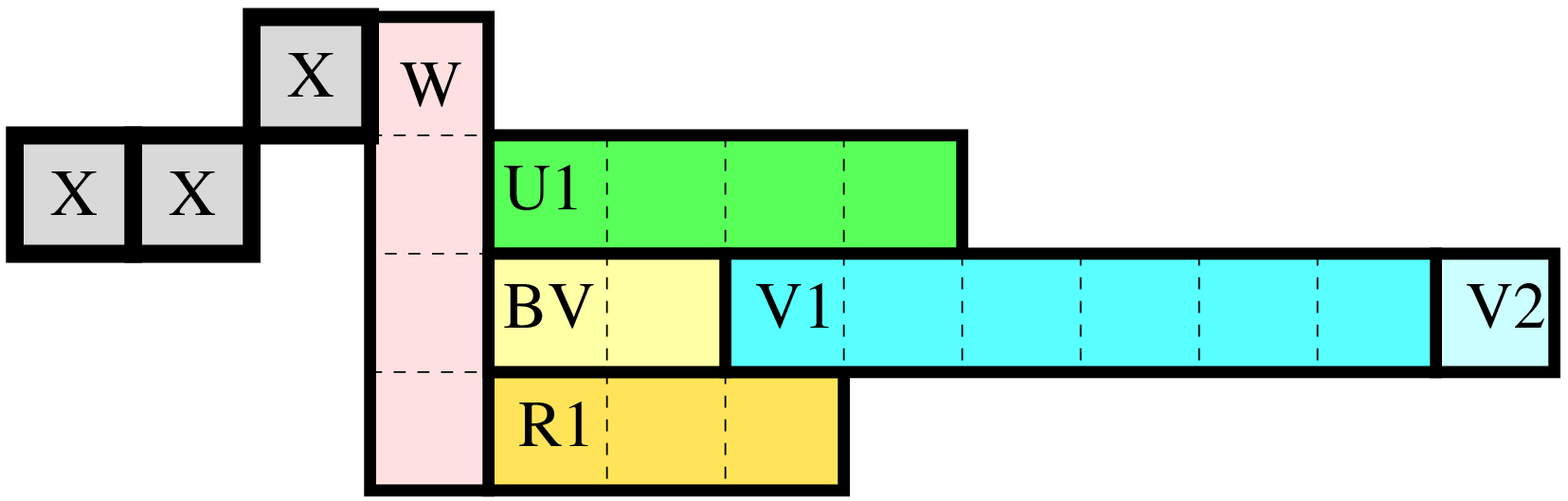}}
\hfill}
%\vskip 10pt
%\ligne{\hfill
%\scalebox{0.50}{\includegraphics{etape2_S.eps}}
%\hfill}
\begin{fig}\label{etape3}
When all the data have been shifted by one step to the east.
\end{fig}
}

   First, consider the case of the~$V$-row in which the process is slightly
different. If $W$ sees~$B$ through its eastern side,
then it transforms~$B$ into~$BV$. This state goes from one~$B$ to the next one
until $BV$~can see~$V$ through its western side. Then, $BV$ transforms 
the first~$V$ into~$CV$ which afterwards becomes~$BV$: it will be again~$B$
when turn to the next starting configuration will be in process. Now,
as one~$V$ was removed, it must be created at the other end of the $V$-row.
To this purpose, the~$V$ which sees~$CV$ through its western side becomes~$V1$
and this state propagates step by step to all the elements of the~$V$-row.
When the last~$V$ has changed to~$V1$, its eastern neighbour~$B$ can see~$V1$
through its western side. As a consequence, this~$B$ becomes~$V2$. Now, the
set of~$V1$'s and~$V2$ has the same length as the initial $V$-row.

   Now, let us look at the $U$- and $R$-rows. The shift by one step to the east
is performed in the same way in both cases. The elements of the $U$- and $R$-rows
are changed to~$U1$ and to~$R1$ respectively. But this change is not performed
in the same way as with the $V$-row. The reason is that in this case, we do not
mark the last element because the row must be uniform after the change.
We proceed as follows, considering the $U$-row. Each~$U$ is transformed into~$U0$,
which, at the next time, becomes~$U1$. The propagation is triggered by~$U0$:
when $U$ sees~$U0$ through its western side, it becomes~$U0$. The process starts
with~$W$: when the second~$U$ sees $W$~through its western side, it becomes~$U0$.
The process is stopped by~$B$: when $B$ sees~$U0$ through its western side, it
becomes~$U0$, which restores the~$U$ which was erased by the $W$-column. When
the next~$B$ sees~$U1$ through its western side, it remains~$B$, which stops the
process.  

We can represent these transformations by simple $1D$-rules as they happen
on a line. The format of the rules is \hbox{$\eta_0\eta_g\eta_r\eta^1_0$}, 
where $\eta_0$ is the current state of the cell, $\eta^1_0$ is its new state, 
$\eta_g$ and $\eta_r$ are the states of the left-
right-hand side neighbours respectively. In the case of the $V$-row,
we obtain the following rules:
\vskip 7pt
\ligne{\hfill{\tt V CV V V1},\hskip 7pt {\tt V V1 V V1}, 
\hskip 7pt{\tt V V1 B V1},\hskip 7pt
{\tt B V1 B V2},\hfill}

\ligne{\hfill 
{\tt V1 V V V1}, \hskip 7pt {\tt V1 V B V1}, 
\hskip 7pt {\tt BV V1 V V1},
\hskip 7pt {\tt V1 V1 V V1}, 
\hfill}

\ligne{\hfill 
{\tt V1 V1 V1 V1}, 
\hskip 7pt {\tt V1 V2 B V1}, \hskip 7pt {\tt B V2 B B}.\hfill} 

   The rules of the first row are called {\bf transformation rules}: the state
of the current cell is changed. These rules perform the transformation. The rules
of the next two rows are called {\bf conservative rules} as the current state
is not changed by the application of the rule.

   For the $U$-row, the rules are:

\vskip 7pt
\ligne{\hfill
{\tt U U0 U U0}, \hskip 7pt {\tt U0 U1 U U1}, \hskip 7pt {\tt U W U U0},
\hskip 7pt {\tt U0 W U U1},
\hfill}

\ligne{\hfill
\hskip 7pt {\tt U U0 B U0},
\hskip 7pt {\tt U0 U1 B U1}
\hfill}

\ligne{\hfill
{\tt U1 W U0 U1},
\hskip 7pt {\tt U1 U1 U0 U1},
\hskip 7pt {\tt U1 W U1 U1},
\hskip 7pt {\tt U1 U1 U1 U1}
\hfill}

\ligne{\hfill
{\tt U1 U1 B U1},
\hskip 7pt {\tt B U0 B U1},
\hskip 7pt {\tt B U1 B B}
\hfill}

   Here, we can see that the first two lines consist of transformation rules
and that the next two lines consist of conservation rules. For the $T$-row,
we have the same rules as above, replacing $U$, $U0$ and~$U1$ by~$R$, $R0$
and~$R1$ respectively.

   Before turning to the next stage of the computation, let us remark that
these transformations performed on the~$U$-, $V$- and $R$-rows are performed 
simultaneously. However, they do not start at the same time and, also, they
do not complete at the same time. It is not difficult to see that as long
as $b>a$ and $\mu\geq 0$, when $V2$~appears, the elements of the $U$-row
are all $U1$ and those of the~$R$-row are all $R1$. 
  
\subsubsection{Appending $a$ to~$r$}    
\label{add}

   The appearance of~$V2$ is the end of the shift of the data by one step
to the east. It also triggers the start of the next stage: appending~$a$ to~$r$.
As indicated at the beginning of section~\ref{scenario}.

   The addition is obtained as a sequence of incrementations of the $R$-row
as many times as the length of the~$U$-row. A copy of each element of the~$U$ row
is transported from this elements to the current end of the $R$-row. We presently describe this process.

   When $V2$ appeared, its northern neighbour changes its state from~$B$ to~$C$.
This $C$~is a signal sent on the line of the $U$-row to the eastmost~$U1$ in order
to start the copying process. As $C$ starts its travel step by step to west,
$V2$~changes to~$V3$ in order to produce a signal~$C$. This $V3$~allows the whole
$V$-row to wait the next step raised by the comparison of $a$+$r$ with $\mu$+$b$.

   When traveling to the $U$-row, $C$~obeys very simple rules:
\hbox{\tt B B C C}, \hbox{\tt C B B B}, \hbox{\tt B C B B} until $U1$ is met.
Figure~\ref{etape4_6} illustrates two important configurations: when $C$
and~$V3$ are first present and then when~$C$ reaches the $U$-row with
the effect on the $U$-row.

   When $U1$ is meet by~$C$, it is changed to $U2$, see Figure~\ref{etape4_6},
and this~$U2$ crosses the $U1$'s in 
the same way as $C$ crossed the blanks. Now, the first $U2$ turns to~$U3$, 
which means that the copy is in process. This $U3$ does not affect its 
western $U2$ neighbour and is changed to~$U4$ at the next time. Now, when 
this $U2$ turns back to~$U1$, this~$U1$ sees~$U4$ through its eastern
side, which means that $U1$~has to be copied: it becomes~$U2$,
see  Figure~\ref{etape7}. This new $U2$ moves
again to the west as the previous one. And so, when it sees~$U4$ through its
eastern side, each~$U1$ is changed to~$U4$ in a cycle of three steps:
\hbox{$U1\rightarrow U2\rightarrow U3\rightarrow U4$}. When $U4$~is reached,
the cell remains in that state until the next stage and the occurrence of~$U4$
triggers the same cycle for the western neighbour of the cell. Note that $U3$
introduces a delay between the copies of the elements. This delay is needed in
order to create new copies of~$U1$'s. Without it, $U1$'s would make travel a
single~$U2$.

\vskip 10pt
\vtop{
\ligne{\hfill
\scalebox{0.50}{\includegraphics{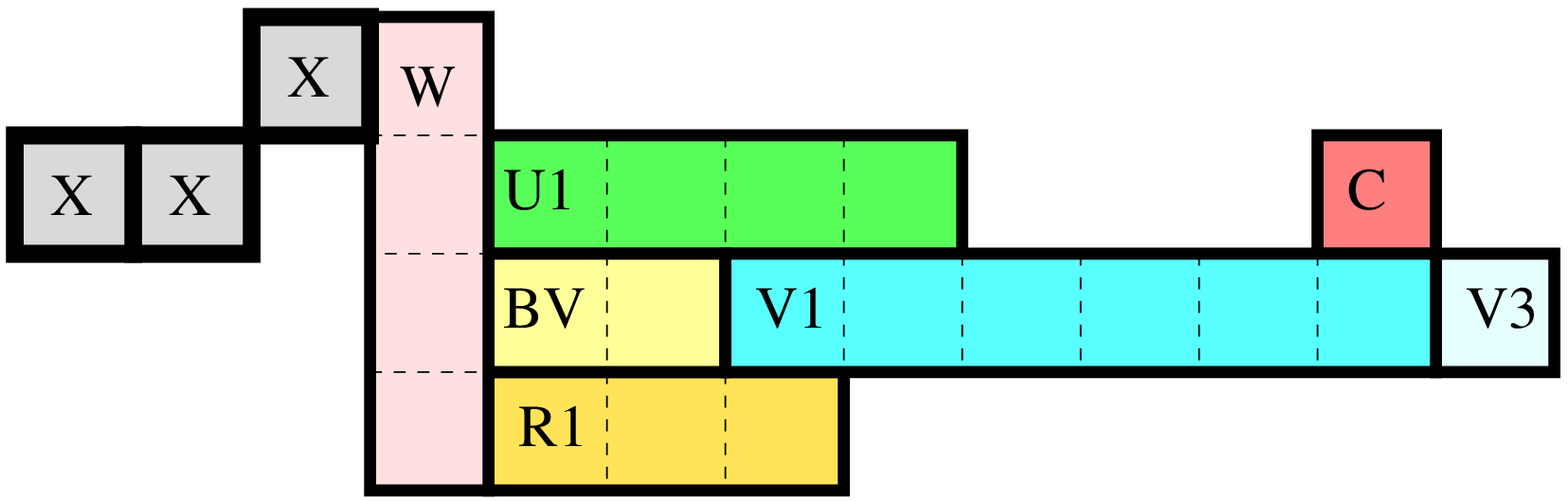}}
\hfill}
\vskip 10pt
\ligne{\hfill
\scalebox{0.50}{\includegraphics{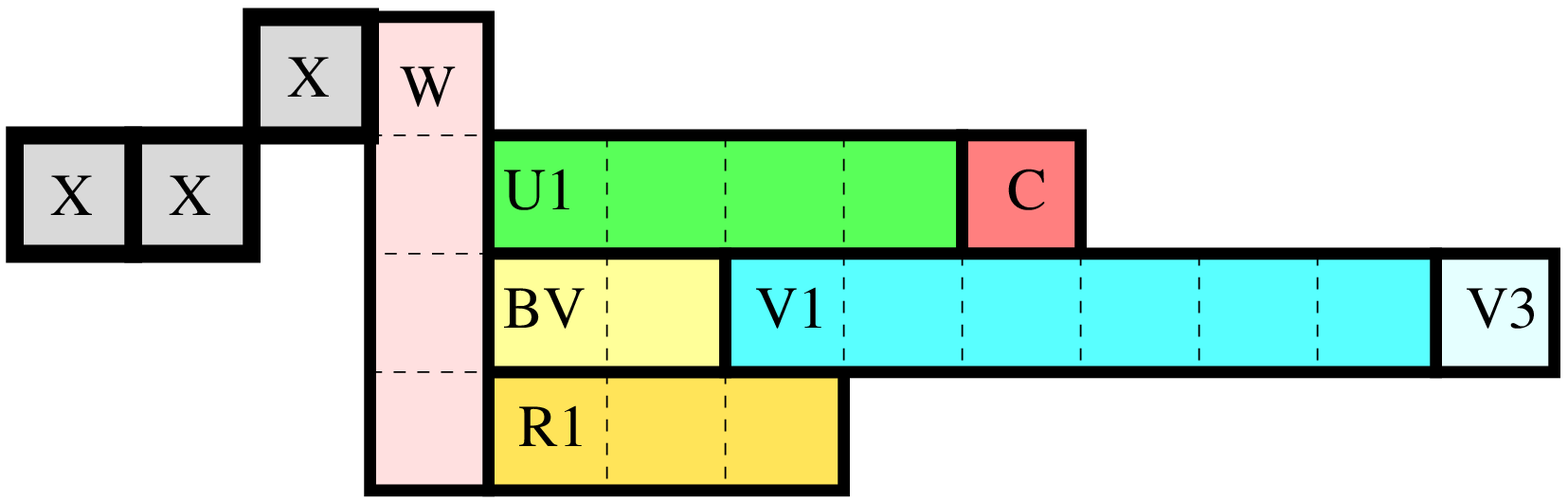}}
\hfill}
\vskip 10pt
\ligne{\hfill
\scalebox{0.50}{\includegraphics{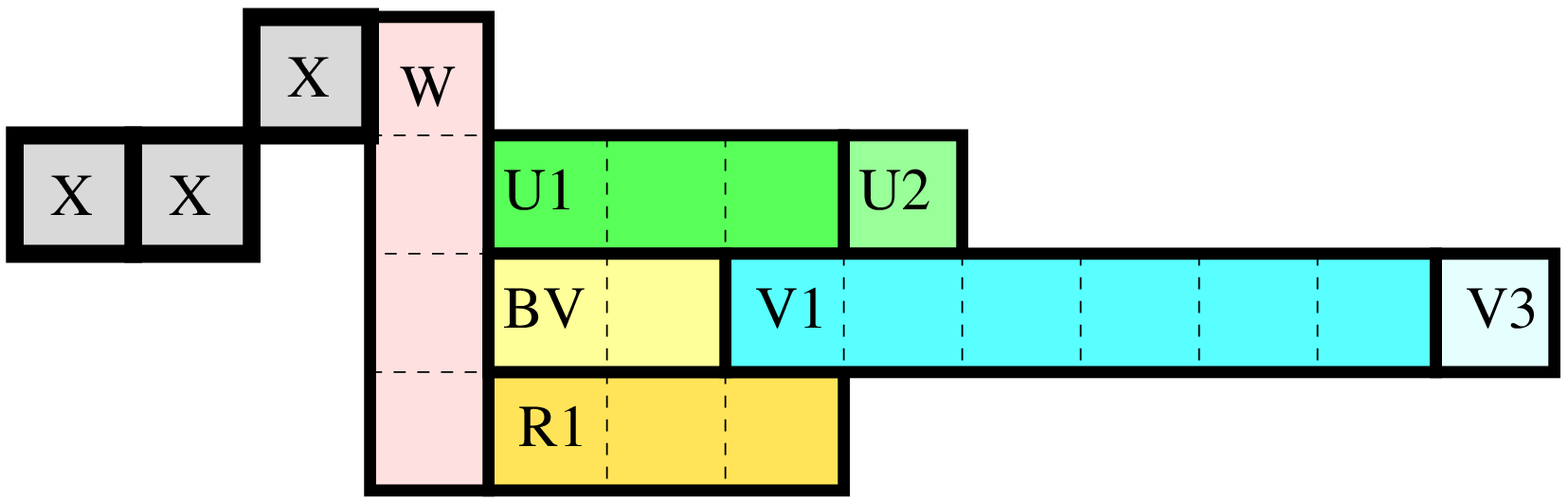}}
\hfill}
\begin{fig}\label{etape4_6}
The appearance of~$C$ together with~$V3$ and the situation when~$C$ reaches
the $U$-row: marking of the rightmost~$U1$ as~$U2$.
\end{fig}
}

   When the traveling~$U2$ reaches the $W$-column, it is transformed into~$R2$:
the corresponding~$W$ of the $W$-column becomes~$R2$ when it sees $U2$ through
its eastern side. Next, $R2$~goes down in the $W$-column in the same way as 
$U2$~moved across the block of $U1$'s. And so,
this $R2$ arrives as the western neighbour of the first~$R1$ of the $R$-row.
Now, $R2$ moves to the east across the~$R1$'s until it reaches the $B$'s:
when the most western~$B$ on the east of the $R$-row sees $R2$ through
its western side, it changes to~$R1$: the corresponding element of~$U$ has been
copied.

   This process goes on as long as the most western~$U4$ triggers the transformation
of its western neighbour~$U1$ into~$U2$. When the block of $U4$~reaches the 
last~$U1$, this~$U1$ is directly transformed into~$U3$ in order to signalize the
$W$-column that it now receives a copy of the last element of the $U$-row. indeed,
when the corresponding element of the $W$-column sees~$U3$ through its eastern
side, it becomes~$R3$, see Figure~\ref{etape8}. From this time, $R3$~travels 
exactly as~$R2$, so that
after a certain time it arrives at a position where it can sees~$B$ through the
eastern side. And now, this $B$-cell knows in which situation we are. 
We study this point in the next sub-subsection.

\vskip 10pt
\vtop{
\vskip 10pt
\ligne{\hfill
\scalebox{0.50}{\includegraphics{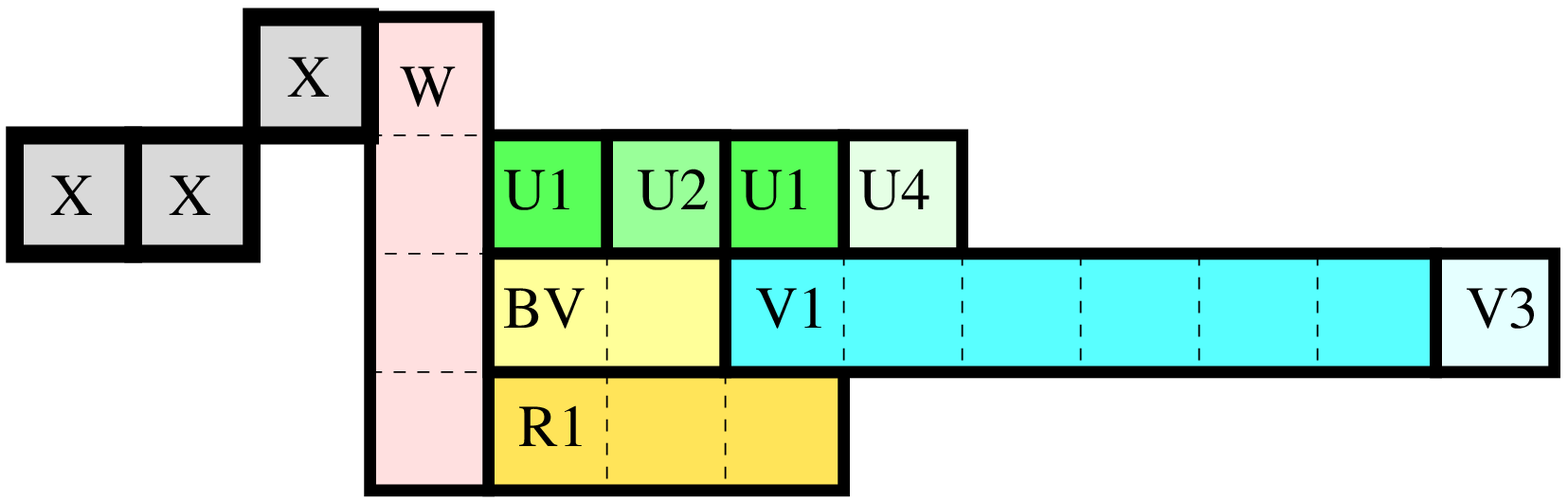}}
\hfill}
\begin{fig}\label{etape7}
The marking of the elements of the $U$-row for copying them.
\end{fig}
}

   Consider a copy of~$U1$ which moves to the west through the remaining $U1$'s 
of the $U$-row. The first one which meets the $W$-column transforms it into $R1$'s.
In order to keep track of the copy, $W$ is first transformed into~$R2$ which then
turns to~$R1$. Now, $R2$ travels through $W$'s and $R1$'s as $C$ through the blanks.
Simply, it goes to the south or to the east. The next $U2$'s which meet the $W$-column
first fall across $R1$ which is thus transformed into~$R2$ in order to convey the
copy further, this very cell becoming $R1$ back at the next time. 

\vskip 10pt
\vtop{
\vskip 10pt
\ligne{\hfill
%\scalebox{0.50}{\includegraphics{etape8_S.eps}}
\scalebox{0.50}{\includegraphics{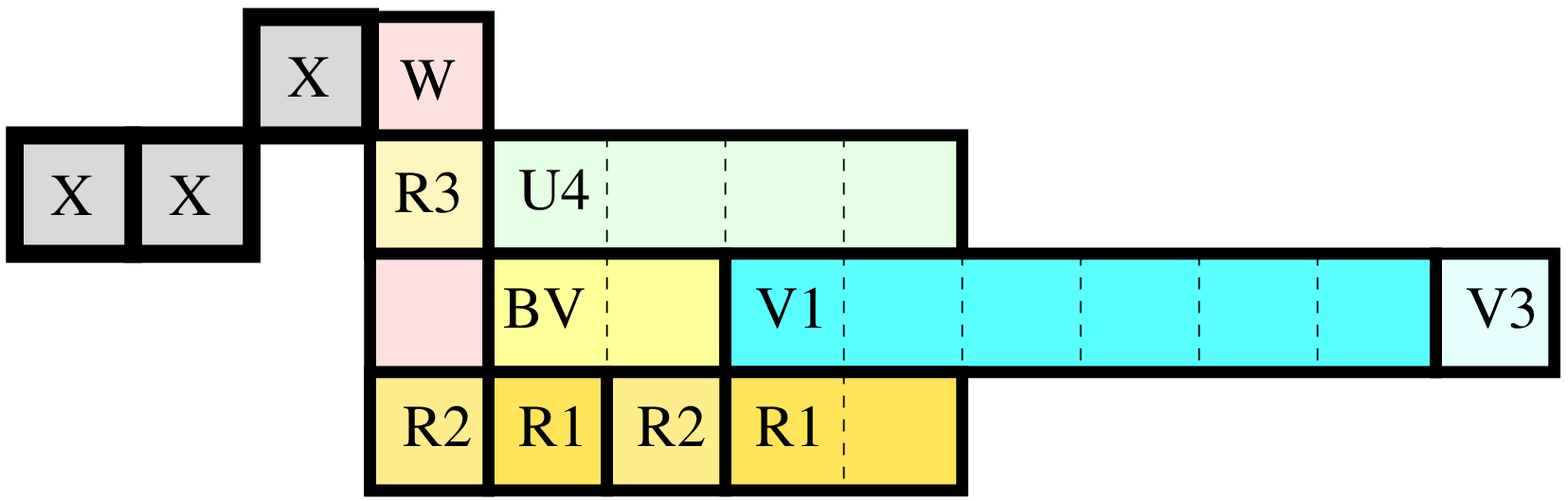}}
\hfill}
\begin{fig}\label{etape8}
When all $U$'s has been transformed into~$U4$, $R3$ starts its travel to the end 
of the $R1$'s.
\end{fig}
}

    When $U3$ meets the $W$-column, $R1$ is then transformed into~$R3$ which behaves
on the path of $R1$'s as~$R2$. 

   When $R3$ arrives as a western neighbour of a~$B$, this means that 
\hbox{$a$+$r$} is materialized and the comparison with \hbox{$\mu$+$b$} can take
place.

\subsubsection{Comparison with \hbox{\rm $\mu$+$b$} and subsequent actions}

   Indeed, when~$B$ sees~$R3$ through its western side, the state of its northern
neighbour indicates him whether \hbox{$a$+$r < \mu$+$b$}, 
\hbox{$a$+$r = \mu$+$b$} or \hbox{$a$+$r > \mu$+$b$}. In the first case,
the northern neighbour is~$V1$, in the second case it is~$V3$ and, in the third
one, it is~$B$.  Then the blank cell
becomes $RR$, $Z$ or~$R$ respectively.

\vskip 10pt
\vtop{
\vskip 10pt
\ligne{\hfill
%\scalebox{0.50}{\includegraphics{etape9_S.eps}}
\scalebox{0.50}{\includegraphics{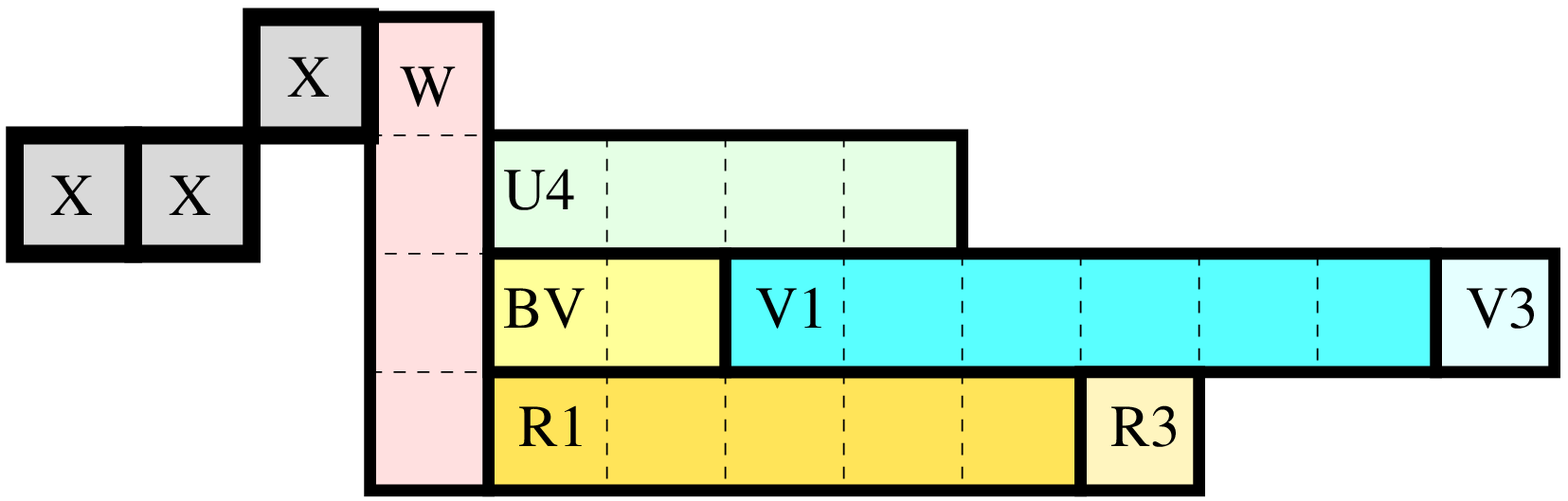}}
\hfill}
\vskip 10pt
\ligne{\hfill
\scalebox{0.50}{\includegraphics{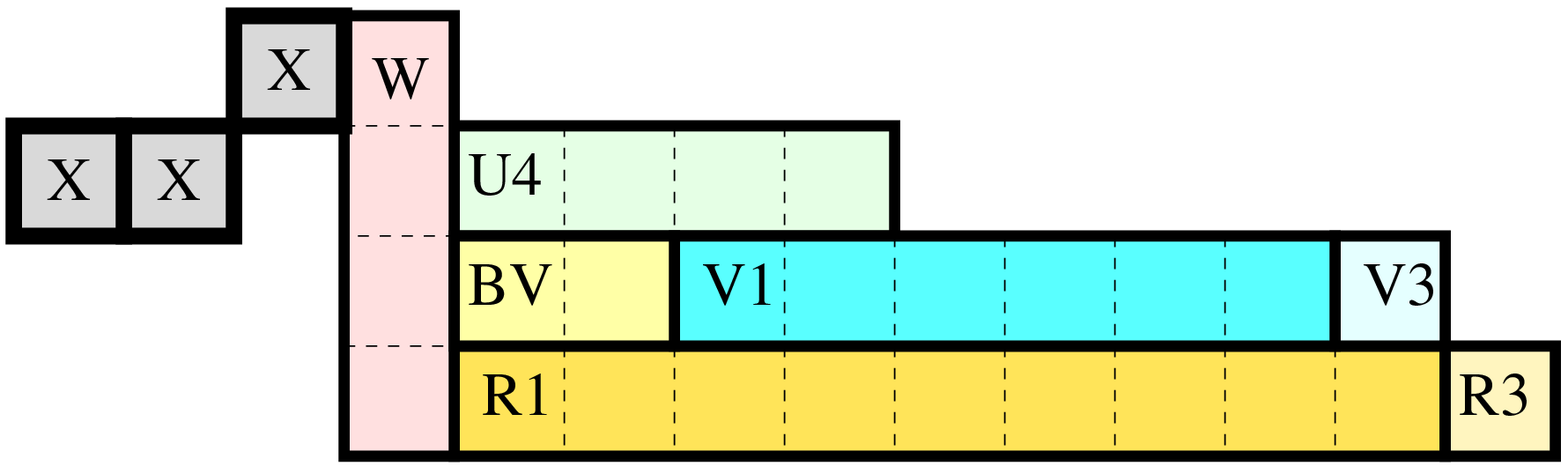}}
\hfill}
\vskip 10pt
\ligne{\hfill
\scalebox{0.50}{\includegraphics{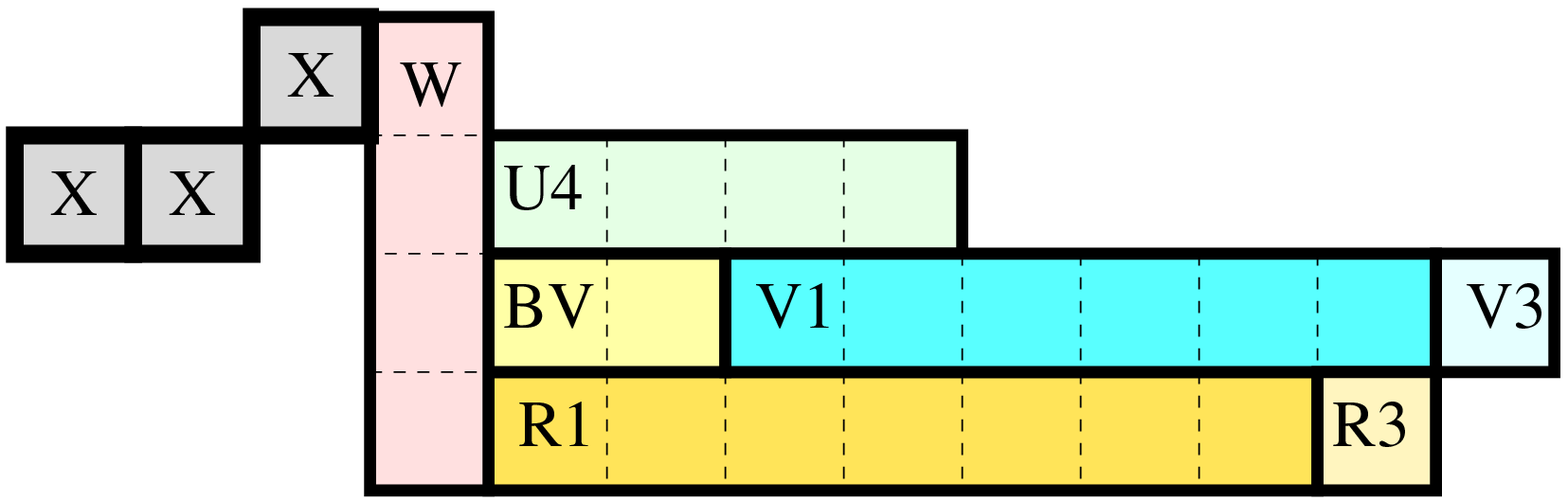}}
\hfill}
\begin{fig}\label{etape9}
Here, $R3$ arrives at the blank. The figure represents all the possible cases,
depending on what is seen by the eastern neighbour~$B$ of~$R3$ through
its northern side.
First row: $B$ sees~$V1$, hence \hbox{$a$$+$$r < \mu$$+$$b$}. 
Second row: $B$ sees the blank, hence \hbox{$a$$+$$r > \mu$$+$$b$}.
Third row: $B$ sees~$V3$, hence \hbox{$a$$+$$r = \mu$$+$$b$}. Of course,
this concerns different cycles.
\end{fig}
}

   Figures~\ref{etape9} illustrates the three cases.
We successively consider in each case what is the transformation from this
situation to the next starting configuration.

   First, we consider the case when \hbox{$a$$+$$r < \mu$+$b$} as, in this
case, it is not needed to subtract~$b$ from the result of the computation
of \hbox{$a$$+$$r$}.

\subsubsection{The case $a$$+$$r < \mu$+$b$}

   And so in this case, the new data is correct. We have simply to erase the 
marks in order to get true $U$-, $V$- and $R$-rows. The first idea would be 
that $RR$~dispatches the transformation of~$V1$ back to~$V$ and the of~$U4$ 
to~$U$ by {\it contamination}. And then the signal
sent from~$RR$ would reach the bottom of the former $W$-column, a signal would
go up in order to place the new~$X$ at the right place and a new cycle
could start. But this propagation process could be long if 
$b$~is very big, so
that a new cycle could start before the complete restoration of all~$U$'s
and $V$'s. In order to avoid such a situation, $RR$~triggers a signal to the right
which will circumscribe the configuration by looking at the end of the $V$-row, then
go back to the $U$-row and inspect it from the just above row, so that the switch
to the next cycle will be obtained when the initial $U4$~of the $U$-row sees
the signal coming from this circumscribing motion through its northern side.

\vtop{
\vskip 10pt
\ligne{\hfill
\scalebox{0.50}{\includegraphics{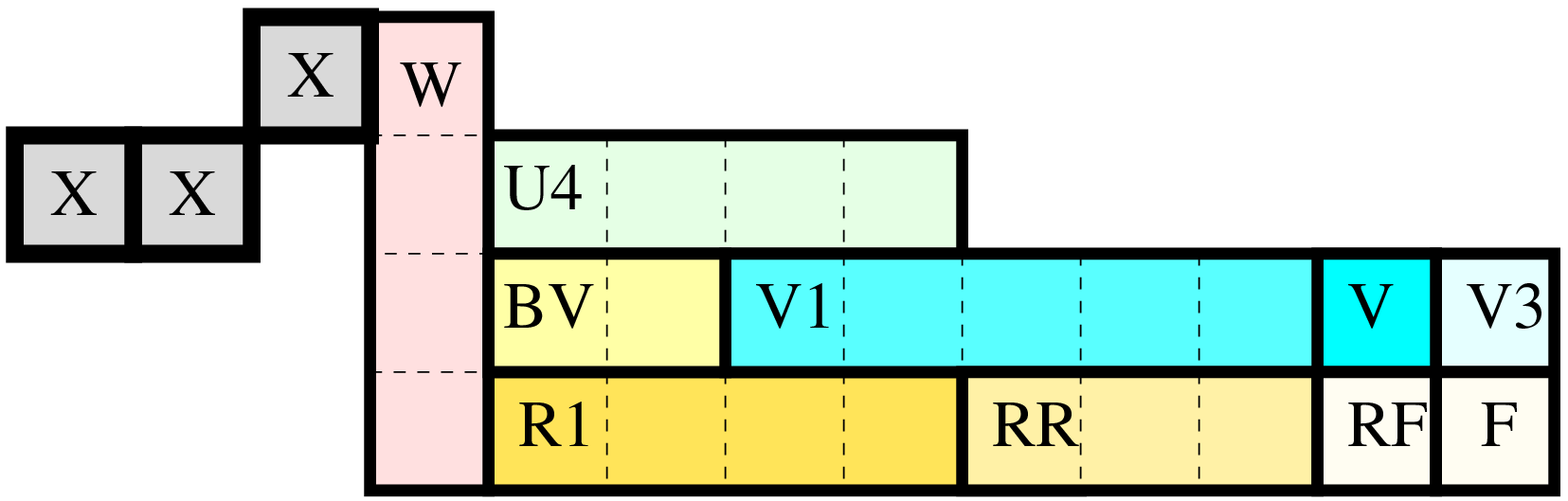}}
\hfill}
\begin{fig}\label{etape10}
The signal~$RF$ arrives to the cell before~$V3$: this creates~$F$ in the south 
of~$V3$ and~$V$ in its west.
\end{fig}
}

In full details, $RR$ propagates to the left, transforming each~$R1$ into~$RR$.
It also propagates to the right, transforming each blank into~$RF$ until $V3$~is
seen through the northern side. When this happens, the blank cell becomes~$F$.
Now, the cell~$V1$ which can see~$V3$ also sees~$RF$ through its southern side:
this triggers the transformation of~$V1$ back to~$V$,
see Figure~\ref{etape10}. And this situation is
now repeated for each cell~$V1$ which sees~$V$ trough its eastern side and 
either~$RF$ or~$RR$ through its southern side. This also transforms $BV$~back
to~$B$. Note that when $V$~was transformed
into~$V1$ and~$B$ to~$BV$, their possible southern neighbours were~$R$ and~$B$.
But this transformation of~$V1$ back to~$V$ also transforms~$RF$ back to the 
blank and~$RR$ back to~$R$: it is enough that the considered $RF$- or $R$-cell 
sees~$V$ through its northern side. 
At the same time, $V$~also triggers the transformation of~$U4$ into~$U$.
when $U4$ sees~$V$ through its southern side and~$U4$ through its western side, 
it becomes~$U$. Accordingly, the first cell of the $U$-row of the new data is
still in state~$U4$. 

   Indeed, when the blank cell which is the southern neighbours of~$V3$ 
sees~$RF$ through its western side, it becomes~$F$ which is the signal of
the termination of the computation for this cycle. Now the automaton enters 
the last stage of the cycle: it removes all marks. We have seen how the
turn to a starting configuration is triggered in the $V$-, $R$- and~$U$-rows.
As the length of these rows may be very different, it is important to create
a synchronization point so that when the new cycle starts there is no part of
the data in the letters of another stage: this would ruin the computation.
The synchronization is obtained by a signal which will be issued from~$V3$
which circumscribe the data and by the first~$U$ of the $U$-row: this latter
cell which is in~$U4$ at the moment we consider remains in this state as long
as it does not see the signal as~$FF$ through its northern side.

   In details this happens as follows: when~$V3$ sees~$F$ through its southern 
side, it becomes~$VF$. At the next time, its northern blank neighbour 
becomes~$UF$ and at the following time, the northern blank neighbour 
of~$UF$ becomes~$FF$, see Figure~\ref{etape12}. We can
see that~$FF$ is on a line which is just above the $U$-row. After its creation,
$FF$ moves on this line to the west, by one step at each time.

\vtop{
\vskip 10pt
\ligne{\hfill
\scalebox{0.50}{\includegraphics{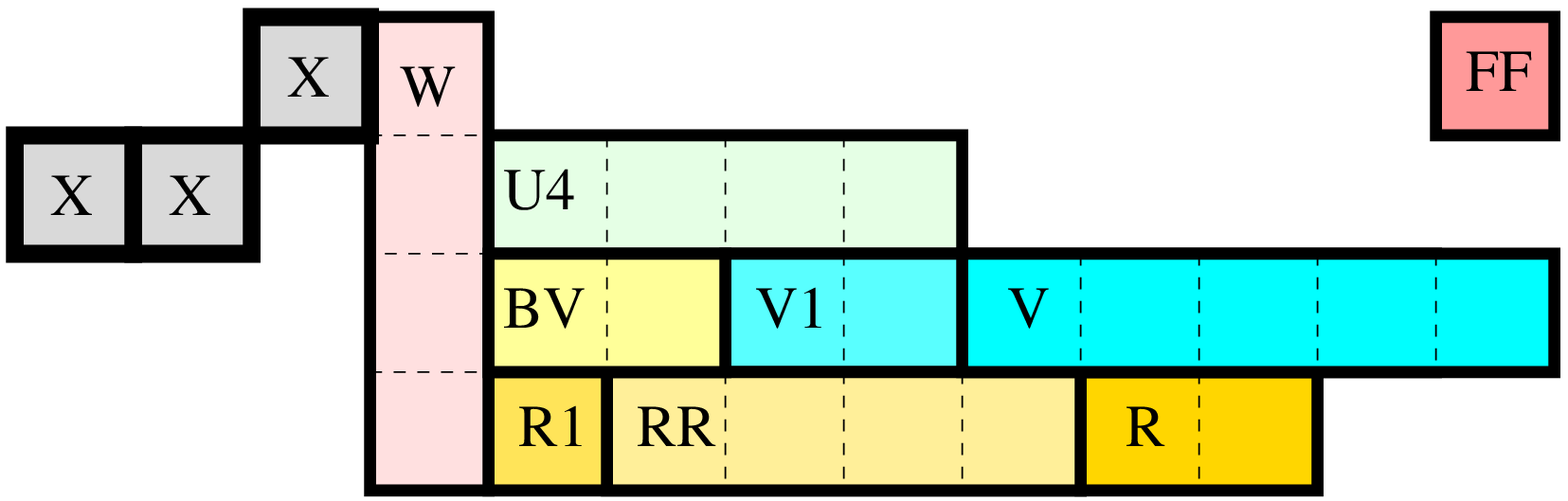}}
\hfill}
\begin{fig}\label{etape12}
The signal~$FF$ is created. It will travel to the~$W$ which stands on the line
above the $U$-row. Note that the transformation of~$V1$ back to~$V$ already 
started and that the transformation of~$R1$ into~$RR$ is almost completed.
\end{fig}
}

   During this time, the transformation of~$RF$ to~$B$ and then of~$RR$ to~$R$
arrives at the $W$-column. Note that before, $RR$~has transformed the $R1$-states
of the $R$-row into~$RR$. When this propagation of~$RR$ to the west reaches
the $W$-column, the $W$ on the line of the $R$-row sees~$RR$ through its
eastern side. At this time, it becomes~$W1$. Now, when~$W1$ sees~$R$ through
its eastern side, it remains~$W1$. But, the propagation of~$V$'s and then of 
possible~$B$'s on the $V$-row is ahead the propagation of~$R$'s by just one step.
And so, when the $W$ on the $V$-row sees~$B$ through its eastern side, it 
becomes~$W1$. At this moment we have two $W1$'s one as the northern neighbour of
the other. At the next time, the southern $W1$~vanishes, turning to~$B$. But 
the northern~$W1$ contaminates its northern neighbour which turns from~$W$ 
to~$W1$. So that we have again this configuration of two consecutive cells 
in~$W1$ on the $W$-column. And so, the southern~$W1$ again vanishes, 
turning to~$B$. However, the northern~$W1$ which is on the $U$-row is now
the western neighbour of~$U4$. This presence of~$U4$, still waiting for~$FF$,
keeps the western neighbour in the state~$W1$, see Figure~\ref{etape13}. 

\vtop{
\vskip 10pt
\ligne{\hfill
\scalebox{0.50}{\includegraphics{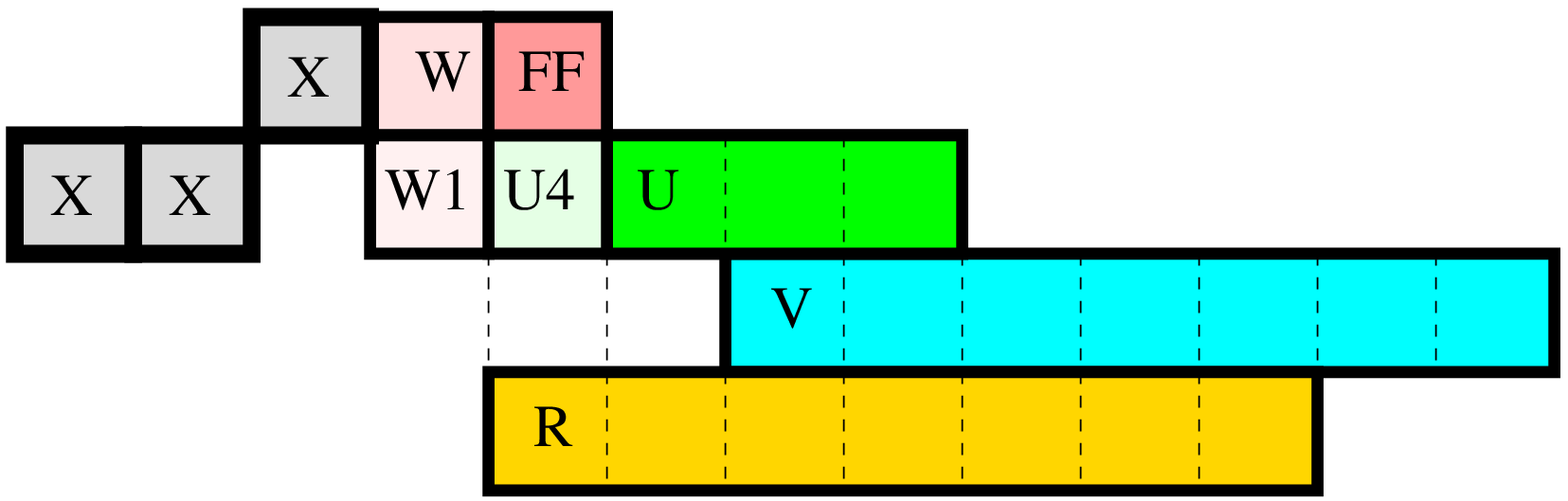}}
\hfill}
\begin{fig}\label{etape13}
Now, the signal~$FF$ arrives on the north of~$U4$. This is the signal of the 
very last steps of the current cycle. The second next step will be the
starting configuration of the new cycle, see Figure~{\rm\ref{etape14}}.
\end{fig}
}

   At last, when~$U4$ sees~$FF$ through its northern side, it becomes~$U$.
Now, from the starting configuration, the northern neighbour of~$W1$ is
in the state~$W$ and its eastern neighbour is~$FF$: from this situation,
this~$W$ knows that the restoration of the data is completed and so it turns
to~$X$, appending the new pixel to those which are already constructed.
But the next time, we have the first~$W$ of the new $W$-column triggered by
the new configuration and~$W1$ vanishes, turning to~$B$, as it sees~$X$ through
its northern side. 

   Note that this evolution of the computation explains why we take as starting 
configuration the configuration where there is a~$W$ seeing both the last
written~$X$ and the first~$U$ of the $U$-row, see Figure~\ref{etape14}.

\vtop{
\vskip 10pt
\ligne{\hfill
%\scalebox{0.50}{\includegraphics{etape14bis_S.eps}}
\scalebox{0.50}{\includegraphics{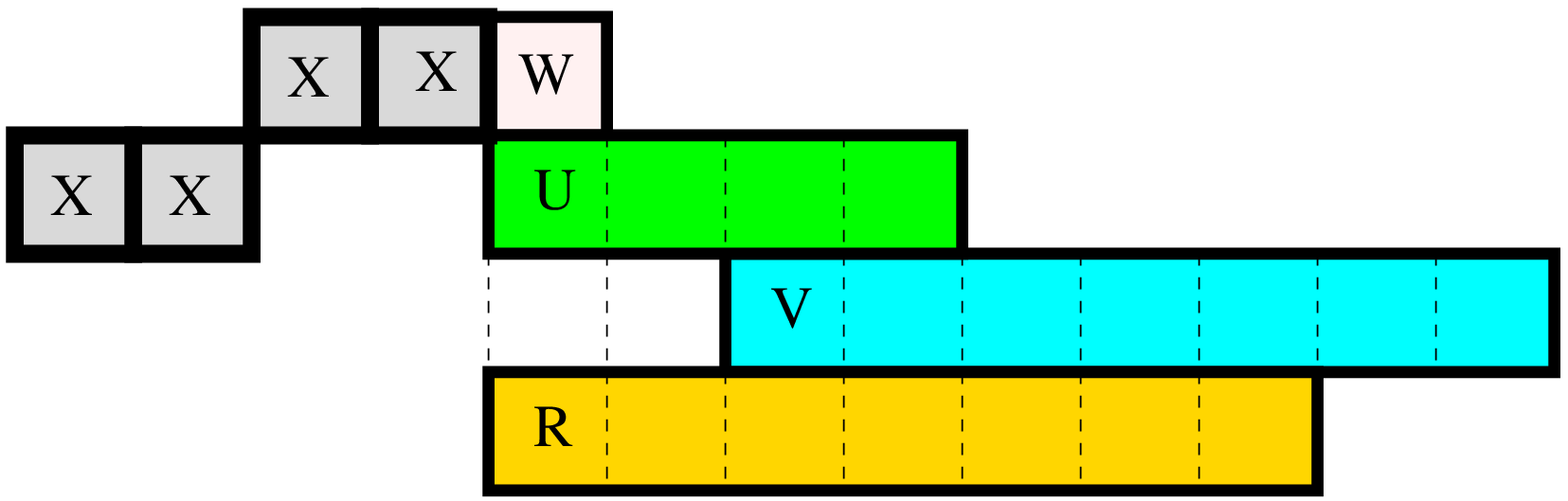}}
\hfill}
\begin{fig}\label{etape14}
The starting configuration of the new cycle. Note that it is very similar
to the starting configuration of Figure~{\rm\ref{kconfigpos}}.
\end{fig}
}

\subsubsection{The case $a$$+$$r \geq \mu$+$b$}
\label{notsmaller}

   When $a$$+$$r \geq \mu$+$b$, the automaton works in the same way in the case
when $a$$+$$r = \mu$+$b$ as well as in the case when $a$$+$$r > \mu$+$b$. The 
starting is
different as different cells are involved in the detection of the situation.

\vskip 5pt
   $\underline{\hbox{\rm The case when $a$$+$$r > \mu$+$b$}}$. 
\vskip 5pt
We know that in this case, the blank
which sees~$R3$ for the first time through its western side becomes~$R$. Now,
this $R$~propagates to the left, until~$V3$ is seen. As the $R$-row consists
of cells in~$R1$ except the last one which is~$R$, each~$R1$ which sees~$R$ 
through is eastern side and, at the same time, the blank through its northern 
side becomes~$R$. Now, the cell~$R1$ which sees~$R$ through the eastern side 
but, at the
same time, sees~$V3$ through the northern one, this cell becomes~$Z$. It is now
plain that the number of~$R$'s on the right hand side is \hbox{$a$$+$$r$$-$$b$}
which is less than~$b$.

\vtop{
\vskip 10pt
\ligne{\hfill
%\scalebox{0.50}{\includegraphics{etape9_S.eps}}
\scalebox{0.50}{\includegraphics{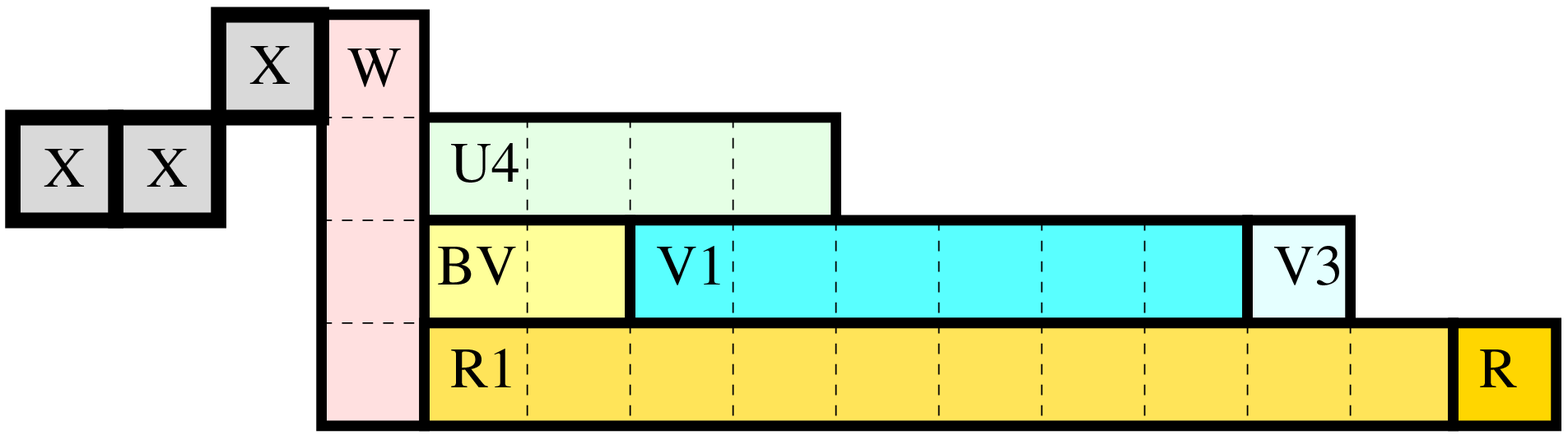}}
\hfill}
\vskip 5pt
\ligne{\hfill
\scalebox{0.50}{\includegraphics{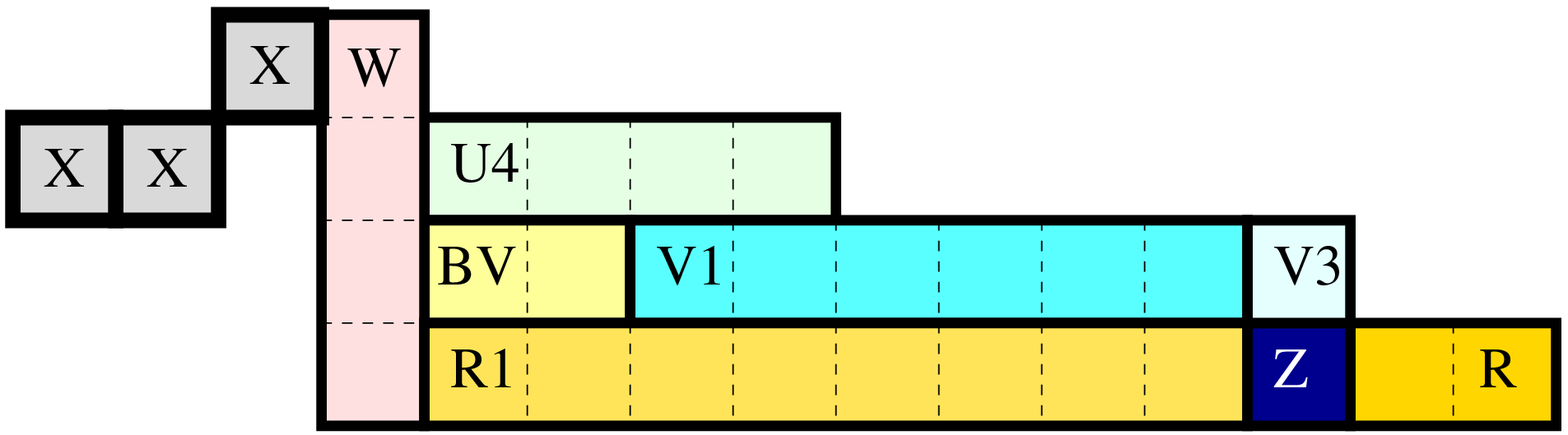}}
\hfill}
\vskip 10pt
\begin{fig}\label{etape10G}
Case $a+r > \mu+b$.
Above: the first $R$ after $R_3$ is written.
Below, here, two steps later, the southern neighbour of~$V_3$
has seen $R$ through its eastern side, so it became~$Z$.
\end{fig}
}

\vtop{
\vskip 10pt
\ligne{\hfill
%\scalebox{0.50}{\includegraphics{etape9_S.eps}}
\scalebox{0.50}{\includegraphics{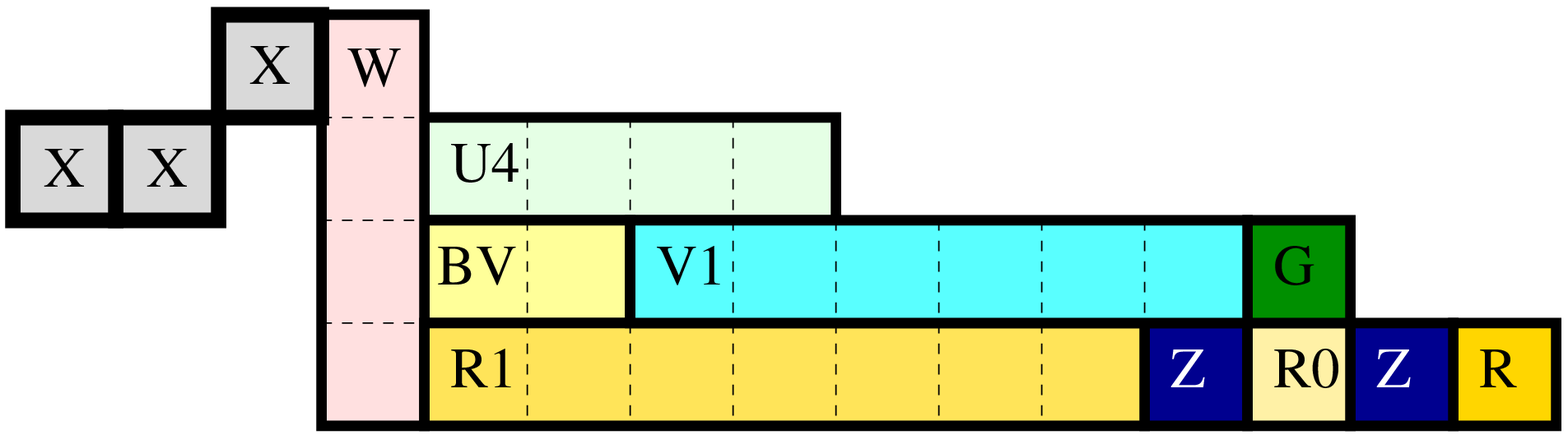}}
\hfill}
\vskip 5pt
\ligne{\hfill
\scalebox{0.50}{\includegraphics{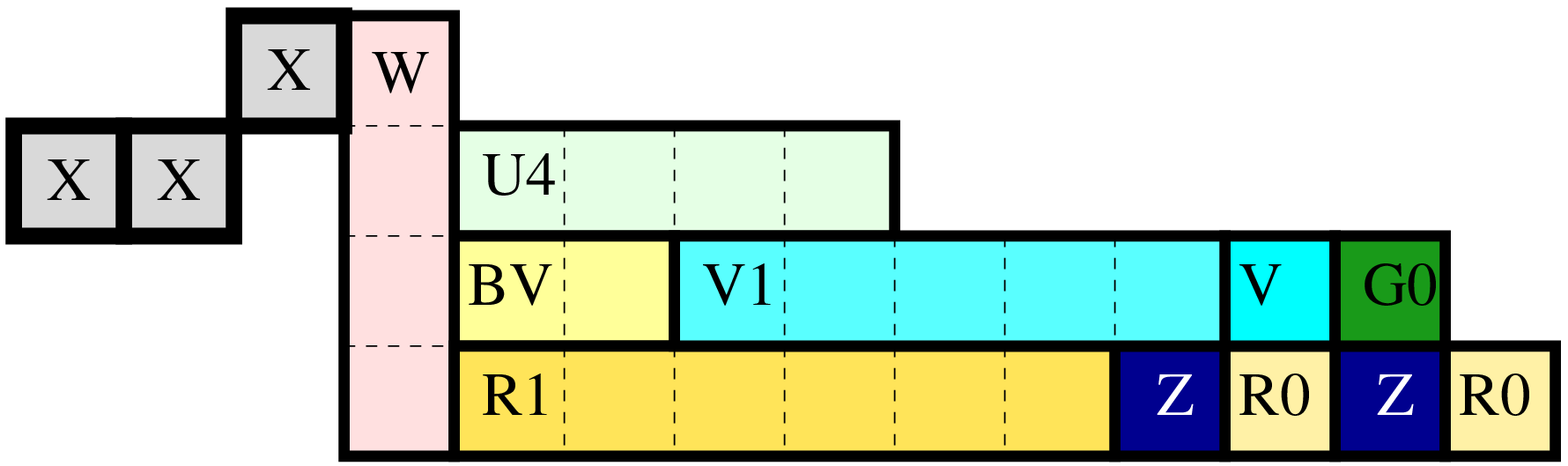}}
\hskip 20pt\hfill}
\vskip 10pt
\begin{fig}\label{etape11G}
Case $a+r > \mu+b$.
Above: $V_3$ was just changed to~$G$.
Below: next step, $G$~is changed to $G_0$ while the $R$'s to the right-hand
side of~$Z$ are changed to $ZR_0$.
\end{fig}
}

   Now, we have to erase the $R1$'s which are on the left-hand side of the 
leftmost~$Z$ and to keep the number of those which are to its right-hand side.
   This problem is solved as follows: $Z$~moves to the left, erasing the~$R1$'s
as long as the concerned~$R1$'s see~$V1$ or~$V$ through their northern side:
these $R1$'s are one by one transformed into~$Z$. But, at the same time, 
$Z$~drags to the left the block of~$R$'s which stand on its right-hand side. 
To do this, $Z$~sends a copy of itself to the right: when $R$~is met, it 
changes to~$R0$. From the position of the leftmost~$Z$ to the first blank on 
its right-hand side, each cell has the following cycle of transformations: 
\hbox{$Z\rightarrow R0\rightarrow Z$}. The cycle starts with the
change \hbox{$R\rightarrow Z$}, and it stops when a transformation
\hbox{$R0\rightarrow B$} happens. In fact, when $R$ sees~$Z$ through its
western side and, at the same time it sees~$R$ through the eastern side, it 
becomes~$Z$, and the just mentioned cycle starts. Now, when $R$ sees~$Z$
through the western side and the blank through the eastern side, then $R$~becomes
blank, and the cycle stops. When all $R$'s have been turned to~$R0$ by this
transformation of~$Z$, the end of the $R$-row is a word of the form
\hbox{$(ZR0)^\rho$} where $\rho$~is the new value of~$r$. This word moves to
the left by one step at each time, see~Figure~\ref{etape11G}.

   During this process, $G0$~remains unchanged as long as it can see~$R_0$
or~$Z$ through its southern side. Now, as soon as it sees~$B$, this means that
there is no more copy of the pattern $ZR0$ on the right-hand side of the~$B$
seen by~$G0$. As a consequence, we can start the process which will allow
to move the data by one step upward.

\vtop{
\vskip 10pt
\ligne{\hfill
%\scalebox{0.50}{\includegraphics{etape9_S.eps}}
\scalebox{0.50}{\includegraphics{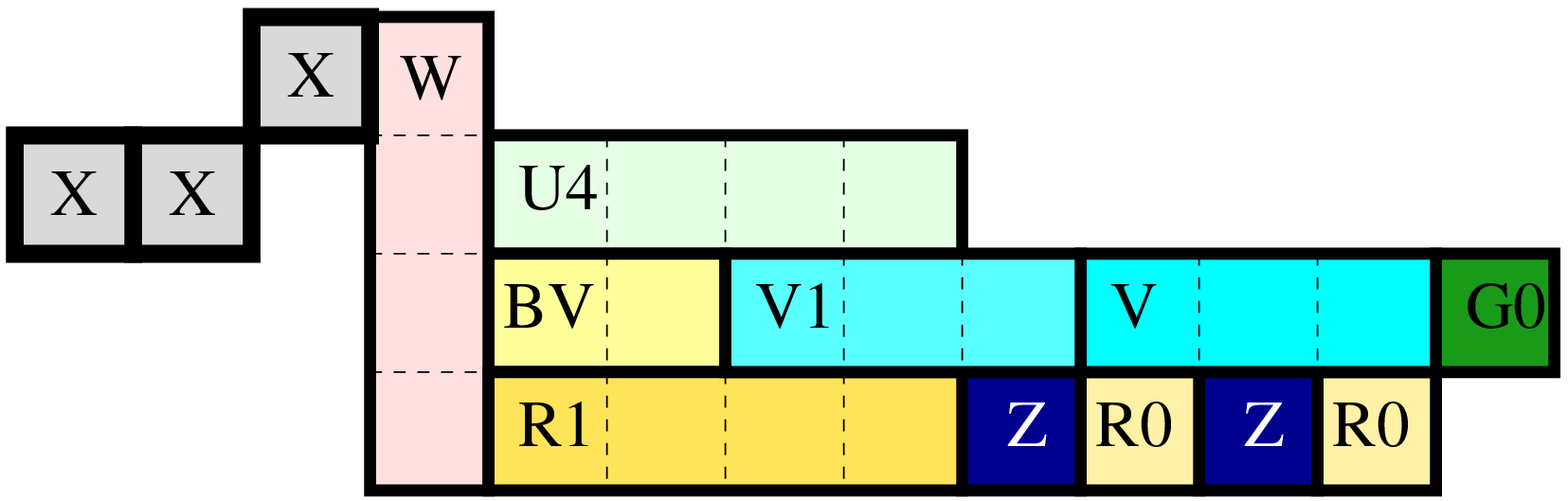}}
\hfill}
\vskip 5pt
\ligne{\hfill
\scalebox{0.50}{\includegraphics{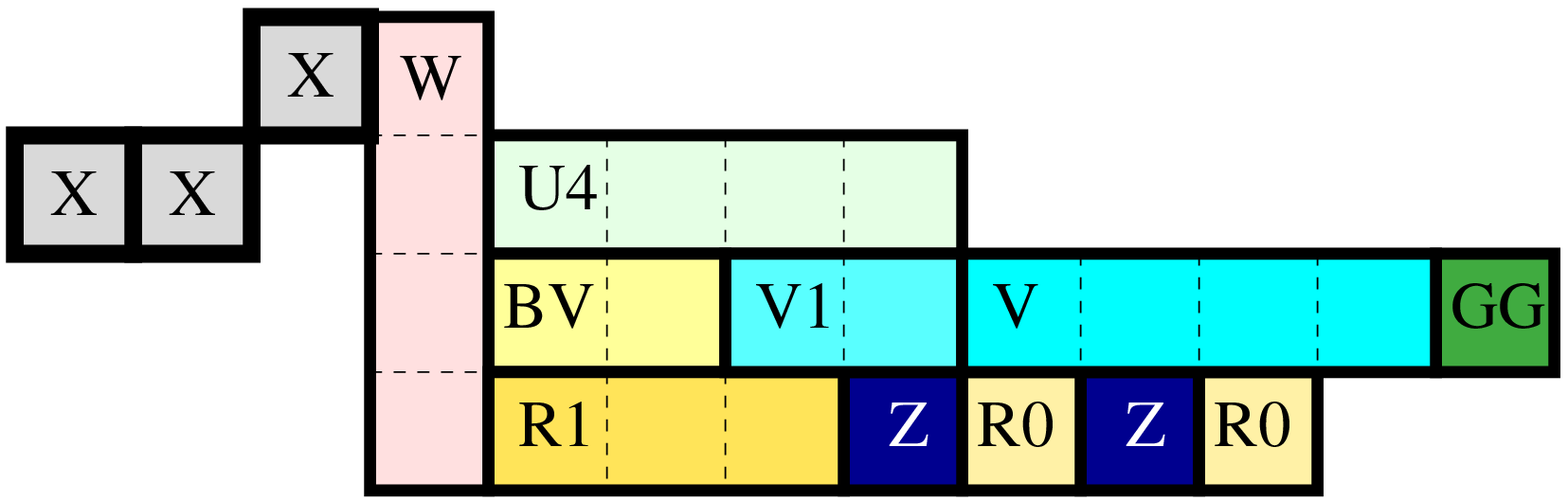}}
\hfill}
\vskip 10pt
\begin{fig}\label{etape12G}
Case $a+r > \mu+b$.
Above: the last step when $G_0$ is present.
Below: next step, $G0$~is changed to $GG$; there are only blanks
to the right-hand side of the rightmost $R0$.
\end{fig}
}

   Note that during this process, $V1$'s are turned to~$V$. This is made
possible by the fact that $V1$ seeing~$V$ through its eastern and~$Z$
through its southern side becomes~$V$. Now, as $V$ sees~$Z$, $R$ or~$R0$
when it is to the right-hand side of the leftmost~$V$, these $V$'s are stable.

\vtop{
\vskip 10pt
\ligne{\hfill
%\scalebox{0.50}{\includegraphics{etape9_S.eps}}
\scalebox{0.50}{\includegraphics{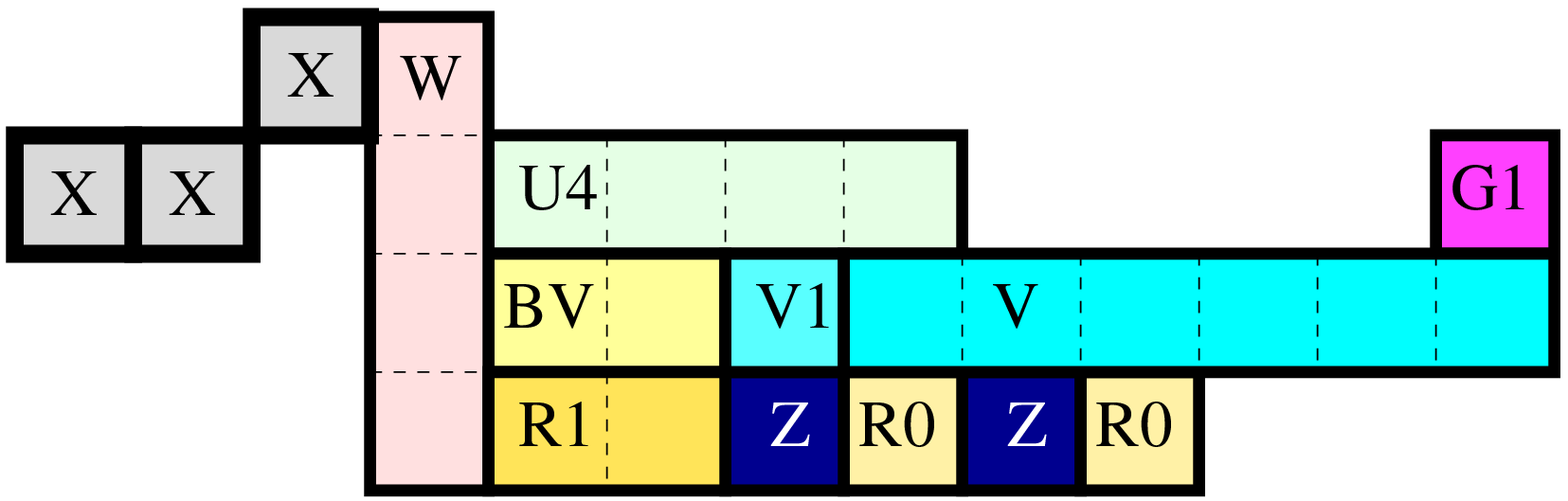}}
\hfill}
\vskip 5pt
\ligne{\hfill
\scalebox{0.50}{\includegraphics{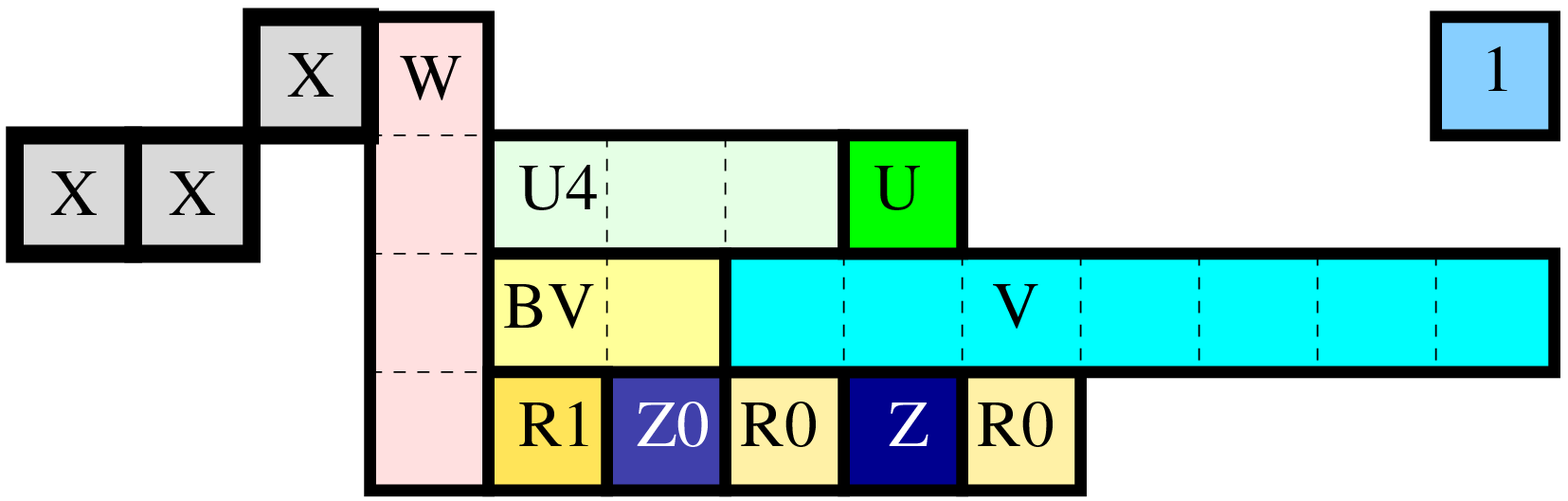}}
\hfill}
\vskip 10pt
\begin{fig}\label{etape13G}
Case $a+r > \mu+b$.
Above: $GG$ disappeared, leaving~$V$ on its place and triggering the 
transformation of its northern neighbour from~$B$ to~$G1$.
Below: the first appearance of~$1$ which is one of the signals used for 
lifting the data by one step upward.
\end{fig}
}

   When the southern neighbour of the rightmost~$BV$ can see~$Z$ through its
eastern side, it becomes~$Z0$. This change to~$Z0$ sends a signal to the right
by the successive transformation of~$R0$ into~$R$. By the constant shift
to the west of the $R0$'s and the transformation of the leftmost~$R_0$ into~$R$
by seeing the rightmost~$R$ through its western side, all $R0$'s are transformed
into~$R$ and, at the same time, the new block of~$R$'s moves by one step to
the west at each time. The result is that, at some point, $Z_0$ can see the bottom
of the $W$-column through its western side. During this time, the occurrence
of~$Z0$ allows the cellular automaton to transform the $BV$'s back to~$B$.
These two processes are a bit squeezed in Figure~\ref{etape14G} but 
a careful comparison of the configurations in Figures~\ref{etape13G} 
and~\ref{etape14G} shows that things happen as just described above.
Now, when the bottom of the $W$-column sees $Z0$~through its eastern side,
it becomes~$W1$. This $W1$ goes up along the column, transforming the $W$'s 
to~$B$'s until $W1$ can see~$U4$ through its eastern side. Then, $W1$ stops
at this place until $U4$ disappears, a certain time later, see 
Figure~\ref{etape16G}.
  
   In the meanwhile, at the other end of the data, things are turning to
the process which raises the data by one step upward.

   Remember that $G0$ remained unchanged until it can see~$B$ through its 
southern side. This happens when the migration of the block of $(ZR0)^{\rho}$
arrives to such a situation. Then, $G0$ becomes~$GG$, see Figure~\ref{etape12G}.
At the next step, the northern neighbour of~$GG$ turns from~$B$ to~$G1$ and
$GG$~itself changes to~$V$. The transformation of the $V1$'s and~$V3$ to $V$'s 
is completed in this part of the $V$-row while, at the other end, the progressive
transformation of~$V1$ to~$V$ is still going on, triggered by the leftmost~$Z$,
as already noticed.      

    Then, the northern neighbour of~$G1$ turns from~$B$ to~$1$, see 
Figure~\ref{etape13G}.

\vtop{
\vskip 10pt
\ligne{\hfill
%\scalebox{0.50}{\includegraphics{etape9_S.eps}}
\scalebox{0.50}{\includegraphics{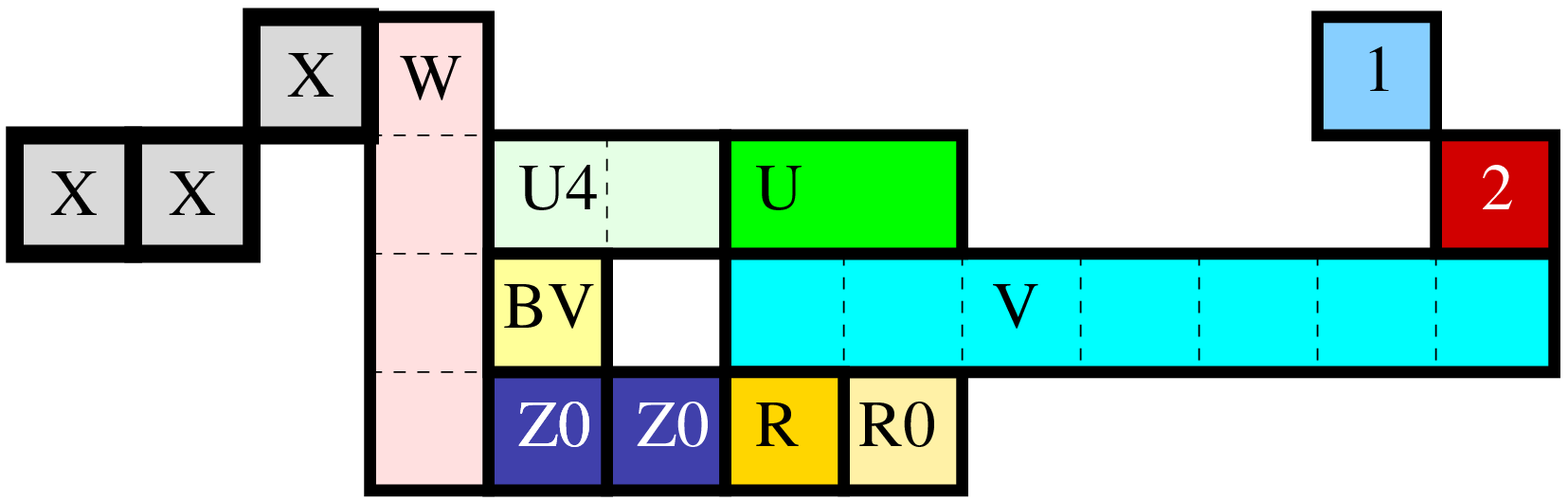}}
\hfill}
\vskip 5pt
\ligne{\hfill
\scalebox{0.50}{\includegraphics{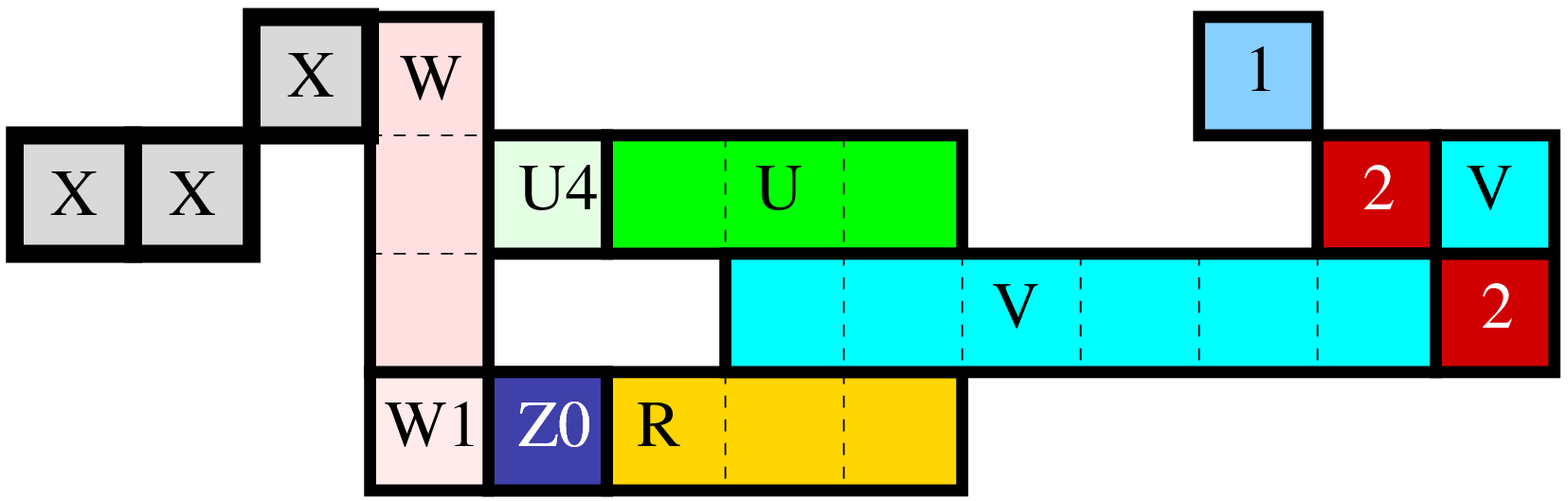}}
\hfill}
\vskip 10pt
\begin{fig}\label{etape14G}
Case $a+r > \mu+b$.
The mechanism of lifting the data by one step upward.
\vskip 0pt
Above: $1$~already moved to the west to prepare the lifting of the
next symbol; to the south of the previous place of~$1$, $2$ appears.
Below: again~$1$ moved to the west by one step; again, the southern neighbour
of its previous position became~$2$; the previous~$2$ performed the lifting
of~$V$ looking now for a possible final lifting of~$R$ or~$B$.
\end{fig}
}

This~1 triggers the mechanism of raising the whole set of data by one step
upward. Note that~1 is on the row which is just above the $U$-row. The mechanism
is as follows: 1~moves by one step to the left and, on its former place, 
it copies the state it sees through its southern side and, at the same time, 
it transforms its southern neighbour into~2. Note that, by the construction 
itself, 2~necessarily sees~$V$ through its southern side. Now, 2~does the half 
what 1~does: it does not move, neither to the left or to the right, but it copies what it sees through its southern
side and it transforms its southern neighbour to~2 if this neighbour is 
neither~$B$ nor~$R$ or if its northern neighbour is~$U$. This means that the 
blank can be moved by one step upward once and that, afterwards, it stops and 
erases state~2. This also means that when the southern neighbour of~$2$ is~$R$,
$2$~raises this~$R$ but do not make it replaced by~$2$.
  
As 1~moves to the west, this means that step by step, the configuration is 
raised by one step upward, with a delay of two steps for the $R$-row. 
Because of this delay, when $W$~sees $1$~through its eastern side, $W$~becomes
$WW$~which in its turn becomes~$W3$. After this delay, the last~$R$ has
been raised, see Figure~\ref{etape16G}. And so, $W1$~seeing~2 through its 
eastern side vanishes and~$W3$ turns to~$B$. Now, the occurrence of~$W3$
triggers the writing of the next pixel~$X$ at the right place, {\it i.e.}
the cell of its northern neighbour, see Figure~\ref{etape17G}. Consequently, 
the next configuration is the starting
configuration of the next cycle of computation, see Figure~\ref{etape17G}
again.

\vtop{
\vskip 10pt
\ligne{\hfill
%\scalebox{0.50}{\includegraphics{etape9_S.eps}}
\scalebox{0.50}{\includegraphics{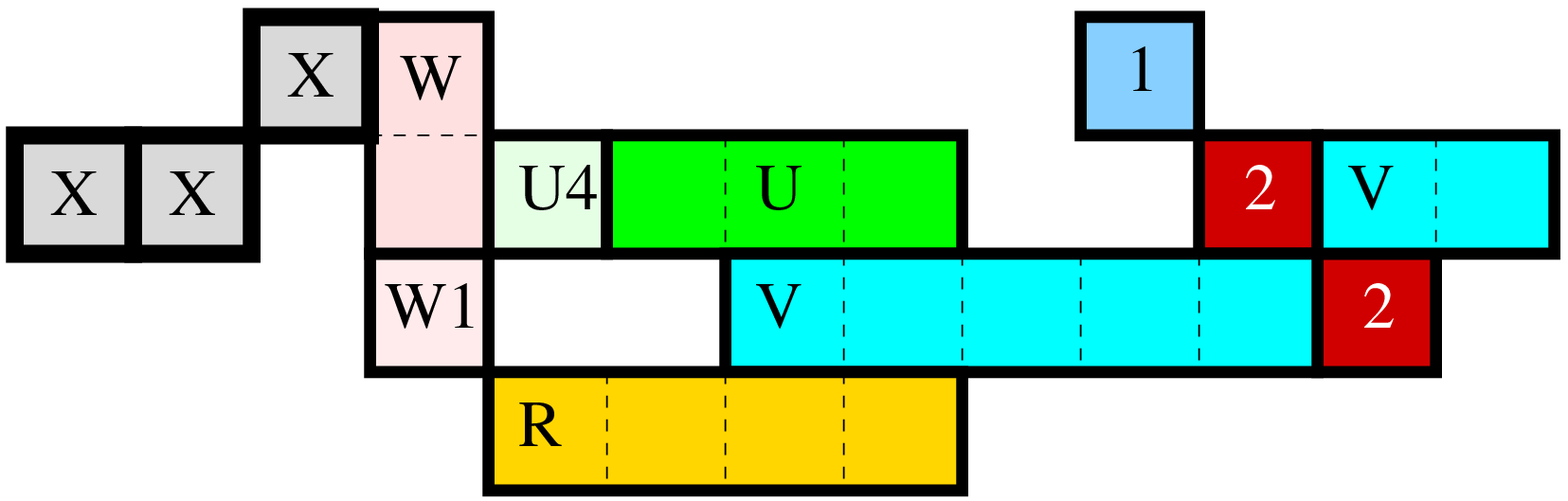}}
\hfill}
\vskip 5pt
\ligne{\hfill
\scalebox{0.50}{\includegraphics{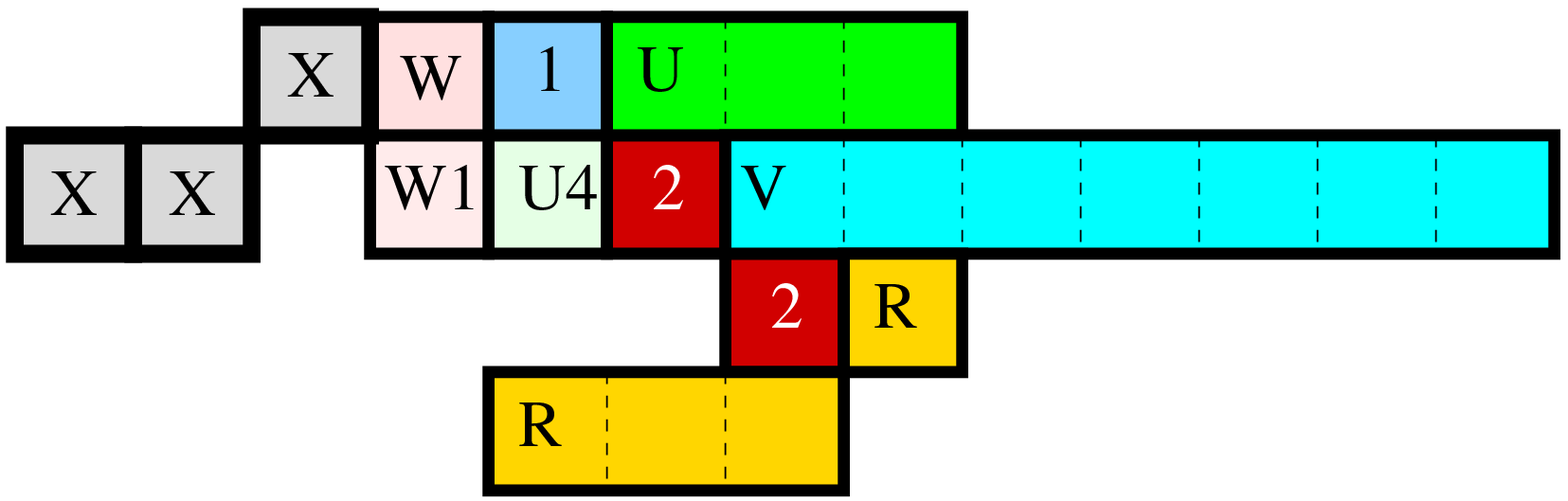}}
\hfill}
\vskip 10pt
\begin{fig}\label{etape15G}
Case $a+r > \mu+b$.
The mechanism of lifting the data by one step upward.
\vskip 0pt
Above: this time, the $R$- and $V$-rows are restored; in the $U$-row, except the
leftmost~$U_4$, all others have been turned to~$U$. We can see the disposition
of~$1$ and~$2$'s for moving the data by one step upward.
Below: the signal~$1$ arrives at its last point, it disappears at the next step,
see Figure~{\rm\ref{etape16G}}; the $R$-row is being to be moved upward; this 
is performed for the $V$-row and for all cells of the $U$-row, except the first
element, still in~$U4$.
\end{fig}
}

\vtop{
\vskip 10pt
\ligne{\hfill
%\scalebox{0.50}{\includegraphics{etape9_S.eps}}
\scalebox{0.50}{\includegraphics{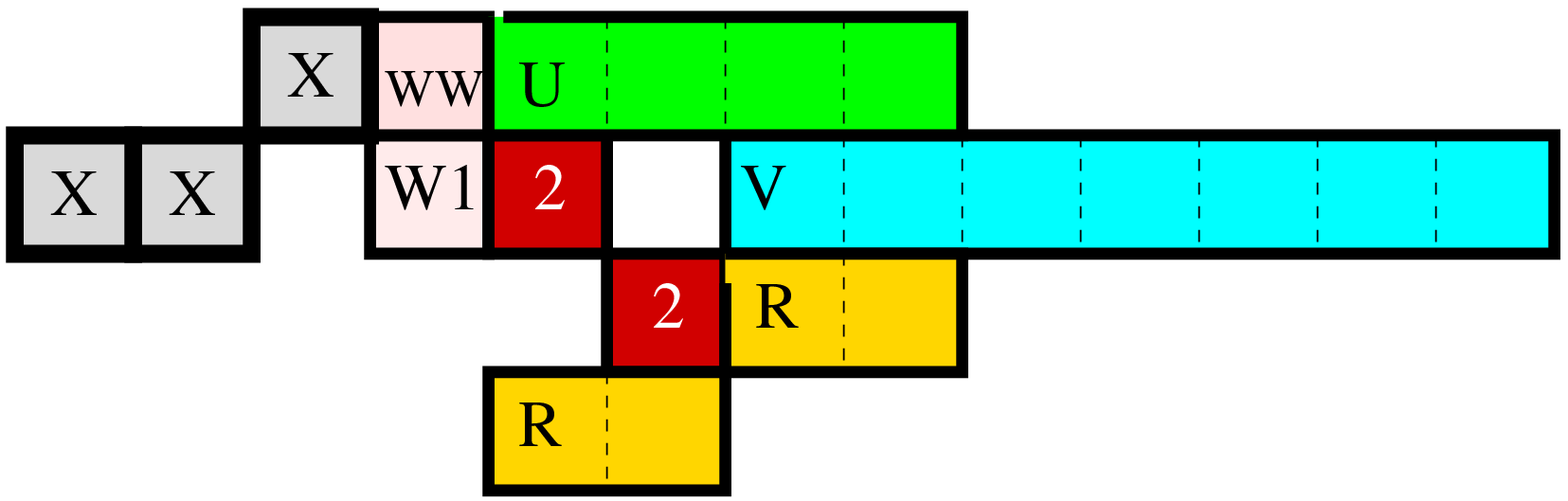}}
\hfill}
\vskip 5pt
\ligne{\hfill
\scalebox{0.50}{\includegraphics{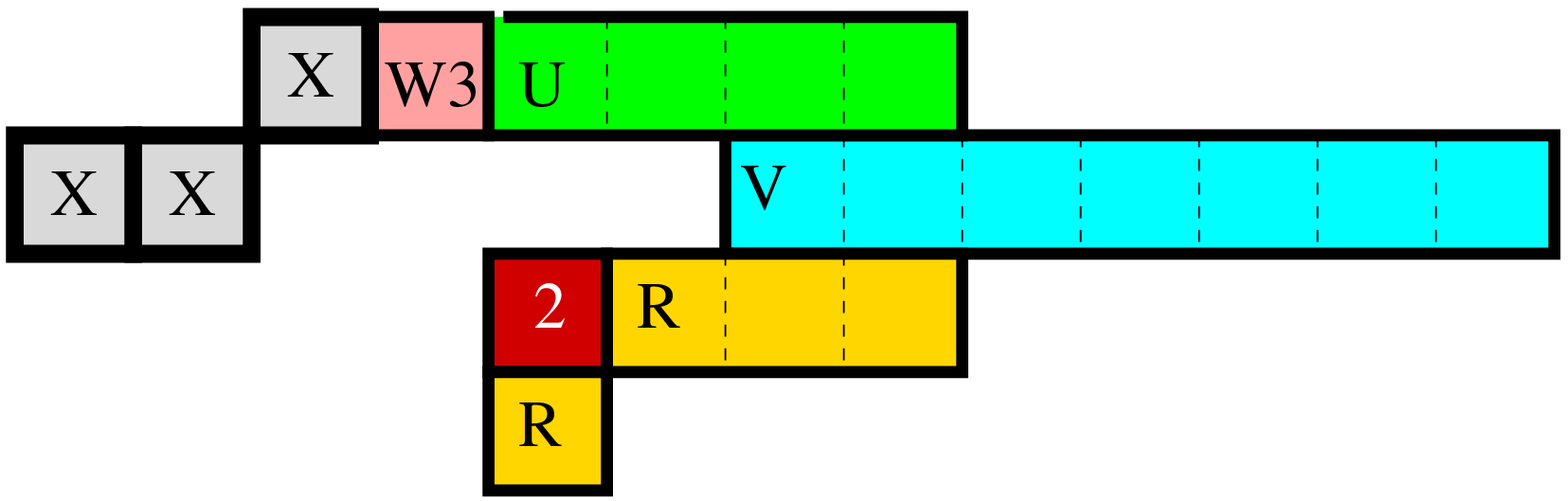}}
\hfill}
\vskip 10pt
\begin{fig}\label{etape16G}
Case $a+r > \mu+b$.
The end of the process.
\vskip 0pt
Above: $U4$ has now been turned to~$U$; two~$¶$'s have still to be lifted.
This will be performed for the right-hand one at the next step and for the
last one the second step after the present one.
Below: $W1$ disappeared and $WW$~has been changed to~$W3$. The last~$R$ remains 
to be lifted.
\end{fig}
}

\vtop{
\vskip 10pt
\ligne{\hfill
%\scalebox{0.50}{\includegraphics{etape9_S.eps}}
\scalebox{0.50}{\includegraphics{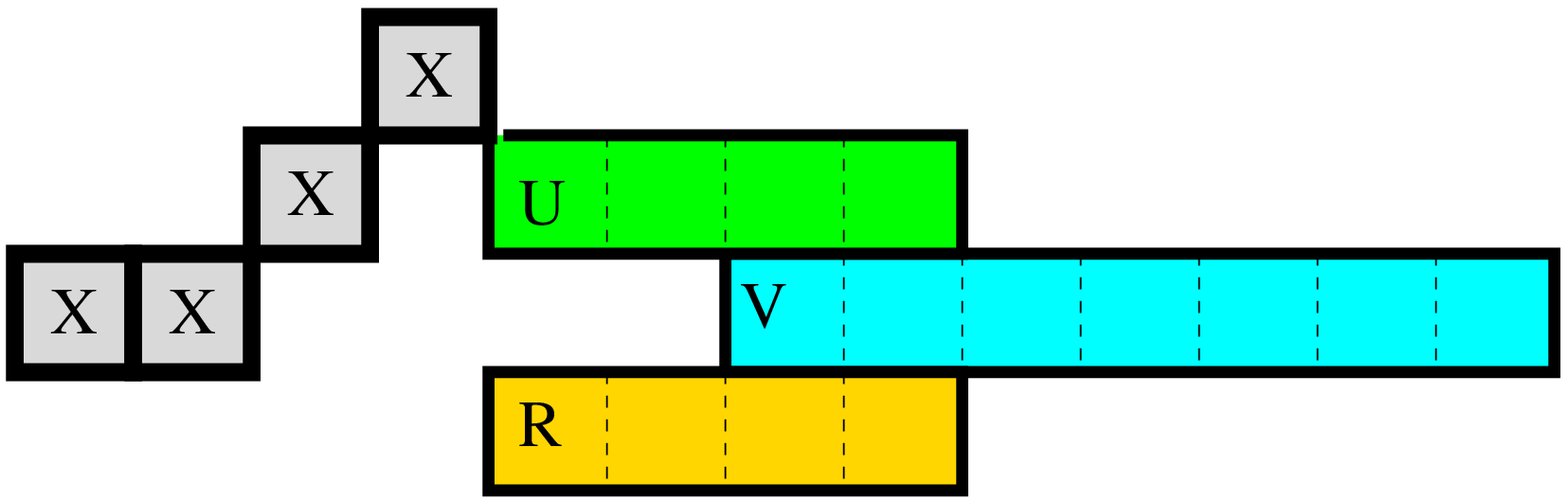}}
\hfill}
\vskip 5pt
\ligne{\hfill
\scalebox{0.50}{\includegraphics{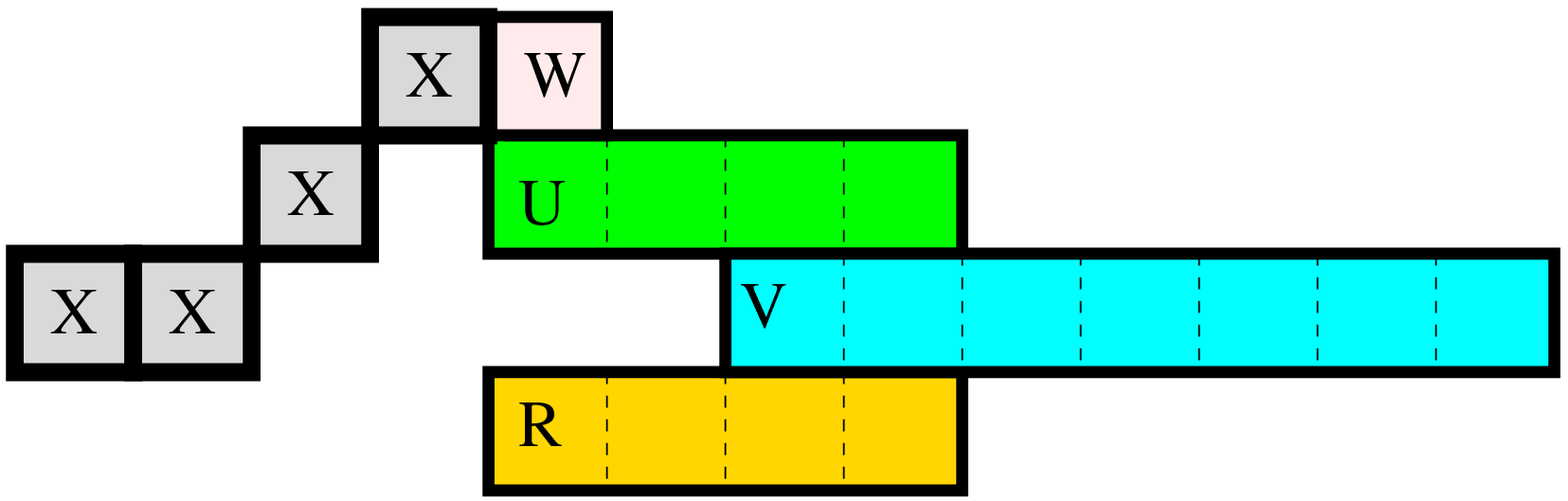}}
\hfill}
\vskip 10pt
\begin{fig}\label{etape17G}
Case $a+r > \mu+b$.
The turn to the starting configuration of a new cycle. 
\vskip 0pt
Above: this time, the new data is at the right place and the new pixel has been
written.
Below: The next step: it is the first step of the new cycle.
\end{fig}
}

\vskip 5pt
   $\underline{\hbox{\rm The case when $a$$+$$r = \mu$+$b$}}$. 
\vskip 5pt

   Remember that this situation is detected by the
fact that the blank which is the southern neighbour of~$V3$ sees~$R3$ through
its western side. Then this blank cell becomes~$Z$. The action to the left of~$Z$
is the same as previously: the cells which are on the left-hand side of~$Z$
cannot see what is on the right-hand side of~$Z$. Similarly, this is the same 
for the cells which are exactly on a row above~$Z$. From the rules
for the case when \hbox{$a$+$r> \mu$+$b$}, we conclude that this~$Z$ moves
to the west by one step. Now, in the case when \hbox{$a$+$r> \mu$+$b$},
at that time, the southern neighbour of~$G$ is~$R1$ or~$R$. Here, it is~$B$.
This is why this~$B$ becomes~$Z$, providing us with the pattern of two
consecutive~$Z$, see Figure~\ref{etape10E}. This patterns reduces the handling 
of the right-hand side
of~$Z$ to nothing has there are only blank cells. Now, on the left-hand side,
the leftmost~$Z$ behaves as previously, both for the $R$- and the $V$-rows.
The second~$Z$ has simply to follow the first one by a similar motion to the
west by one step at each time. In the meanwhile, as we had the change directly
from~$G$ to~$GG$, the evolution on this side of the configuration is the
same as in the case when \hbox{$a$+$r> \mu$+$b$}. In particular, signals~1 and~2
appear in order to lift the data by one step upward, see~Figure~\ref{etape11E}. 

The block~$ZZ$ goes on to the west until it reaches the area where the $BV$'s
are. When $Z$~can see~$BV$ through its northern side, it becomes~$Z0$
which triggers the transformation of~$BV$ to~$B$ as in the case
when \hbox{$a$+$r> \mu$+$b$}, see Figures~\ref{etape11E} and~\ref{etape12E}. 
In Figure~\ref{etape12E}, the leftmost~$Z0$ can see the bottom of the 
$W$-column through its western side. At this moment, almost all $BV$'s are
turned to~$B$ and almost all needed $R$'s have been restored. Starting from the
next configuration, see Figure~\ref{etape13E}, the rules of the
case when \hbox{$a$+$r> \mu$+$b$} allow the automaton to complete the 
computation.

\vskip 10pt
\vtop{
\ligne{\hfill
%\scalebox{0.50}{\includegraphics{etape9_S.eps}}
\scalebox{0.50}{\includegraphics{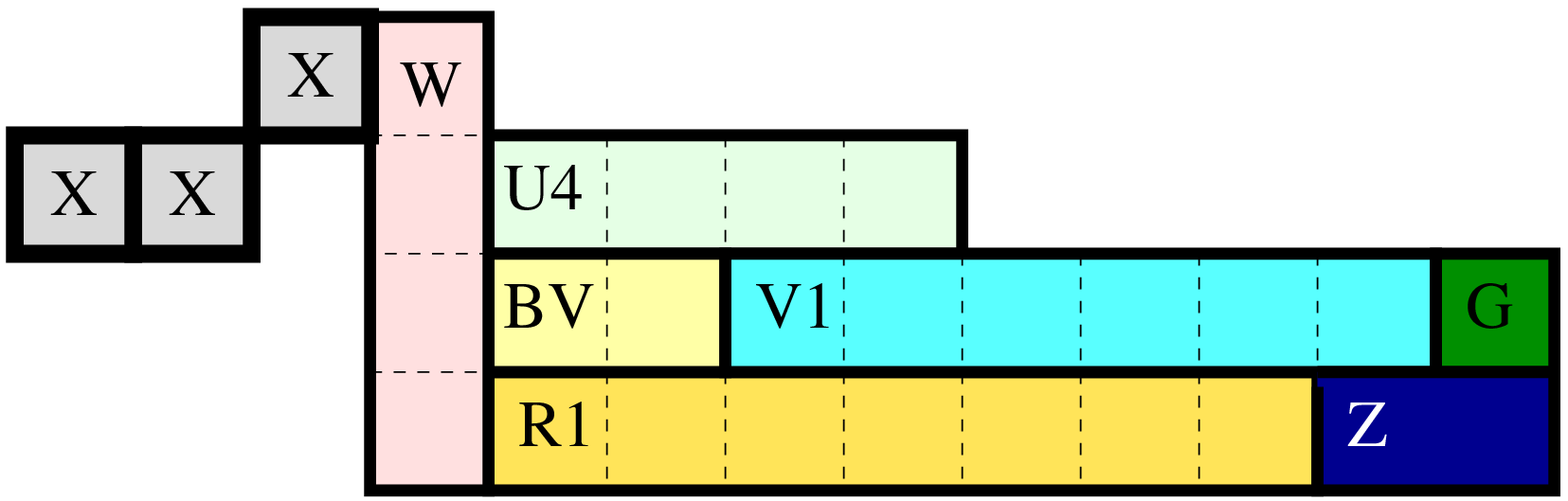}}
\hfill}
\vskip 5pt
\ligne{\hfill
\scalebox{0.50}{\includegraphics{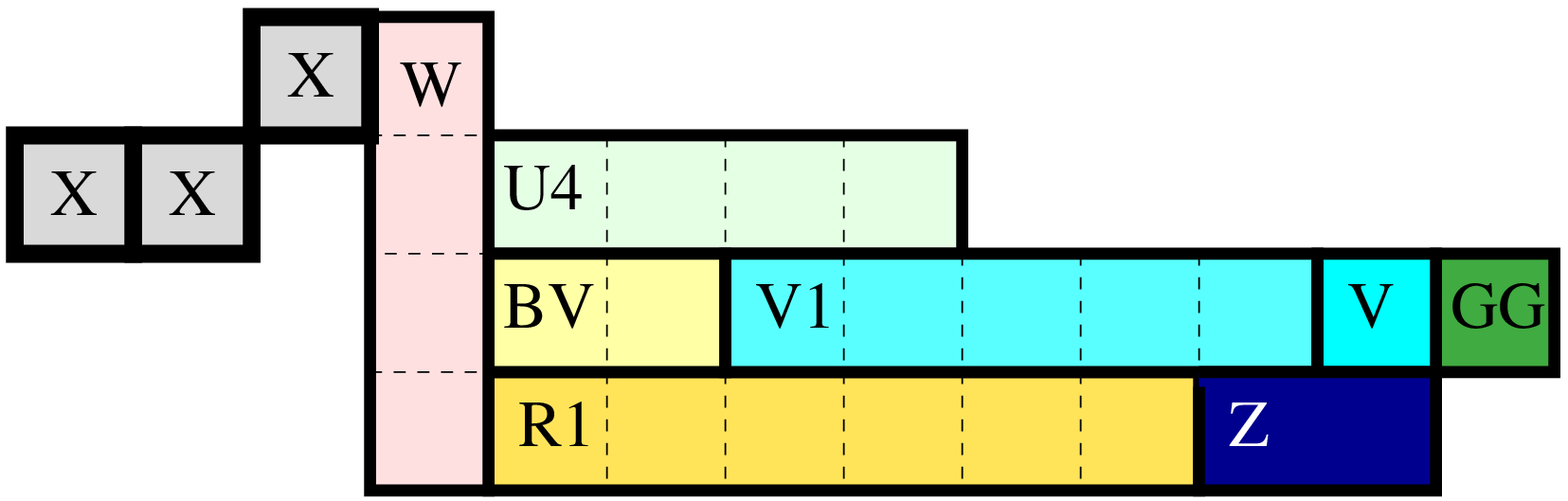}}
\hfill}
\vskip 10pt
\begin{fig}\label{etape10E}
Case $a$+$r = \mu$+$b$. 
Initialization of the $ZZ$ pattern which clears the remainder.
\vskip 0pt
Above: this configuration is the one which occurs at the time just after
the one illustrated by Figure~{\rm\ref{etape9}}. 
Below: the next step: $ZZ$ moved by one step to the west; note that $V3$
has changed to~$G$ and that $G$~has directly changed to~$GG$.
\end{fig}
}

\vskip 10pt
\vtop{
\ligne{\hfill
%\scalebox{0.50}{\includegraphics{etape9_S.eps}}
\scalebox{0.50}{\includegraphics{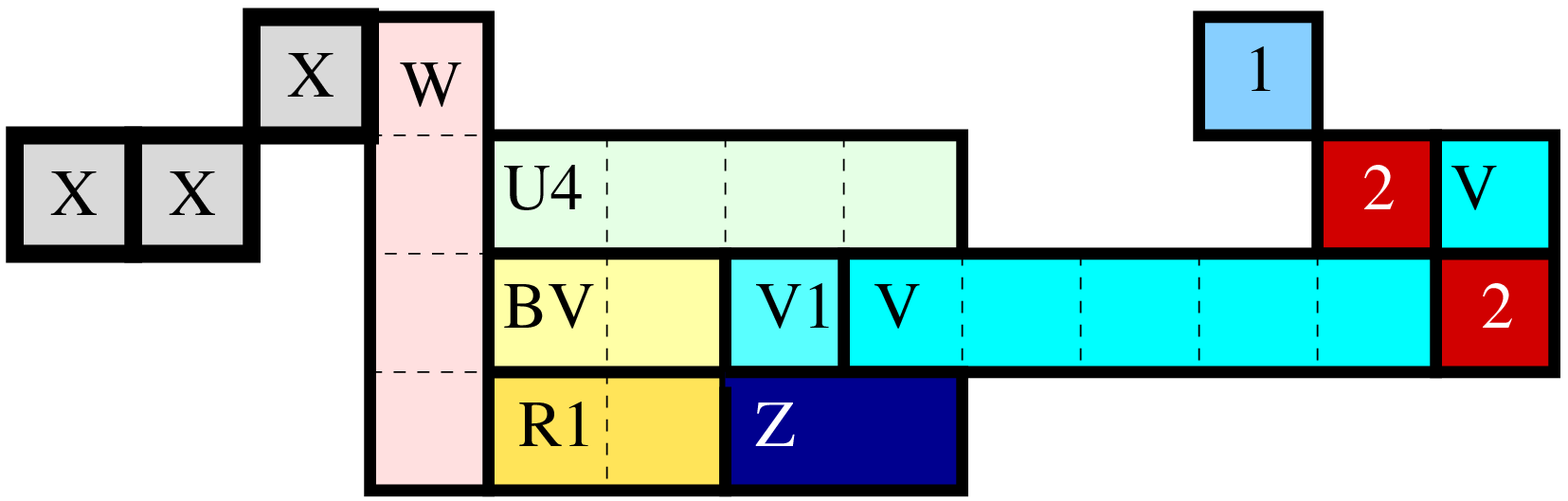}}
\hfill}
\vskip 5pt
\ligne{\hfill
\scalebox{0.50}{\includegraphics{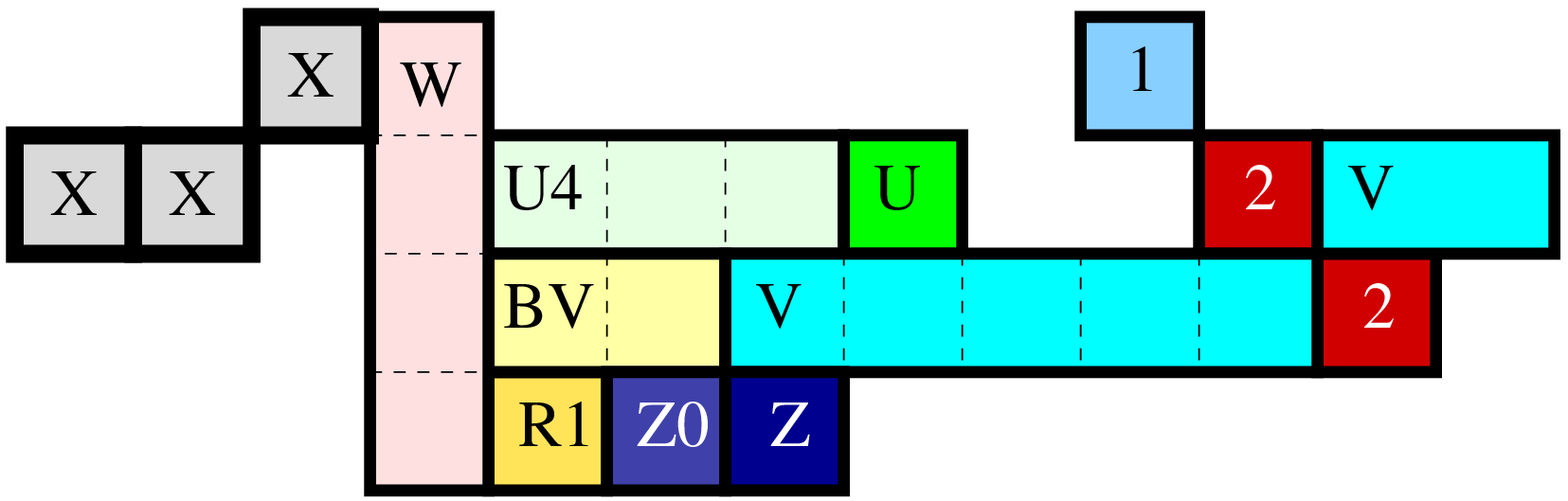}}
\hfill}
\vskip 10pt
\begin{fig}\label{etape11E}
Case $a$+$r = \mu$+$b$. 
When the $ZZ$ pattern arrives at its destination.
\vskip 0pt
Above: the pattern reaches the $BV$ area.
Below: the next step: occurrence of~$Z0$.
\end{fig}
}

\vskip 10pt
\vtop{
\ligne{\hfill
%\scalebox{0.50}{\includegraphics{etape9_S.eps}}
\scalebox{0.50}{\includegraphics{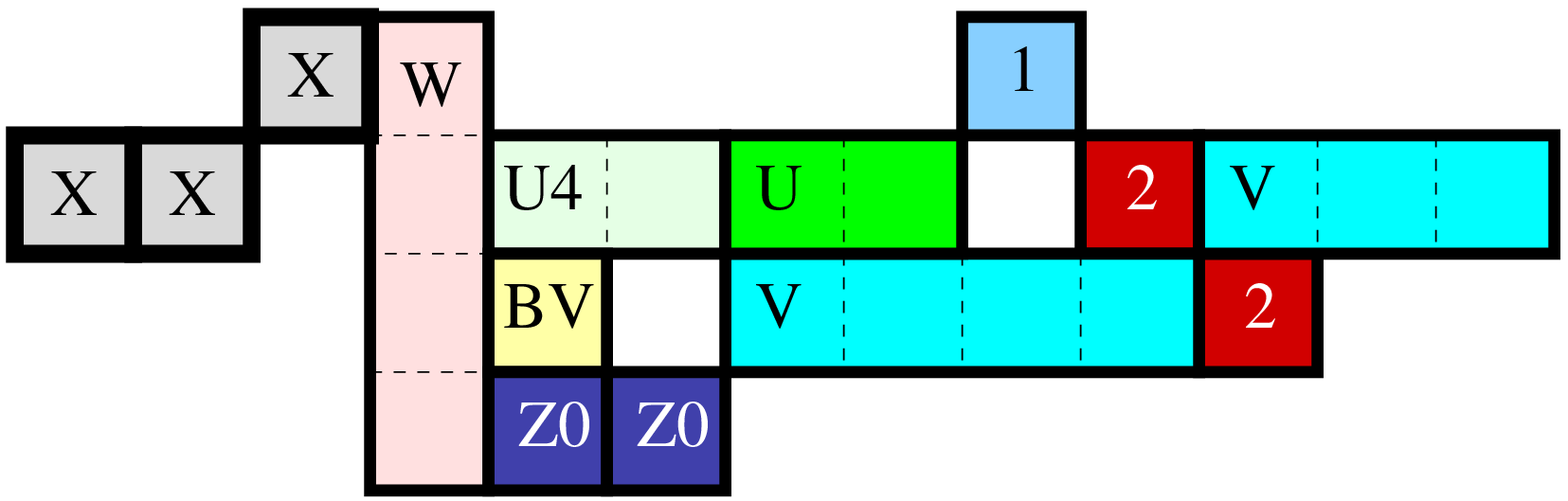}}
\hfill}
\vskip 5pt
\ligne{\hfill
\scalebox{0.50}{\includegraphics{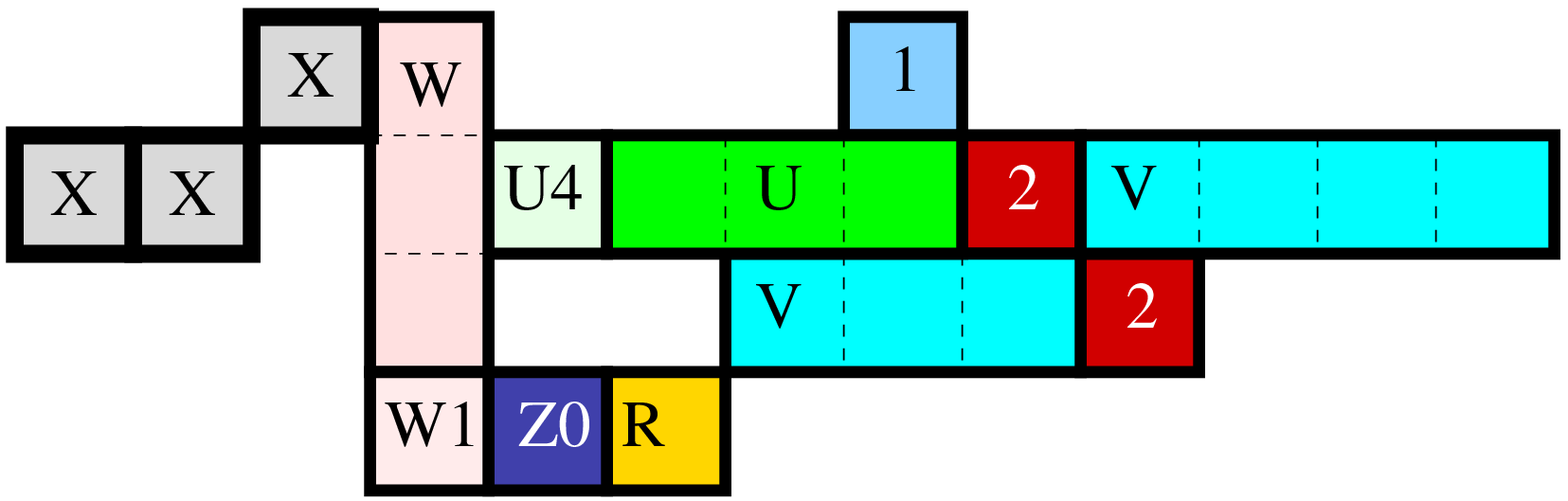}}
\hfill}
\vskip 10pt
\begin{fig}\label{etape12E}
Case $a$+$r = \mu$+$b$. 
The $Z0Z0$ pattern arrives at its destination.
\vskip 0pt
Above: The first $BV$ has just been just changed to~$B$.
The $V$- and the $U$-rows are being to be lifted.
Below: One $Z0$ disappears, corresponding to the change of the 
bottommost~$W$ to~$W0$. Note that the $R$-row below the blank is starting to 
be restored.
\end{fig}
}

\vskip 10pt
\vtop{
\ligne{\hfill
\scalebox{0.50}{\includegraphics{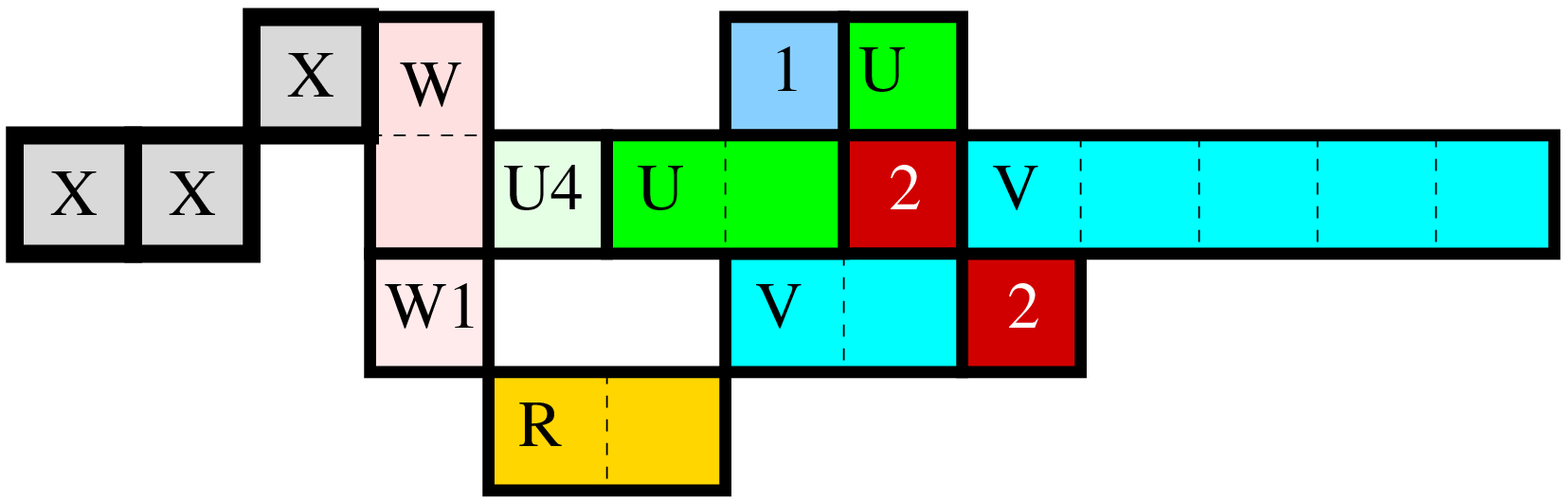}}
\hfill}
\vskip 10pt
\begin{fig}\label{etape13E}
Case $a$+$r = \mu$+$b$. 
Starting from this configuration, the rules of the 
case \hbox{$a$+$r > \mu$+$b$} allow the automaton to complete the computation.
\end{fig}
}

%%the next step: occurrence of~$Z0$.

\subsection{The case when $\mu <0$}
\label{scenarneg}

   In the case when $\mu<0$, we try to keep to the previous scenario as
much as possible. In order to do this, we change the implementation of the
data. This new display is illustrated in Figure~\ref{etape0neg}. In the new
display, first, the vertical~$v$ of the left-hand side border of 
the $V$-row coincide with the vertical line~$\chi$ which passes through the 
right-hand side of the rightmost~$X$, which is the most recent written pixel 
of the discrete line. Second, the vertical~$u$ of the left-hand side border
of the $U$-row is obtained by shifting~$v$ to the east by 
$\vert\mu\vert$~squares, see Figure~\ref{etape0neg}.

   In this situation, the construction of the $W$-column is a bit different than
in the case when $\mu>=0$. Indeed, when $\mu<0$, the $W$~which is still created
as the eastern neighbour of the ultimate~$X$ has the blank as its southern 
neighbour. This makes it possible that there is no~$R$-row in the case 
when $r=0$. In this situation, the erasing of the leftmost~$R$ by~$W$ makes
no difference with the writing of~$W$ on a blank cell. And so we decide to
mark the situation when $r=1$ by the writing of~$WR$ instead of~$W$. Indeed,
the southern neighbour of the $W$~which is just written on the $V$-row
knows whether $r=0$, $r=1$ or $r\geq2$. If its blank or of its has an~$R$ as
its eastern neighbour, it may be replaced by~$W$, as $W$~will distinguish
between the case $r=0$ and~$r\geq2$. If the southern neighbour of~$W$ is an~$R$,
this~$R$ knows whether it is alone or not: this is why it can select~$WR$
or~$W$ respectively.
\vskip 10pt
\vtop{
\ligne{\hfill
\scalebox{0.50}{\includegraphics{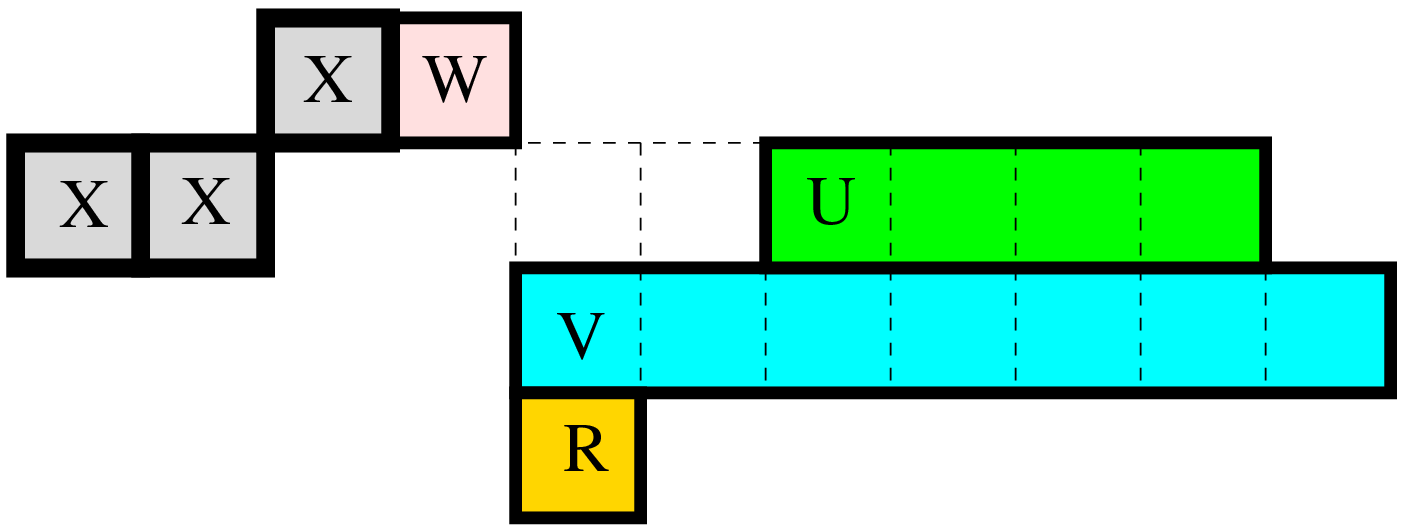}}
\hfill}
\vskip 10pt
\begin{fig}\label{etape0neg}
\leurre
Case $\mu < 0$, the initial data in a starting configuration. Here, $a=4$,
$b=7$, $r=1$ and~$\mu=$$-$$2$.
\end{fig}
}

   Another difference consists in the making of the $U$- and the $V$-rows 
respectively. Here, the situation is somehow symmetric to the one we had in
the study of the case when $\mu\geq 0$. In particular, the marking of the
zones of $U$'s and~$V$'s is the same, but as the blank occurs on the $U$-row,
we have that~$BV$ and~$CV$ are replaced by~$BU$ and~$CU$ respectively. We also 
have that the copies of elements of the $U$-row crosses a blank zone,
which raises no problem. The new situation is illustrated by 
Figure~\ref{etape1neg}. We can see that when the elements to be copied reaches
the $BU$-area, it crosses it as~$R2$. When it reaches the $W$-column, the
rules for the case when $\mu\geq0$ apply and allow to perform what is needed
in the $R$-row.

   With this point, we can see that afterwards, the motion is like the case 
when~$\mu\geq0$. In particular, the comparison of $a$+$r$ with $\mu$+$b$
makes use of the same rules as previously. From the display of the data,
we compare $a$+$r$ with~$b$ directly, as the $R$-row is aligned with 
the~$V$-row. And so the three possible cases are exactly determined in the
same way as previously as from the level of the $R$-row, any cell can see 
what happens on the level of the $V$-row only.

   Note that in the case when the right-hand side limit of the $U$-row
would be to the east of the eastmost element of the $V$-row, this induces
a small change in the scenario. Instead of going upwards along of the
column of~$V3$, the various signals which are triggered by~$V3$ would
go to the east on the $V$-row, until they can see the eastmost element
of the $U$-row and there, they would again behave as in the case when
$\mu\geq0$. However, there are two points where some tuning is needed.
In the case when the eastmost $U$~lies further to the east than the
eastmost~$V$, the final signal, $FF$ or~$1$ would trigger the 
transformation of~$U4$ to~$U$. Some care has to be observed when
this signal arrives at the column of~$V3$ in order that going further to the
west, things happen as they do in the case when $\mu\geq0$. This can be
performed by additional rules and the situation is clearly determined by the
fact that the northern neighbour of~$V3$ is~$U4$.

\vskip 10pt
\vtop{
\ligne{\hfill
\scalebox{0.50}{\includegraphics{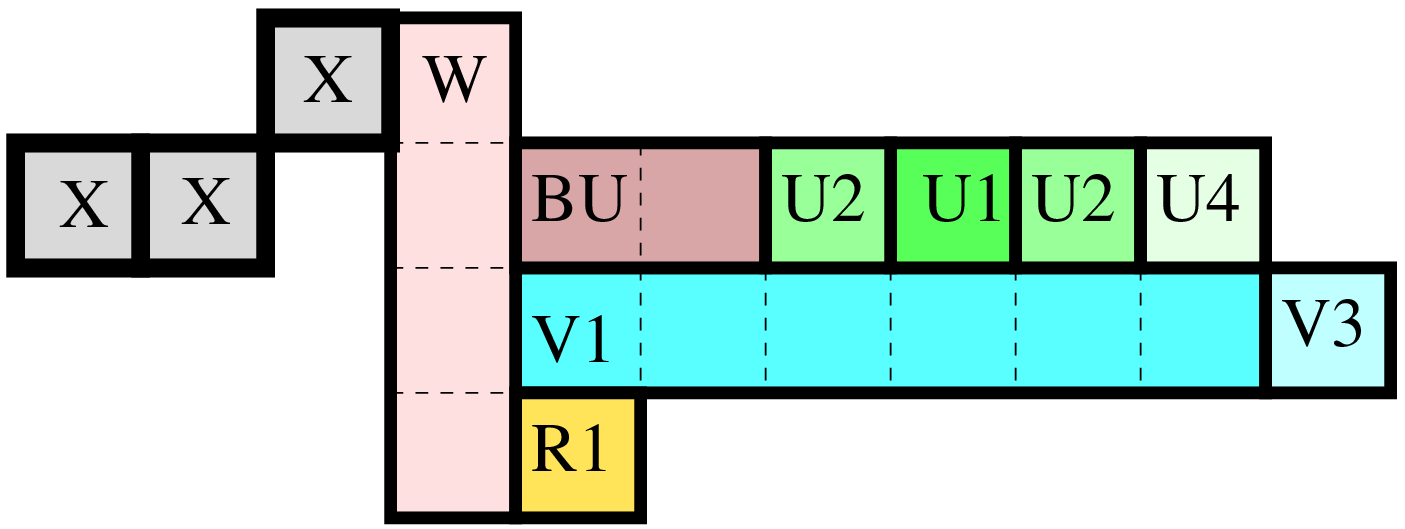}}
\hfill}
\ligne{\hfill
\scalebox{0.50}{\includegraphics{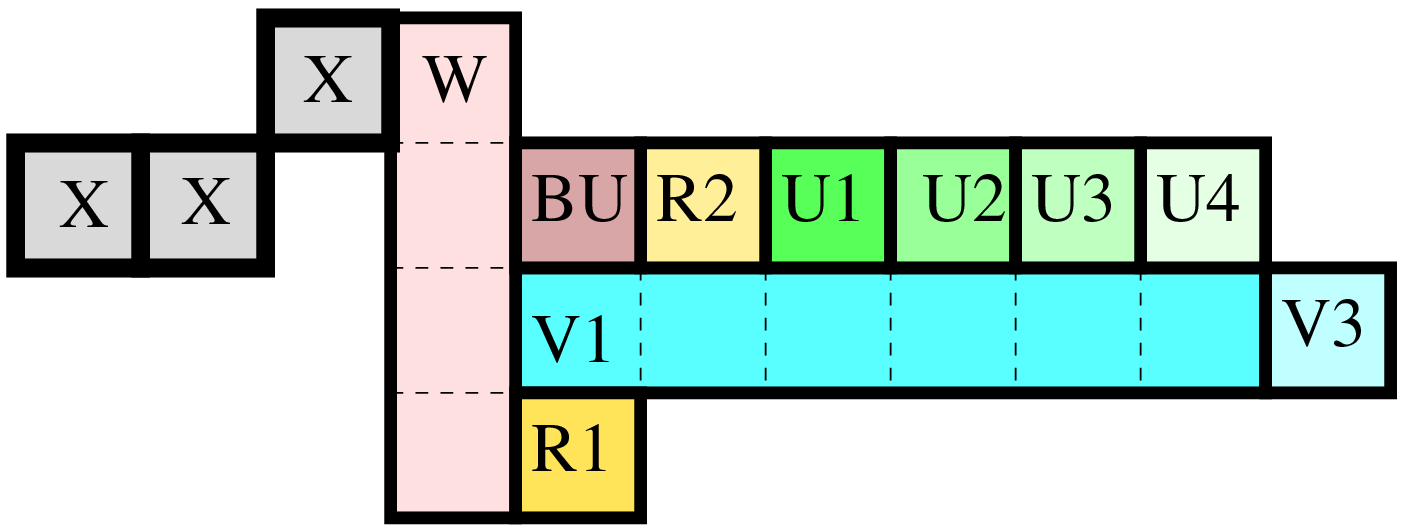}}
\hfill}
\ligne{\hfill
\scalebox{0.50}{\includegraphics{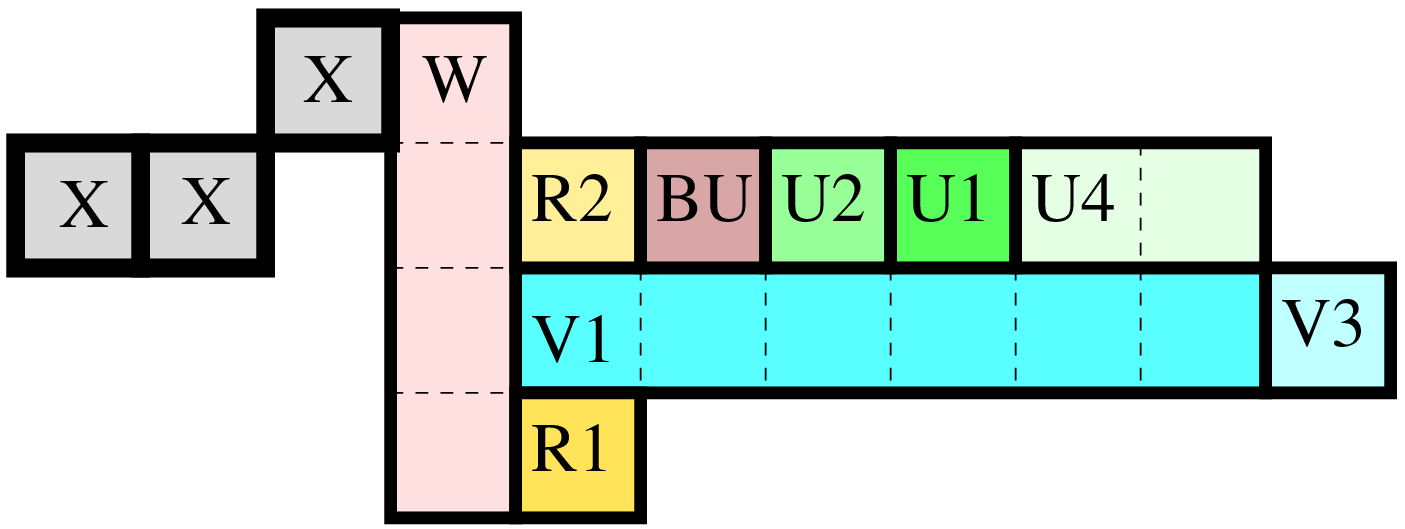}}
\hfill}
\ligne{\hfill
\scalebox{0.50}{\includegraphics{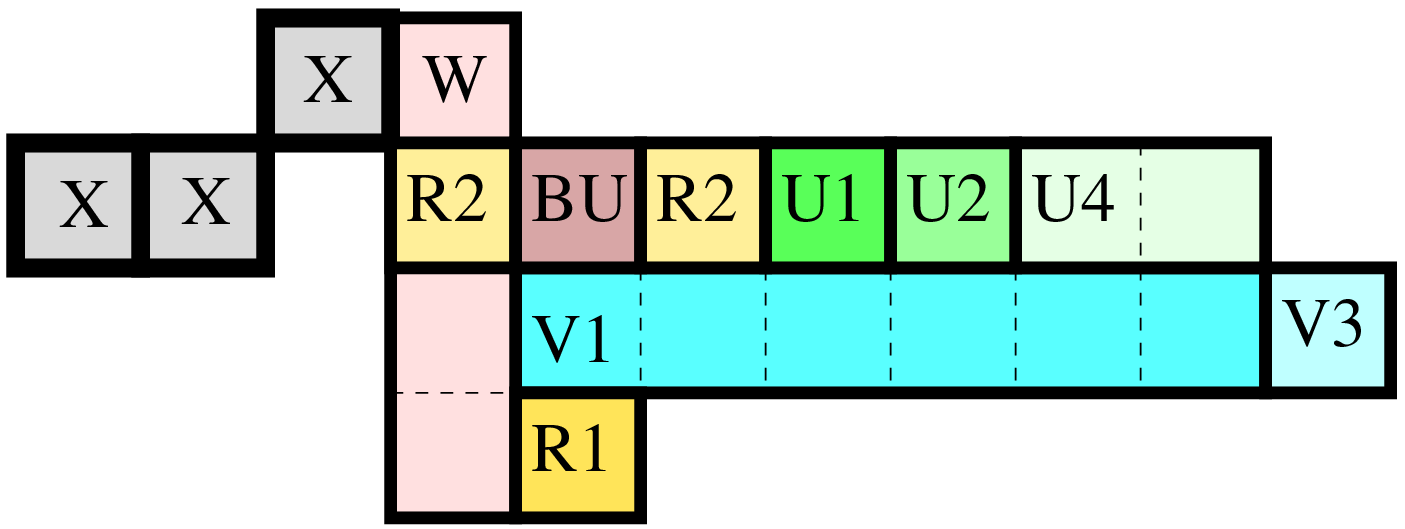}}
\hfill}
\vskip 10pt
\begin{fig}\label{etape1neg}
\leurre
Case $\mu < 0$: copying the elements of the $U$-row. Here, the copies through
the $BU$-area.
\end{fig}
}

   The second point we have to look at is the case when $a$+$r<b$, regardless
of the respective position of the eastmost~$U$ and the eastmost~$V$. Due to
the fact that we have $BU$'s between the $W$-column and the westmost~$U$,
the step when the cycle of computation has to turn to a new cycle must be
somehow different.

  As illustrated by Figure~\ref{etape2neg}, the scenario is the following.
In the final part of the cycle of computations when $a$+$r<b$, the westmost
element of the $U$-row is still in the state~$U4$, waiting for the signal~$FF$
in order to be changed to~$U$. When $FF$ arrives as the northern neighbour 
of~$U4$, $FF$ goes on to the west, but $U4$ becomes~$U2$: this is to prevent
a transformation of the new~$U$ into~$CV$ as~$BU$ has not yet been changed.
Then, when $FF$ is the northern neighbour of~$BU$, this~$BU$ is changed
to~$B0$ and $U2$~turns to~$U$ as $V$~is a southern neighbour of~$U2$. 
Later, $FF$ and $B0$ go by one step to the west, the column of~$FF$ begin 
ahead of that of~$B0$ by one step: $B0$~leaves a blank in the cell it
previously occupied. This motion goes on until $FF$ can see~$W$
through its western side. Then, $B0$ still advances by one step and $FF$
becomes~$B0$ and as $W$~can see~$FF$ through its eastern side, it becomes~$X$. 
Thus, we have a small column of~$B0$ against what remains of the $W$-column.
At the next step, both $B0$ vanish, leaving a blank in their places: this
configuration is the last one of the cycle: at the next step, we have the 
starting configuration of the new cycle.

\vskip 10pt
\vtop{
\ligne{\hfill
\scalebox{0.50}{\includegraphics{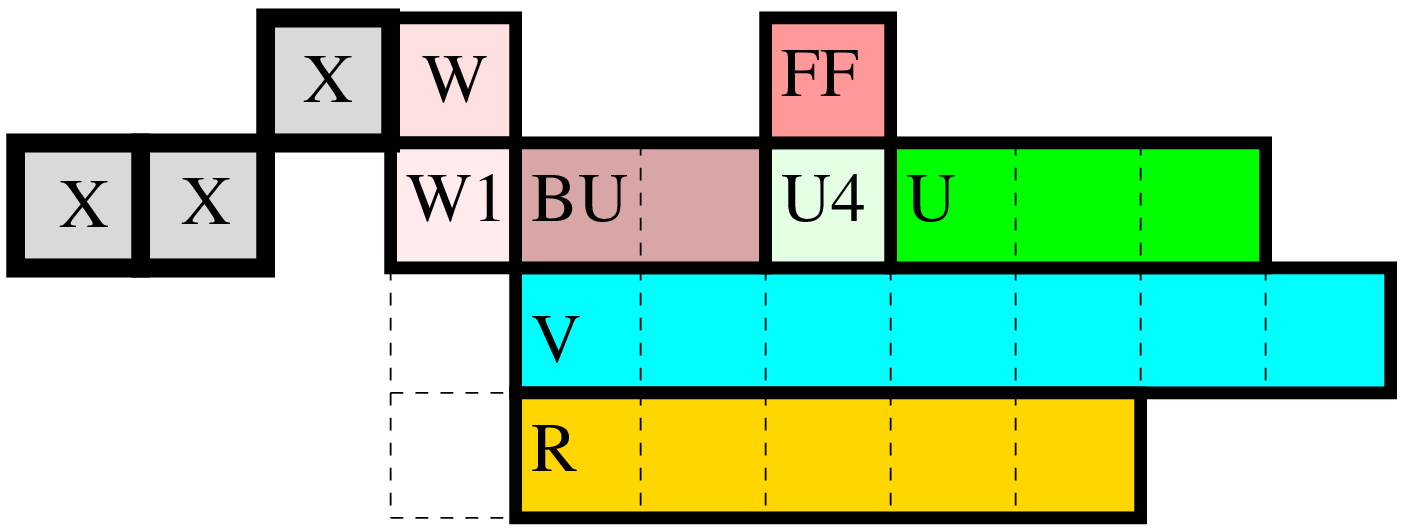}}
\hfill}
\ligne{\hfill
\scalebox{0.50}{\includegraphics{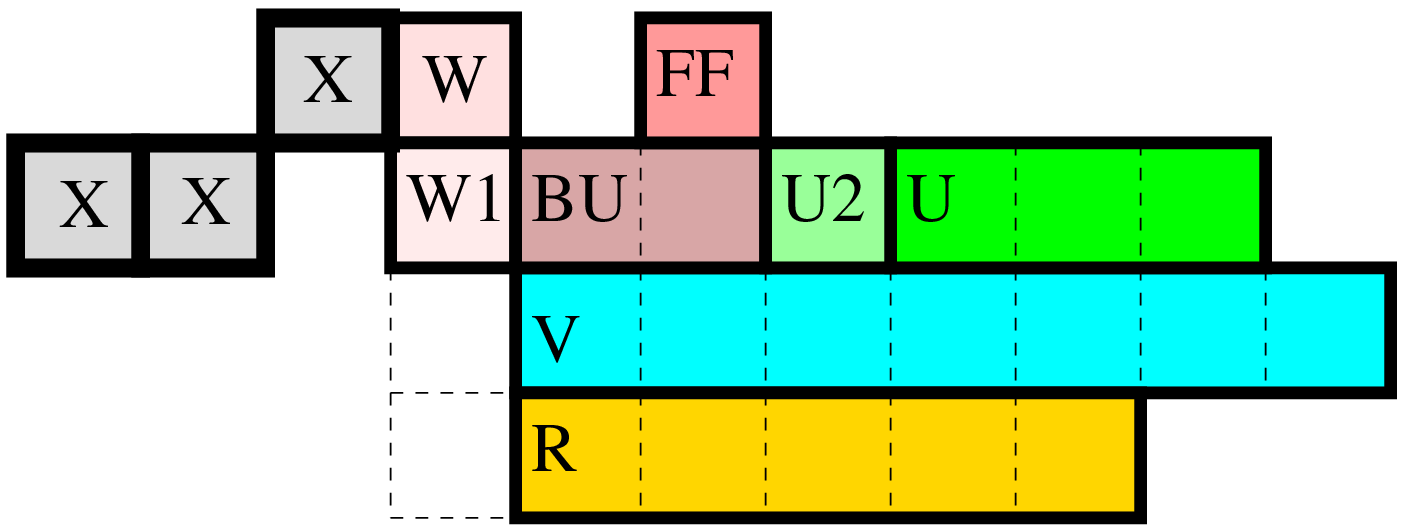}}
\hfill}
\ligne{\hfill
\scalebox{0.50}{\includegraphics{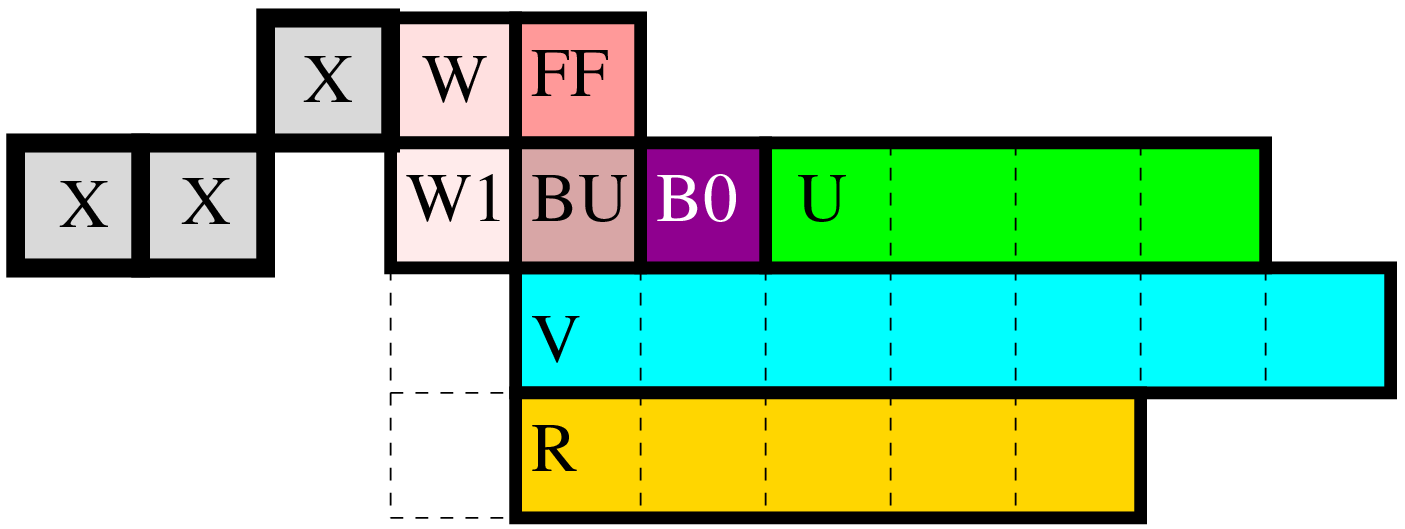}}
\hfill}
\ligne{\hfill
\scalebox{0.50}{\includegraphics{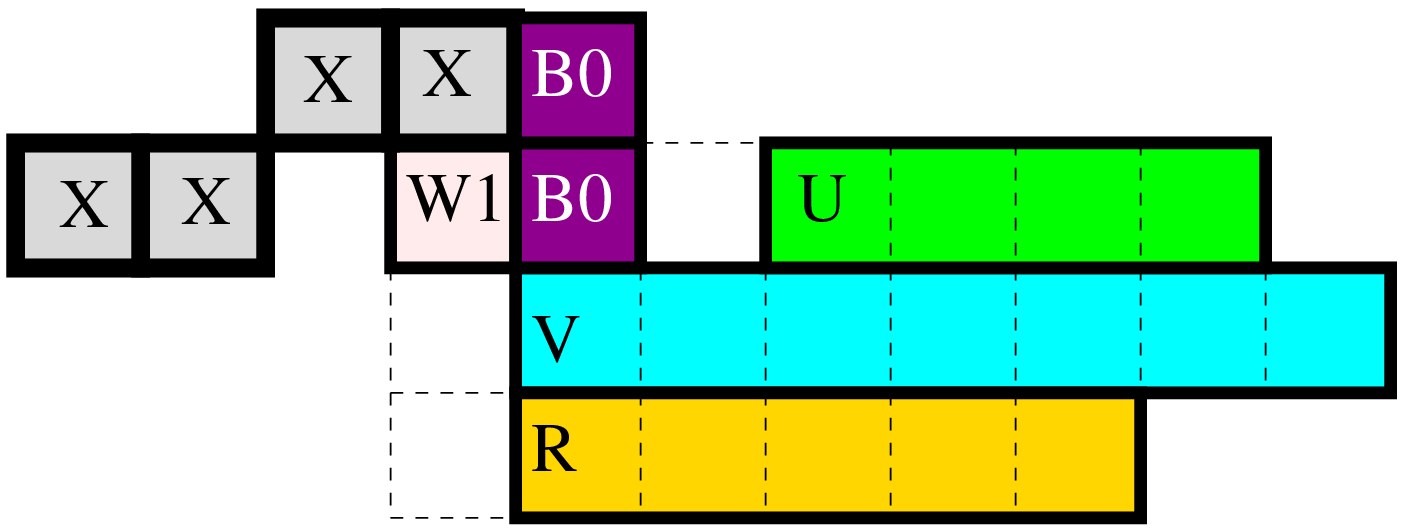}}
\hfill}
\ligne{\hfill
\scalebox{0.50}{\includegraphics{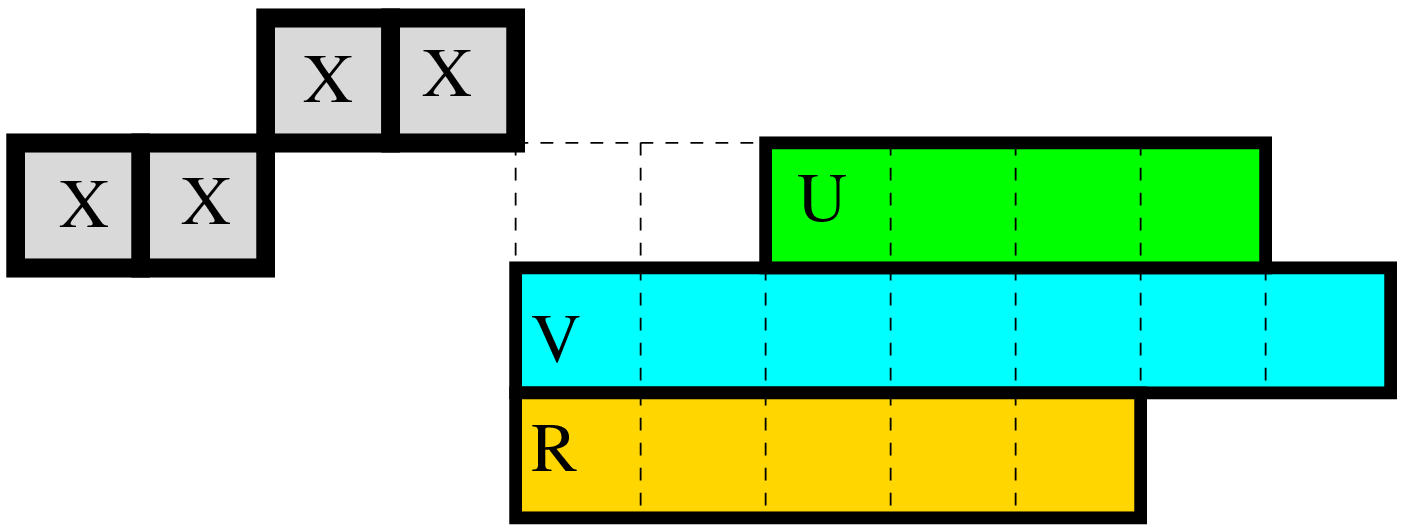}}
\hfill}
\vskip 10pt
\begin{fig}\label{etape2neg}
\leurre
Case $\mu < 0$: the end of the cycle of computations when $a$+$r<b$.
The next step after the lower picture is the first step of a new cycle of
computations.
\end{fig}
}

   We conclude this study of the case when $\mu<0$ by the following important
remark. All illustrations of this section,from Figure~\ref{etape0neg} 
to Figure~\ref{etape2neg} are performed under the assumption that the eastmost
element of the $U$-row is in the column which is not to the east of the column of the
eastmost element of the $V$-row. This is the case in the situation when $\mu\geq0$
as we assume that $a\leq b$. Now, here, if we number by~0 the column of the $W$-column,
the place of the eastmost element of~$U$ after the shift by one step to the east
is $\vert\mu\vert$+$a$ which may be bigger than~$b$, the number of the column of the
eastmost element of the $V$-row. However, as we may assume that 
$\vert\mu\vert< b$, the westmost element of the $U$-row sees a ~$V$ through its
southern neighbour.
In this case, we have to change a bit the strategy.
If the automaton realizes that \hbox{$\vert\mu\vert$+$a>b$}, it will change the mark 
of the eastmost element of the $V$-row which will be~$V4$ instead of~$V2$. The end of 
the configuration will be determined by the eastmost element of the $U$-row.

   The changes are the following. 
   First, the copying of the $U$-row is triggered later: $V2$ cannot issue~$C$ on the
line of the $U$-row, so that it issues it on the $V$-row. The signal~$C$ moves on this
line to the east, as long as it sees~$U$ through its northern side. When its blank 
eastern neighbour sees also a blank through its northern neighbour, it becomes~$C$ and, 
at the next time, this northern neighbour also becomes~$C$ leaving a mark~$H0$
on the line of the $V$-row. Now, as $C$~is now on the $U$-row and as it sees the 
eastmost element of the $U$-row, the standard scenario can take place.
   
   During the second and the third stage, $V4$ allows to perform the comparison
of~$a$+$r$ with~$b$. If \hbox{$a$+$r<b$}, the coloration with $RF$ will go until
$H0$ is seen, so that what happened before with $G0$ will happen with~$H0$.
If  \hbox{$a$+$r\geq b$}, again the colorations described in the present sub-section
can take place with $H0$ playing the role of~$G0$. Also, the lifting of the data
can be performed by signals~1 and~2 as described previously.

\subsection{For all the cases}
\label{allcases}

   It is not very difficult to adapt the above scenario when $a$ and~$b$ do not
satisfy the condition $0 < a\leq b$, with $b>0$. 

   First, let us assume that both $a$ and~$b$ are non-negative integers. We have
just to see what to do when $a>b$. In this condition, we are in the other half of
the quarter of the plane defined by the condition $x\geq 0$ and $y\geq 0$.
Now, it is not difficult to see that if we exchange $x$ and~$y$, a discrete line
below the first diagonal line is  transformed into a discrete line above the
diagonal. However, it is not enough to perform a reflection in the first 
diagonal line which means exchanging the role of~$a$ and~$b$. We have
to also change the value of~$\mu$. We have to remember that in full generality,
the equation of a na\"\i ve discrete line is of the form
$\mu\leq ax-by < \mu+\max\{\vert a\vert,\vert b\vert\}$. If we exchange $x$ 
and~$y$ we get $\mu\leq ay-bx < \mu+\max\{\vert a\vert,\vert b\vert\}$,
which means, changing the signs:
$-\mu-\max\{\vert a\vert,\vert b\vert\} < bx-ay \leq -\mu$. In order to
get the same form, using that the inequalities apply to integers:
$-\mu-\max\{\vert a\vert,\vert b\vert\}+1 \leq bx-ay < -\mu+1$.
Accordingly, if we exchange the role of~$x$ and~$y$, we have also to replace
$\mu$ by $-\mu+1$.

   This means that we apply the reflection in the first diagonal to the data 
too and that we take into account the change for~$\mu$. And so, the
data are placed in columns along the $y$-axis and 
the construction of the line is still performed by advancing northwards or 
eastwards as previously, but the meaning is opposite: we go upwards when
$bx-ay<r$ and we go to the east in the other cases.
Note that the data are now to west of the line instead of being to their 
eastern side.

   From this, it is easy to perform the construction in the other quarters.
As the linear form occurring in the inequation is always $ax-by$, the sign 
of the coefficients defines the quarter of the plane where the line has to be constructed. 
Next, in the appropriate quarter, the comparison between $\vert a\vert$ 
and~$\vert b\vert$ defines which is the place of the line with respect to 
the bisector of the angle defined by the quarter. More details about this 
implementation will be given in Section~\ref{rules}.

   Now we have all the information needed for the construction of the rules.

\section{{\Large The rules}}
\label{rules}

    Remember that the form of the rules is defined by the diagram illustrated by 
the left-hand side picture of Figure~\ref{voisinages}. We shall represent the 
rules of the automaton in the following format:

\ligne{\hfill
$\eta_0\eta_1\eta_2\eta_3\eta_4\eta^1_0$
\hfill}   

\noindent
where $\eta_0$~is the current state of the cell, $\eta_i$, $i\in\{1..4\}$
the states of the cells and~$\eta^1_0$ the new state of the cell.
Remember that this numbering of the neighbours is given to the cells in
increasing numbers while counter-clockwise turning around the cell, 1~being
the number of the northern neighbour. Accordingly, the correspondence can be
given by the following diagram:
\vskip 10pt
\ligne{\hfill
\hbox to 30pt{\hfill$\eta_1$\hfill}
\hbox to 30pt{\hfill$\eta_2$\hfill}
\hbox to 30pt{\hfill$\eta_3$\hfill}
\hbox to 30pt{\hfill$\eta_4$\hfill}
\hfill}
\vskip 2pt
\ligne{\hfill\tt
\hbox to 30pt{\hfill north\hfill}
\hbox to 30pt{\hfill west\hfill}
\hbox to 30pt{\hfill south\hfill}
\hbox to 30pt{\hfill east\hfill}
\hfill}
\vskip 10pt

\subsection{General conditions}
\label{genconds}

   In order to define the rules, we start from the configurations indicated 
in Section~\ref{scenario}. Our first observation is that we have
two kinds of rules: the {\bf conservative} ones and the {\bf active} ones.
A conservative rule is a rule in which the new state of the cell is the 
same as the current one. An active rule is the opposite: the new state is 
different from the current one. This remark is important: the active rules 
are derived from the propagation of the various signals described in the 
scenario and the conservative ones are needed for keeping a part of the 
configuration unchanged as long as it is needed.

   Another point which we have to take into account is that the scenario involves
situations which induces a lot of rules due to the discrete nature of the cellular
automaton. It is not possible to describe here all the rules induced by these
particular cases. In fact, the particular cases can be described by a few
parameters. We have four parameters which determine the initial configuration:
$a$,$b$, $r$ and~$\mu$. What we shall call the {\bf general case} and for
which we shall see the rules in this section, are the initial configurations
in which $a$, $b$, $r$, $\mu$, as well as \hbox{$\vert a$$-$$b\vert$},
\hbox{$\vert a$$-$$r\vert$}, \hbox{$\vert a$$-$$\vert\mu\vert\vert$}, 
\hbox{$\vert b$$-$$r\vert$}, \hbox{$\vert b$$-$$\vert\mu\vert\vert$}
and \hbox{$\vert r$$-$$\vert\mu\vert\vert$} are large. In practice,
this means that the particular cases are defined by the configurations
when at least one of these parameters are less than~4. When the other 
parameters are at least~4, the rules are the same for all cases defined by
a fixed value of the considered parameter. As will be clear from the figures
of Section~\ref{scenario}, the rules needed for the particular cases introduce
shortcuts leading from one phase of the cycle to the next one. An example is
given by the figures of Section~\ref{scenario} where, for instance, 
$\vert\mu\vert=2$, a particular case. As an example, the situation concerning
the $BV$- or $BU$-areas at the beginning of the area and the situation
concerning the end of the area address consecutive steps in the computation:
any rule regarding a cell of the area implies a neighbouring cell which does not
belong to the area. And so, there are specific rules accordingly.

Now, during the construction of the rules, as our automaton is deterministic,
we have to always check the following condition: if two rules
$\eta_0\eta_1\eta_2\eta_3\eta_4\eta^1_0$ and
$\omega_0\omega_1\omega_2\omega_3\omega_4\omega^1_0$ 
satisfy $\eta_i=\omega_i$ when $i\in\{0..4\}$, then $\eta^1_0=\omega^1_0$.
If this condition is satisfied for all pairs of rules, we say that the rules 
are {\bf compatible}. If the condition is not satisfied by a pair of rules
$\rho_1$ and~$\rho_2$, we say that $\rho_1$ and~$\rho_2$ are 
{\bf incompatible} or that $\rho_i$ is in {\bf contradiction} with
$\rho_j$, where $\{i,j\}=\{1,2\}$. 

   According to the scenario, we first derive the rules for moving the data 
by one state to the right.

\subsubsection{Conservative rules}
\label{consrules}

   We start with the conservative rules, as most of the configuration remains 
unchanged during the first steps of the computation.

   Remember that the rule for the blank, namely \hbox{\small\tt B B B B B B} is 
a conservative rule.
We have another group of conservative rules linked to the state~$X$: once it is 
written, it is never replaced by another state. We say that $X$~is a 
{\bf non-erasing} state.  For such a state we write a {\bf meta-rule} which 
allows to gather several rules
under the same pattern: \hbox{\small\tt X $\eta_1$ $\eta_2$ $\eta_3$ $\eta_4$ X}.

   We can distinguish several groups of conservative rules: the blank cells which
are a neighbour of the data. Here too, we can devise meta-rules for two groups of 
blank cells: those which are to the west of the configuration and those which 
are to its south. Indeed, from the scenario, assuming $0< a\leq b$, we know 
that the configuration moves to the east or to the north, never in the other 
directions. The corresponding meta-rules
are: \hbox{\small\tt B B B B $\eta$ B}, \hbox{\small\tt B $\eta$ B B B B} and 
\hbox{\small\tt B $\eta_a$ B B $\eta_b$ B}. 
However, the other neighbours
of the data are also unchanged, except the cell which sees~$X$ through its 
western side. Accordingly, we also have the following meta-rules:
\hbox{\small\tt B B $\eta$ B B B}, when $\eta\in\{U,V,R\}$  and 
\hbox{\small\tt B B B $\eta$ B B}
when $\eta\in\{U,V,R,X\}$. Now, due to the relative positions of the $U$-, $V$-
and $R$-rows and the position of the $U$-row with respect to~$X$, we have other
conservative-rules. We have \hbox{\small\tt B B B X X B} and 
\hbox{\small\tt B X X B B B}, as the neighbours
of the discrete line are unchanged, except the already mentioned situation.
%We have \hbox{\small\tt B B U V B B} when $a<b$ and 
%\hbox{\small\tt B V R B B B } when $0<r$ as $r<b$ is
%always true. 
Besides almost blank neighbours of the data, the cells of the data
are also applied conservative rules, as long as the signals of the computation
did not reach them. Consequently, we have the following conservative rules 
with~$U$-, $V$- and~$R$-cells: \hbox{\small\tt U B U V U U} as $a\leq b$ 
when $\mu\geq 0$. Note that when $a=1$, we have
\hbox{\small\tt U B B V B U}, an example of a conservative rule in a particular
case. 

\subsubsection{Active rules: general principles}
\label{genactive}

   If we look at the scenario, many motions are linear: a few symbols are
moving on a row or a column, always in the same direction as long as this
motion is needed during the stage of the cycle in which it occurs. 
For such a motion, remember what we did in Sub-subsection~\ref{shift},
were we have written the corresponding $1D$-rules.
As an example, consider a motion on a row. Then, if the motion goes
to the east, for instance, we can write 
\hbox{$\eta_0\eta_2\eta_4\eta^1_0$}.
As an example, consider $R2$ moving on a row of~$R1$'s. We have
two motion $1D$-rules: \hbox{$R1\,R2\,R1\,R2$}, \hbox{$R2\,R1\,R1\,R1$} 
and the conservative $1D$-rule: \hbox{$R1\,R1\,R2\,R1$} which says that
the $R1$~which sees $R2$~going away remains~$R1$.
These rules are written \hbox{$R1\,\eta_1\,R2\,\eta_3\,R1\,R2$}
\hbox{$R2\,\eta_1\,R1\,\eta_3\,R1\,R1$}
and \hbox{$R1\,\eta_1\,R1\,\eta_3\,R2\,R1$}. From the initial
configuration, we know that $\eta_3$~is always~$B$. Now, $\eta_1$, which
is the state seen by the cell through its northern side, may take a priori
a lot of values: $B$, $BV$, $CV$, $V$, $V1$, $V2$ or $V3$.
In fact, if we carefully the scenario, when $R2$~crosses a row of~$R1$'s,
the $U$~is progressively transformed in a row of~$U4$'s and all~$V$'s
of the $V$-row are transformed into~$V1$ and there is an additional~$V3$
at the eastern end of the $V$-row. Accordingly, $\eta_1=BV$, 
$\eta_1=V1$, $\eta_1=V3$ and $\eta_1=B$ are possible and only them.

Another
situation is the {\bf coloration} of an interval on a row or a column.
This coloration consists in replacing one colour by another, step by step,
from one end of the interval to the other. As an example, take the coloration
of the $V$'s of the $V$-row into~$V1$'s. Once the coloring started, it
works on the basis of two $1D$-rules: \hbox{$V\,V1\,V\,V1$} and
\hbox{$V1\,V1\,V\,V1$}, contamination and persistence respectively.
The full rules are \hbox{$V\,\eta_1V1\,\eta_3\,V\,V1$} and 
\hbox{$V1\,\eta_1\,V1\,\eta_3\,V\,V1$} respectively. Now, later in 
the cycle, we have the opposite transformation,with the $1D$-rules
\hbox{$V1\,V1\,V\,V$} and \hbox{$V\,V1\,V\,V$}. Here, the contamination rule
is in contradiction with the persistence rule of the previous case. Now,
the full rules are \hbox{$V1\,\eta_1\,V1\,\eta_3\,V\,V$}
and \hbox{$V\,\eta_1\,V1\,\eta_3\,V\,V$} respectively. Accordingly, if
the couple $\eta_1$, $\eta_3$ used in one direction is different from
the couple $\eta_1$, $\eta_3$ used in the opposite direction,then the
rules are compatible. We shall intensively use this principle.

\subsection{The rules for the general case of the scenario}
\label{rulesgenscen}

   With the help of the above guidelines, we turn to the description of
the active rules needed by the execution of the scenario we described in
Section~\ref{scenario}. This means that we assume that $a<b$. We also
consider the case when $\mu<0$ but, in this latter case, the rules
which we indicate here do not cover the case when 
\hbox{$\vert\mu\vert$+$a>b$}. In most cases, the rules implied for an action
are active. We mention conservative rules when they are needed for the understanding
of a coloration process. We shall not mention the conservative rules generated
by a passive part of the configuration during a given stage. 

\subsubsection{Rules for the $W$-column and motion of the data by one step to
the east}
\label{Wcol1step}

  When $\mu\geq0$, the first active rule is given by 
\hbox{\small\tt B B X U B W}, which opens the starting configuration. 
Now, the presence of~$W$ as an eastern neighbour of the anchor triggers the 
construction of the $W$-column which first replaces the first element of the
$U$-, $V$- and $R$-rows by~$W$. In the case when $\mu\geq0$, $W$
replaces the first elements of the row. If the $W$-column erases the
single~$R$, then $R$~is replaces by~$WR$ for one step and then $W$ 
replaces~$WR$. This happens when $r=0$ and $\mu=1$ or when
$r=1$ and $\mu=0$. The construction of the $W$-column induces the 
following rules:
\hbox{\small\tt U W B V U W}, %\hbox{\small\tt U W B V B W} when $a=1$, 
\hbox{\small\tt V W B R V W}, \hbox{\small\tt V W B B V W2} when $r=0$,
and \hbox{\small\tt R W B B R W}. The already mentioned case when
$r$+$m=1$ entails the rule \hbox{\small\tt R W B B B WR}. 
%\hbox{\small\tt R W B B B W2} when $r=1$ and 
%\hbox{\small\tt B W B B B W1} when $r=0$.

Now, $W$~also triggers the marking of the $U$-, $V$- and $R$-rows.
In Sub-subsection~\ref{shift}, we mentioned the $1D$-rules used
in this case. Applying the principles of Sub-subsection~\ref{genactive},
we get the following active rules:

\vskip 4pt
\ligne{\hskip 50pt\tt
\hbox to 60pt{U U0 U U0\hfill}\hskip 10pt$\Rightarrow$\hskip 20pt 
\hbox{\small U B U0 V U U0},\hskip 7pt 
\hbox{\small U B U0 B U U0}
\hfill}
\ligne{\hskip 50pt\tt
\hbox to 60pt{U W U U0\hfill}\hskip 10pt$\Rightarrow$\hskip 20pt
\hbox{\small U B W B U U0}, \hskip 7pt  
\hbox{\small U B W V U U0}
\hfill}
\ligne{\hskip 50pt\tt
\hbox to 60pt{U U0 B U0\hfill}\hskip 10pt$\Rightarrow$\hskip 20pt
\hskip 7pt {\small U B U0 V B U0}
\hfill}
\vskip 4pt

The first line shows the general rule which has two basic variants:
$\eta_3=B$ and $\eta_3=V$. The second line indicates the rules at the
ends of the interval of transformation. There the two variants for $\eta_3$
when $\eta_2=W$ and there is a single case when $\eta_4=B$, the end of the
$U$-row.

\vskip 4pt
\ligne{\hskip 50pt\tt
\hbox to 60pt{U0 U1 U U1\hfill}\hskip 10pt$\Rightarrow$\hskip 20pt 
\hbox{\small U0 B U1 V U U1},\hskip 7pt 
\hbox{\small U0 B U1 B U U1} 
\hfill}
\ligne{\hskip 50pt\tt
\hbox to 60pt{U0 W U U1\hfill}\hskip 10pt$\Rightarrow$\hskip 20pt 
\hbox{\small U0 B W V U U1}\hskip 7pt 
\hfill}
\ligne{\hskip 50pt\tt
\hbox to 60pt{U0 U1 B U1\hfill}\hskip 10pt$\Rightarrow$\hskip 20pt
\hskip 7pt {\small U0 B U1 V B U1}
\hfill}
\ligne{\hskip 50pt\tt
\hbox to 60pt{U0 W U U1\hfill}\hskip 10pt$\Rightarrow$\hskip 20pt
\hskip 7pt {\small U0 B W B U U1}
\hfill}
\vskip 4pt
Now, we have the transformations of the basic
$1D$-rule \hbox{$U0\,U1\,U\,U1$} and its variants \hbox{$U0\,W\,U\,U1$}
with  \hbox{$U0\,U1\,B\,U1$} and for the ends of the interval. 

   In Sub-subsection~\ref{shift}, we also mentioned conservative $1D$-rules
associated with the transformation of the $U$-row. We leave as an exercise for
the reader to develop these $1D$-rules into rules for our automaton. Similarly, we leave the writing of the rules needed for the $R$-row as their $1D$-analogs are 
obtained from the $1D$-rules for the $U$-row by changing~$U$ to~$R$, keeping the
same additional digits.

   As mentioned in Section~\ref{scenario}, the construction of the $U$- and
the $V$-rows do not follow the same lines. Indeed, in the $V$-row, the last 
element is identified as the last one, which is not the case, neither for the 
$U$-row nor for the $R$-one. Now, this makes things easier as pure coloration
rules are involved, those which we indicated in Sub-subsection~\ref{genactive}.

\subsubsection{Rules for adding~$a$ to~$r$ and for deciding whether
to subtract~$b$ or not}
\label{addstep}

   As known from Section~\ref{scenario}, adding~$a$ to~$r$ consists in
copying one by one the elements of the $U$-row in a parallel way.

   We know that this process starts when the signal~$C$ emitted by the eastmost 
element of the~$V$-row when it is in the state~$V2$ reaches the eastmost~$U1$
of the $U$-row. This $U1$ becomes~$U2$, whence the rule
\hbox{\small\tt U1 B U1 V1 C U2}. Each cell $U1$ of the $U$-row evolves according
to the cycle: \hbox{$U1\rightarrow U2\rightarrow U3\rightarrow U4$}. When
the cell reaches the state~$U4$, it remains in this state until an appropriate
signal appears. The cell remains in the state~$U1$ until its eastern neighbour
becomes~$U4$: at this moment, the above cycle starts. In the period when
the cell~$U1$ remains in this state, it simply passes each copy~$U2$ of an
already $U$ changed to~$U4$ according to the mechanism which we indicated
in Sub-subsection~\ref{genactive}. In $1D$-rules, this can be written as:
\vskip 4pt
\ligne{\hfill\tt
\hbox{U1  U1  U2  U2},\hskip 15pt
\hbox{U1  U2  U2  U2},\hskip 15pt
\hbox{U2  U1  U1  U1},
\hfill}

\ligne{\hfill\tt
\hbox{U1  U2  U4  U2},\hskip 15pt
\hbox{U2  U1  U4  U3},\hskip 15pt
\hbox{U3  U2  U4  U4},\hskip 15pt
\hbox{U4  U1  U4  U4}
\hfill}
\vskip 4pt
The first line corresponds to the transportation of~$U2$ to the west across 
the $U1$'s. The second line describes the cycle for~$U1$. The first rule of the
second line shows that the cycle is triggered when~$U4$ is the eastern neighbour
of the cell containing~$U1$, and the other rules describe the whole cycle.
Of course, additional rules, essentially conservative ones are needed and we
leave them to the reader as an exercise. To facilitate it, we indicate how
the $1D$-rules become rules of the automaton:

\ligne{\hfill\tt
\hbox{U1  U1  U2  U2}\hskip 10pt$\Rightarrow$\hskip 10pt
\hbox{\small U1 B U1 BV U2 U2},\hskip 7pt
\hbox{\small U1 B U1 V1 U2 U2},\hskip 7pt
\hfill}
\ligne{\hfill\tt
\hbox{U1  U2  U2  U2}\hskip 10pt$\Rightarrow$\hskip 10pt
\hbox{\small U1  B   U2  BV  U2  U2},\hskip 7pt
\hbox{\small U1  B   U2  V1  U2  U2},\hskip 7pt
\hfill}
\ligne{\hfill\tt
\hbox{U2  U1  U1  U1}\hskip 10pt$\Rightarrow$\hskip 10pt
\hbox{\small U2  B   U1  BV  U1  U1},\hskip 7pt
\hbox{\small U2  B   U1  V1  U1  U1},\hskip 7pt
\hfill}

\ligne{\hfill\tt
\hbox{U1  U2  U4  U2}\hskip 10pt$\Rightarrow$\hskip 10pt
\hbox{\small U1  B   U2  BV  U4  U2},\hskip 7pt
\hbox{\small U1  B   U2  V1  U4  U2},\hskip 7pt
\hfill}
\ligne{\hfill\tt
\hbox{U2  U1  U4  U3}\hskip 10pt$\Rightarrow$\hskip 10pt
\hbox{\small U2  B   U1  BV  U4  U3},\hskip 7pt
\hbox{\small U2  B   U1  V1  U4  U3},\hskip 7pt
\hfill}
\ligne{\hfill\tt
\hbox{U3  U2  U4  U4}\hskip 10pt$\Rightarrow$\hskip 10pt
\hbox{\small U3  B   U2  BV  U4  U4},\hskip 7pt
\hbox{\small U3  B   U2  V1  U4  U4},\hskip 7pt
\hfill}
\ligne{\hfill\tt
\hbox{U4  U1  U4  U4}\hskip 10pt$\Rightarrow$\hskip 10pt
\hbox{\small U4  B   U1  BV  U4  U4},\hskip 7pt
\hbox{\small U4  B   U1  V1  U4  U4},\hskip 7pt
\hfill}
\vskip 4pt
   The transportation of the copy of a $U$-element in the $W$-column
follows similar principles. This time, the copy travels as~$R2$ and the 
$1D$-rules are this time of the form \hbox{$\eta_0\eta_1\eta_3\eta^1_0$}
as the northern and western neighbours are primarily concerned:
\vskip 4pt
\ligne{\hfill\tt
\hbox{W R2 W R2},\hskip 7pt \hbox{R2 W W W}
\hfill}
\vskip 4pt
\noindent
giving rise to the rules:
\vskip 4pt
\ligne{\hskip 50pt\tt
\hbox to 60pt{W R2 W R2\hfill}\hskip 10pt$\Rightarrow$\hskip 20pt
\hbox{\small W R2 B W BV R2}
\hfill}
\ligne{\hskip 50pt\tt
\hbox to 60pt{R2 W W W\hfill}\hskip 10pt$\Rightarrow$\hskip 20pt
\hbox{\small R2 W B W U1 W},\hskip 7pt
\hbox{\small R2 W B W BV W}
\hfill}
\vskip 4pt
The rules for the ends of the $W$-column are:
\vskip 4pt
\ligne{\hskip 50pt\tt
\hbox{\small W W B W U2 R2},\hskip 7pt 
\hbox{\small R2 W B W U1 W}
\hfill}
\ligne{\hskip 50pt\tt
\hbox{\small W R2 B B R1 R2},\hskip 7pt
\hbox{\small W R2 B B R2 R2},\hskip 7pt
\hbox{\small R2 W B B R1 W} 
\hfill}
\vskip 4pt
\noindent
where the first line deals with the corner of the trajectory of the copy on
the level of the $U$-row; the second line deals with the other corner on 
the level of the $V$-row.

   We have seen that the transformation of the $R$-row is analogous to that
of the $U$-row and we know that the transportation of~$R2$ along the $R1$'s of
the $R$-row has be seen as an example in Sub-subsection~\ref{genactive}.
The travel of~$R3$ which represents the copy of the last $U$-element is similar to
that of~$R2$: it is enough to replace~$R2$ by~$R3$ in the corresponding rules.

Now, we arrive to the rules corresponding to the comparison of $a$+$r$
with $\mu$+$b$, illustrated by Figure~\ref{etape9}. These instructions are:
\vskip 4pt
\ligne{\hfill\small\tt
\hbox{B V1 R3 B B RR}, \hskip 15pt
\hbox{B V3 R3 B B Z}, \hskip 15pt
\hbox{B B R3 B B R}
\hfill}
\vskip 4pt
\noindent
with, from the left to the right: the case when $a$+$r<\mu$+$b$,
$a$+$r=\mu$+$b$ and $a$+$r>\mu$+$b$ respectively.

\subsubsection{Rules for the case when $a$+$r<\mu$+$b$}
\label{smallerstep}

   From the scenario, we know that in this case, there are two parallel
coloration processes on the level of the $R$-row: one to the left, transforming 
all $R1$'s to~$RR$ and one to the right, transforming all blanks to~$RF$.
For the coloration with $RR$, the rules are of the form
\hbox{\small\tt R1 $\eta_1$ R1 B RR RR} and
\hbox{\small\tt RR $\eta_1$ R1 B RR RR}, with $\eta_1=V1$ or $\eta_1=BV$. 
For the coloration with~$RF$ the rules are:
\hbox{\small\tt B V1 RF B B RF},
\hbox{\small\tt RF V1 RF B B RF}
and \hbox{\small\tt RF V1 RF B RF RF}.

   Now, we are interested by two phenomena: what are the rules when the 
$RR$-coloration reaches the $W$-column and what are the rules when $RF$
arrives to the column of the~$V1$ which is the western neighbour of~$V3$.

   When the $RR$-coloration arrives to the $W$-column, the rule
\hbox{\small\tt W W B B RR W1} places $W1$ at the bottom of the column
and this state waits there until a true $R$ appears through the eastern side.
So that we have time to see what are the rules at the other end.

   In this case, we know that the end of the $RF$-coloration is achieved 
when the blank which is the southern neighbour of~$V3$ can see~$RF$
through its western side. This is detected by the rule:
\hbox{\small\tt B V3 RF B B F}. This $F$-signal triggers a sequence of 
transformations along this column given by the following rules:
\vskip 4pt
\ligne{\hfill
\hbox{\small\tt V3 B V F B VF},\hskip 15pt  
\hbox{\small\tt B B B VF B UF},\hskip 15pt 
\hbox{\small\tt B B B UF B FF},
\hfill}
\ligne{\hfill
\hbox{\small\tt VF B V B B V},\hskip 15pt 
\hbox{\small\tt UF B B V B B}.
\hfill}
\vskip 4pt
The rules of the first line indicate that $V3$ triggers~$UF$ in the
column and on the upper row which itself triggers~$FF$ in the column and on
the upper row, which means that $FF$ is on the level of the last~$X$ written
by the automaton. The second line tells us that $VF$ leaves~$V$ on its place
when it vanishes and that $UF$ leaves a blank.

Now, we may wonder why the rule on~$V3$ has $\eta_2=V$ and not $\eta_2=V1$?
In fact, when the $V1$~which sees~$V3$ through its eastern side sees~$RF$
through its southern side, it also knows that at the next step $V3$~will
see~$F$ through its southern side. And so, it may start the process of
the back coloration of the $V$-row to~$V$. This is why the instruction
has $\eta_2=V$. This coloration of~$V1$'s back to~$V$ is possible
as the $V1$ seeing~$V$ through its eastern side sees~$RF$, and later~$RR$
through its southern side. And the~$V$ which can see~$V1$ through its western
side can see the same states through its southern side. These contexts are 
different from what was seen by~$V$ and by~$V1$ in the reverse process: 
the~$V$ and~$V1$ which could see each other had both~$R$ as the southern
neighbour. This is why the corresponding rules are compatible.
This remark explains us why $\eta_2=V$ in the rule changing $VF$ to~$V$.

   Now, the occurrence of~$F$ triggers the back coloration of the level of
the $R$-row to its initial configuration: the $RF$'s are transformed to blanks
and the~$RR$'s are replaced by~$R$'s. We leave the writing of the corresponding
rules to the reader as an exercise. We have just to notice that the front
of the transformation to the initial look on the $R$-row is by one column late
with respect to the front of the transformation back to $V$'s on the $V$-row. 
Accordingly, the front on the $V$-row reaches the $W$-column one step before
the front on the $R$-row. On the level of the $V$-row we have the rule
\hbox{\small\tt W W B W1 B W1} so that when the front on the $R$-row
reaches the $w$-column, we have the rule 
\hbox{\small\tt W1 W1 B B R B} which erases the $W1$~which stands on the $R$-row.
We also know that the front of transformation back to~$V$ on the $V$-row
triggers the transformation of~$U4$ back to~$U$ on the $U$-row. This is also 
possible because of the advance of the $V$-transformation by one step
on this new one. So that in the corresponding rules, both for~$U$ and~$U4$
we have~$V$ as the southern neighbour and not~$V1$ or we have $B$~as the southern
neighbour and not~$BV$: the rules are
\hbox{\small\tt U4 B U4 V U U} and \hbox{\small\tt U4 B U4 V B U}.

Now, we can see that the front on the $U$-row arrives to the $W$-column
one step after the arrival of the front on the $V$-row and so, at the same time
when the front on the $R$-row arrives to the $W$-column. This means that $W1$
is present in the $W$-column, on the level of the $V$-row. We have
the rule \hbox{\small\tt W W B W1 U4 W1} so that at the newt time, we have
again two consecutive $W1$ in the $W$-column and so, the lowest one disappears:
\hbox{\small\tt W1 W1 B B B W} and \hbox{\small\tt W1 W1 B B V W} if $\mu=0$. 
Now, at the time~$t$ just after the execution of one of the above rules, $FF$~is 
at one step from the $W$-column. Indeed, $FF$~moves to this~$W$ thanks to the
rules:
\vskip 4pt
\ligne{\hfill 
\hbox{\small\tt B B B B FF FF},\hskip 15pt
\hbox{\small\tt B B B U FF FF},\hskip 15pt
\hbox{\small\tt FF B B B B B},\hskip 15pt
\hbox{\small\tt FF B B U B B}.
\hfill
}
\vskip 4pt
\noindent
Note, that above the $U$'s of the $U$-row, $FF$ does not see~$U4$, but~$U$ as
the front on the $U$-row is ahead the position of~$FF$ by three steps. 
And so, at time~$t$, the rule \hbox{\small\tt B B W U4 FF FF} applies,
leading to the configuration illustrated by Figure~\ref{etape13}.
On the next step, $FF$~disappeared, $U4$ has been changed to~$U$
and the topmost $W$ has turned to~$X$ thanks to the rules:
\vskip 4pt
\ligne{\hfill\tt
\hbox{\small FF B W U4 B B},\hskip 15pt 
\hbox{\small U4 FF W1 B B U},\hskip 15pt
\hbox{\small W B X W1 FF X}
\hfill}
\vskip 4pt
This allows the remaining~$W1$ to also vanish, rule
\hbox{\small\tt W1  X   B   B   U   B}, which is the last step of the cycle.
At this moment, the rule \hbox{\small\tt B B X B B W} applies, producing
the starting configuration of a new cycle.
  
   We have to mention the specific rules for the case $\mu<0$. For the place 
of the $U$-row with respect to the $V$-row, we have symbols $BU$ and~$CU$ during
the copying process in between the $W$-column and the $U$-row. Now,we know that
the rules for these symbols are very similar to those for $BV$ and~$CV$. We have
simply to remember that $\eta_1$ is most often~$B$ but, at the last stage of 
the computation it is~$FF$. The very last part of this stage involves a new
state, $B0$, which appears only at this moment as we have seen in 
Section~\ref{scenario}. This symbol appears when $FF$ leaves the column of~$U4$
and enters the eastmost column of~$BU$. We know that $U4$ becomes~$U2$ before
turning to~$U$ and then, $BU$~becomes $B0$~before turning to~$B$. Indeed,
the main rules are:
\vskip 4pt
\ligne{\hfill\tt
\hbox{\small U4 FF BU V U U2},\hskip 15pt
\hbox{\small FF B B U4 B B},\hskip 15pt
\hbox{\small U2 B BU V U U},
\hfill}
\ligne{\hfill\tt
\hbox{\small BU FF BU V U2 B0},\hskip 15pt
\hbox{\small B0 B BU V B B},
\hfill}
\vskip 4pt
\noindent
as we do not mention the rules needed at the ends of the interval of~$BU$'s.

Now, when $FF$~can see~$W$ through its western side, its southern neighbour~$BU$
becomes~$B0$, continuation of the above rule on~$BU$, and $FF$ itself 
becomes~$B0$, while $W$ becomes~$X$:
\vskip 4pt
\ligne{\hfill\tt
\hbox{\small BU FF W1 V B0 B0},\hskip 15pt
\hbox{\small FF B W BU B B0},\hskip 15pt
\hbox{\small W B X W1 FF X}
\hfill}
\vskip 4pt
At the next step, both $B0$'s disappear and $W1$ also disappear:
\vskip 4pt
\ligne{\hfill\tt
\hbox{\small B0  B0  W1  V   B   B},\hskip 15pt
\hbox{\small B0  B   X   B0  B   B},\hskip 15pt
\hbox{\small W1  X   B   B   U   B},
\hfill}
\vskip 4pt
\noindent
which is the last step of the cycle as already noticed.
 
\subsubsection{Rules for the case when $a$+$r\geq\mu$+$b$}
\label{notsmallerstep}

   As in Section~\ref{scenario}, Sub-subsection~\ref{notsmaller}, we first
consider the case when $a$+$r > \mu$+$b$ and then the case when
$a$+$r = \mu$+$b$ as the latter will appear as a simplified version of the
former.  

When we have the configurations illustrated by Figure~\ref{etape9}, 
we know that the rules which are applied are
\vskip 4pt
\ligne{\hfill\small\tt
\hbox{B B R3 B B R},\hskip 15pt
\hbox{B V3 R3 B B Z},
\hfill}
\vskip 4pt
\noindent
the left-hand side instruction corresponding to the case when
$a$+$r > \mu$+$b$, the right-hand side one corresponding to
$a$+$r = \mu$+$b$.

   In the case when $a+r>\mu$+$b$, we know that the~$R$ written by the 
transformation of~$B$ into~$R$ triggers a coloration of $R1$'s back to~$R$
until the $R1$ which is the southern neighbour of~$V3$ sees~$R$ through its
eastern side. At this moment, this~$R1$ is replaced by~$Z$, rule
\hbox{\small\tt R1 V3 R1 B R Z}, which triggers the subtraction of~$b$ from the
$R$-row. From Section~\ref{scenario}, we know that we have two actions starting
from the appearance of~$Z$. On the left-hand side, $Z$ moves to the west, erasing
the $R1$'s and dragging the block of~$R$'s which are on its right-hand side.
On the right-hand side, the dragging of the block is performed by transforming
$R^\alpha$ into $(ZR0)^\alpha$. The rules for this latter transformation are:
\vskip 4pt
\ligne{\hfill\small\tt
\hbox{R B Z B R Z},\hskip 7pt
\hbox{R B Z B B B},\hskip 7pt
\hbox{Z B R0 B R R0},\hskip 7pt
\hbox{Z B R0 B R0 R0}.
\hfill}
\vskip 4pt  
Together with the two rules about $Z$, there are also rules about~$R0$
which is also transformed into~$Z$. The min rule, in this part of
the configuration is \hbox{\small\tt R0 B Z B Z Z}. Due to the second rule
on~$R$,above, there is a coloration to the west by $ZR0$. This requires
additional instructions taking into account that a greater part of the
$ZR0$-interval is now below the $V$-row. The rules are now:  
\vskip 4pt
\ligne{\hfill\small\tt
\hbox{Z V1 R1 B R0 R0},\hskip 15pt
\hbox{Z G0 R0 B R0 R0},\hskip 15pt
\hbox{Z V R0 B R0 R0},
\hfill}
\ligne{\hfill\small\tt
\hbox{R0 G Z B Z Z},\hskip 15pt
\hbox{R0 G0 Z B Z Z},\hskip 15pt
\hbox{R0 V Z B Z Z}.
\hfill}
\vskip 4pt  
Notice that three rules involve~$G$ and~$G0$. This corresponds to the 
successive transformations of the cell containing~$V3$. First, $V3$ turns to~$G$,
rule \hbox{\small\tt V3 B V1 Z B G}, and then turns to~$G0$,
rule \hbox{\small\tt G B V1 R0 B G0}, remaining in the state~$G0$ until
$G0$~can see~$B$ through its southern side. This will indicate that the shift
of $(ZR0)^\alpha$ is now below the $V$-row. Then, the second part of the 
process can take place and $G0$ becomes~$GG$, rule 
\hbox{\small\tt G0 B V B B GG}. Then, $GG$~turns to~$V$ and, at the same time,
its northern neighbour turns from~$B$ to~$G1$, rules
\hbox{\small\tt GG B V B B V} and
\hbox{\small\tt B B B GG B G1}. At the next step, $G1$~becomes~$B$,
but the northern neighbour of~$G1$ changes from~$B$ to~$1$,
rule \hbox{\small\tt G1  B   B   V   B   B} and
\hbox{\small\tt B B B G1 B 1}. We know that this~1 triggers the process of
lifting the data by one step upwards. Before describing the corresponding
rules, we look at what happens at the other end of the $R$-row.

  First, we note that here, the front of transformation on the $R$-row is
in advance by one step with respect to the front on the $V$-row. Indeed,
this front is materialized by the pattern $R1Z$. Now, the transformation
of~$V1$ to~$V$ on the $V$-row is triggered by the occurrence of~$G$,
rule \hbox{\small\tt V1 B V1 Z G V}. Note that at this moment, $Z$ is the
southern neighbour of this~$V1$. Next, the rules for the coloration back
to~$V$ are similar to those which we have seen in Sub-subsection~\ref{smallerstep}.
However, for these rules $\eta_3$ is different: it is always~$Z$ for the southern
neighbour of the~$V1$ changing to~$V$ and it is~$R0$ for the just restored~$V$.
But for this~$V$, its northern neighbour is~$U4$ as the coloration back to~$U$
on the $U$-row is triggered by the front on the $V$-row: accordingly, the front
on the $U$-row is delayed by one step with respect to that on the $V$-row.
This allows to have rules which are compatible with those of the opposite
coloration on the $V$-row at the beginning of the cycle. 

   And so, $ZR0$ is moving to the west. Now, we have two different situations, 
depending on whether $\mu\geq0$ or $\mu<0$. 

In the first case, when $R1$ 
sees~$BV$ through its northern side and~$Z$ through its eastern one, then
it becomes~$Z0$. This $Z0$ moves to the west and it allows the transformation
of~$BV$ to~$B$, see Section~\ref{scenario}, using basically the
rule \hbox{\small\tt BV U4 BV Z0 B B} until it sees~$W$ through its western
side. Then $W$~is replaced by~$W1$, rule \hbox{\small\tt W W B B Z0 W1}
and at the next step, $Z0$~is replaced by~$R$:
\hbox{\small\tt Z0 B W1 B R R}. The reason of the last rule is that, as
explained in Section~\ref{scenario}, when $Z0$~occurs, it starts a coloration
process to the east which replaces~$R0$ by~$R$ and cancels~$Z$. Just after the
occurrence of~$Z0$, a second one occurs by the application of the rules on $ZR0$.
The rules are:
\vskip 4pt
\ligne{\hfill\small\tt
\hbox{Z0 B Z0 B R R},\hskip 15pt
\hbox{R0 V Z0 B Z R},\hskip 15pt
\hbox{R0 V R B Z R}
\hfill}
\vskip 4pt
When $W1$~occurs, it moves upwards in the $W$-column, leaving~$B$ on its place,
until it sees~$U4$ through its eastern side: it remains there until the
penultimate step of the cycle. The rules are:
\hbox{\small\tt W W B W1 V W1} and \hbox{\small\tt W W B W1 U4 W1}.

When $\mu<0$, the leftmost $Z$~can continuously see~$V1$ through its northern 
side while moving to the west until it sees~$W$ through its western side. So, the
process is a bit simpler in this case. When the leftmost~$Z$ can see~$W$ through
its western side, it is replaced by~$R0$ and this~$W$ is replaced by~$W1$,
rules \hbox{\small\tt Z V1 W B R0 R0} and \hbox{\small\tt W W B B Z W1}.
Now, the pattern $W1R0$ changes to~$BR$, rules
\hbox{\small\tt W1  W   B   B   R0  B} and 
\hbox{\small\tt R0  V   W1  B   Z   R}. From the previous rules, we know
that the $W$ of the $V$-row changes to~$W1$ when it sees~$V$ through its eastern
side which happens at the next step, due to the delay by one step of the front
on the $R$-row with respect to that of~$Z$ on the $R$-row. As the southern neighbour
of this new~$W1$ is~$B$, it disappears, 
rule \hbox{\small\tt W1 W B B V B}, and its northern neighbour turns from~$W$
to~$W1$, due to the presence of~$BU$ through the northern side and of~$W1$
through the southern one, rule \hbox{\small\tt W W B W1 BU W1}. This last~$W1$
remains there until the penultimate step of the cycle. 

   During this time, the signal~1 travels to the west by one step at each time,
rule \hbox{\small\tt 1 B B B B B} and
\hbox{\small\tt B B B B 1 1}. A rule on~1 satisfies the pattern
\hbox{\small\tt 1 B B $\eta_3$ B $\eta_3$}: this means that~1 copies what it
sees through its southern side. Now, this southern neighbour is replaced 
by~2, pattern \hbox{\small\tt $\eta_0$ 1 $\eta_2$ $\eta_3$ $\eta_4$ 2}.
Examples of such rules are given by 
\hbox{\small\tt U 1 U V 2 2} and, when $\mu<0$ also by 
\hbox{\small\tt BU 1 BU V 2 2}, as the eastern neighbour is already
a lifted symbol. We know that~2 behaves like~1, lifting its southern neighbour
but replacing it by~2. This southern neighbour also becomes~2 unless both
its own northern and southern neighbours are~$B$.
In rules, this means that we have \hbox{\small\tt B 2 B B B B}.
This process also restores~$B$ in the place of~$BU$ when $\mu<0$. 
Indeed, in this case, the restoration is performed by~1 thanks to the rule
\hbox{\small\tt 1 B B BU U B}. Remember that when $\mu>0$, the transformation 
from $U4$ to~$U$ and from~$BV$ to~$B$ is triggered by the transformation 
from~$V1$ to~$V$. 

   The shifting of the data by one step upwards is conducted by signal~1.
When 1~can see~$W$ through its western side, we have two configurations,
depending on the sign of~$\mu$, which are slightly different.

We have that $W$ becomes~$WW$, rule \hbox{\small\tt W B X W1 1 WW}, and
that 1~lifts up a symbol. When $\mu\geq0$,
1~lifts up~$U4$, changing it to~$U$, rule \hbox{\small\tt  1 B W U4 B U}
and $U4$~is replaced by~2, rule \hbox{\small\tt U4 1 W1 B 2 2}. 
When $\mu<0$, 1~lifts up~$BU$, changing it to~$B$, rule
\hbox{\small\tt 1 B W BU B B}. 
At the next step, there are no more differences for the active instructions:
$WW$~is replaced by~$W3$, rule \hbox{\small\tt WW B X W1 U W3} and $W1$~is replaced
by~$B$, rule \hbox{\small\tt W1 WW B B 2 B}. The rules involving~2 have still been
in action and, when $W3$~is present, the last remaining~2 is the northern neighbour
of the leftmost~$R$. Accordingly, at the next step, 2~will be replaced by~$R$
using a pattern we have already seen and no~2 will be produced,
rule \hbox{\small\tt R 2 B B B B}.  At the same time,
$W3$~vanishes, rule \hbox{\small\tt W3 B X B U B}, and its northern neighbour
turns from~$B$ to~$X$, writing the new pixel, rule \hbox{\small\tt B B B W3 B X}.
The obtained configuration is the last one of the cycle.

\section{The remaining cases}
\label{remaining}

   As indicated in Section~\ref{scenario}, we cannot give all the details
about the particular cases defined by the conditions on small parameters
or small differences between the parameters. These situations are not difficult
and they are left to the reader. As already mentioned, they can be attached to
the general cases by rules which constitute shortcuts to a situation already
controlled by a general rule. 

   However, we have to go back to what we have depicted, as we had an important
constraint: $a\leq b$. We have dealt with the case $a<b$, but the scenario
fully applies when $a=b$. If we start with $r=0$, as we append~$b$, the comparison
with~$b$ will always detect a situation where~$b$ has to be subtracted from the
computed remainder and so we again have $r=0$. Now, the new pixel is written
at the correct position. As this situation is repeated at each cycle, the pixels
are written on the first diagonal as required, so that there is nothing to do.
Note, that in the execution of the automaton, we never use the fact that~$a$ 
and~$b$ should be coprime numbers, so that we can remove this assumption.

   Here, we shall look at the way we can extend the automaton to the cases
when we do not have $0<a\leq b$. First, we shall successively consider the 
situations when $a=0$, when $0<b<a$ and then the situation when $a$ and~$b$ 
have arbitrary signs.

\subsection{The case $a=0$}

   In this case, the line is a row of $X$'s. If we apply algorithm~\ref{algo1},
we remark that assuming a value of~$r$, appending $a$ to~$r$ does not change the
result. Iterating the cycle will thus lead us to a row of~$X$'s which is the
correct solution.

   The implementation of this solution with our automaton raises a problem.
Indeed, if $a$=0, there is no~$U$ on the $U$-row. This looks like a situation
when $\mu<0$. However, it may happen that $\mu\leq0$. The difference occurs
on the $V$-row where there is at least one~$V$, as we rule out the case when
$a=b=0$ which cannot define a line. If $\mu>0$, the $W$-column meets a~$B$
on the level of the $V$-row. Otherwise, it necessarily meets a~$V$.
Consequently, if during its construction the $W$-column meets a blank 
both on the~$U$- and the $V$-rows, necessarily $a=0$. If it meets a blank
on the $U$-row and a~$V$ on the $V$-row, then the automaton has to explore
the length of the blank area. This length was tacitly assumed to be less 
than~$V$ in Section~\ref{scenario} and also in the previous sub-sections
of Section~\ref{rules}. Now, we may keep this assumption: indeed, $\mu$~is
a parameter which, together with~$b$ defines the point of the $y$-axis 
where the line cuts the axis. By possibly changing the position of the
$x$-axis, we may assume that $\vert\mu\vert<b$. Accordingly, if the automaton
sees that the whole interval of~$V$'s on the $V$-row is covered by blanks,
this means that $a=0$. In this case there is nothing to append to the remainder
and it is enough to write the new pixel. Again, the iteration of such a
cycle will produce the expected row of~$X$'s.

   The situation when $\mu\leq 0$ and $a=0$ is easily detected within the
existing scenario. However, the situation when $\mu>0$ and $a=0$ entails
that the starting configuration remains unchanged, due to the rule
\hbox{\small\tt B W B B B B} used for the stability of the bottom of the 
$W$-row when $r=0$ and $\mu\geq0$. As $a=0$ is a very special configuration,
fixed at the initialization, we may require that, in this case, $\mu\leq0$.

\subsection{The case $0<b<a$}

    In Sub-section~\ref{allcases}, we have defined the general frame for the
study of the case when $0<b<a$. We have seen that the naive discrete
line which is the reflection of the naive discrete 
\hbox{$\mu\leq ax-by< \mu+\max\{\vert a\vert,\vert b\vert\}$} satisfies the
equation 
\hbox{$-\mu-\max\{\vert a\vert,\vert b\vert\}+1 \leq bx-ay < -\mu+1$}.
We have noticed that this leads to exchange the $x$- and the $y$-axes.
  
In Figure~\ref{symetrie}, we can see the change we have to perform.
At first glance, it should be enough to operate the similar change
on the rules. The automaton would then act as required. 
\vskip 10pt
\vtop{
\ligne{\hfill\scalebox{0.70}{\includegraphics{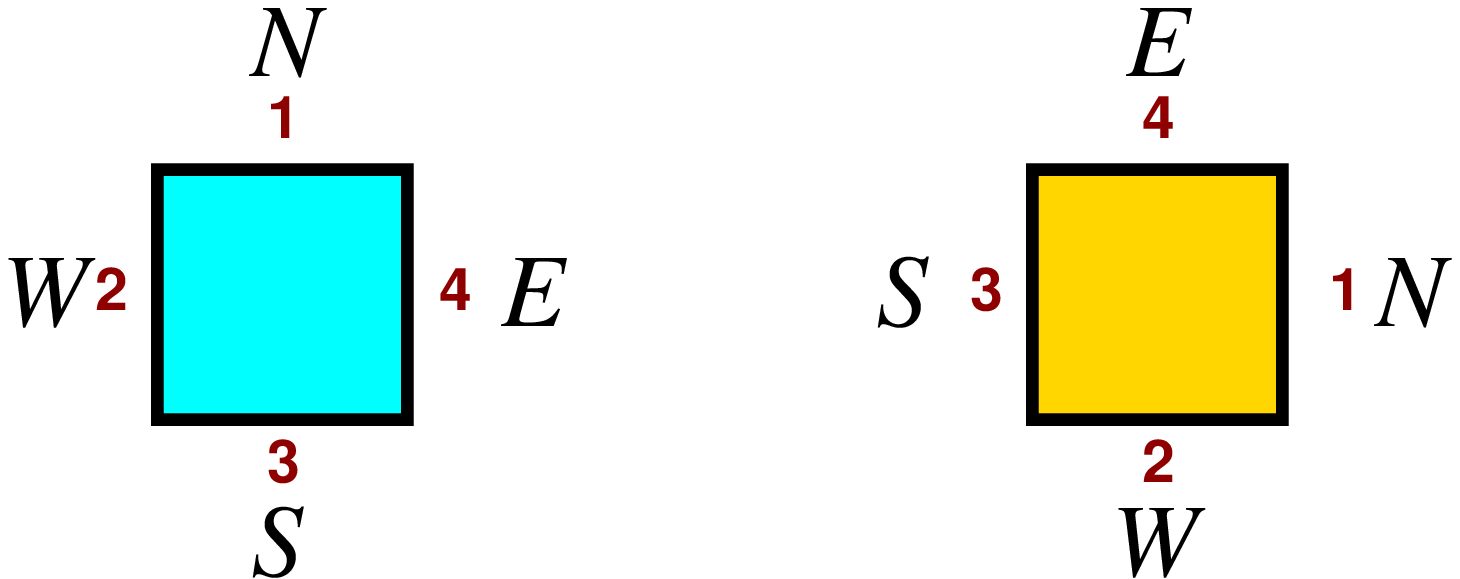}}
\hfill}
\begin{fig}\label{symetrie}
\leurre
Left-hand side: numbering the neighbours according to our conventions. 
Right-hand side: the positions of the information after the reflection in the
fist diagonal.
\end{fig}
}
\vskip 10pt
   In fact, it happens that a rule 
\hbox{\small\tt $\eta_0$ $\eta_1$ $\eta_2$ $\eta_3$ $\eta_4$ $\eta^1_0$}
and a rule
\hbox{\small\tt $\omega_0$ $\omega_1$ $\omega_2$ $\omega_3$ $\omega_4$ 
$\omega^1_0$} satisfy $\omega_i = \eta_{5-i}$ for $i\in\{1..4\}$ and
$\omega_0=\eta_0$ but that $\omega^1_0\not=\eta^1=0$. In such a case,
introducing the new rule would lead to a contradiction. The solution
is to check, for each rule, whether the reflected one exists. If it has
the required sate, it is OK, if not, then the reflected one cannot be
taken. However, the contradiction can be avoided if a state is changed
in the reflected rule. The computer program allows us to detect the rules
whose reflection would produce a contradiction, if appended to the set
of rules. A look at these rules allows us to find which state to replace
in the reflected rule by a new state. With this process, it is not very difficult
to enlarge the table of rules with the ones which are needed for the case
when $0<b<a$.

   And so, we may now consider that our automaton works for any $a,b\geq0$,
$a$+$b>0$. 
 
\subsection{In the other quarters of the plane}

    From this, it is not difficult to extend our automaton in order to construct
any naive discrete line with the condition 
$\vert\mu\vert<\max\{\vert a\vert\,\vert b\vert\}$.

    We know how to initialize the automaton, depending on the signs 
of~$a$ and~$b$ and on the comparison between $\vert a\vert$ 
and~$\vert b\vert$.

\section{Complexity issues}
\label{complexity}

   In this section, we give an estimate of the number of steps performed by
the automaton in a cycle.

   We know that the initial data satisfy the following constraint:
\vskip 4pt
\ligne{\hfill
$\vert a\vert$, $\vert b\vert$, $\vert \mu\vert \leq 
\max\{\vert a\vert,\vert b\vert\}$
\hfill}
\vskip 4pt

   Let $\delta$ be the number of cells between the $W$-column and the
eastmost non-blank cell during the computation of a cycle. From the two 
possible displays discussed in Section~\ref{scenario} and from 
Section~\ref{remaining}, we get that $\delta<3\max\{\vert a\vert,\vert b\vert\}$
as the above constraints are satisfied.

   Also from these sections, we know that we can split a cycle into the
following stages:
\vskip 4pt
\ligne{\hfill
\vtop{\leftskip 0pt\parindent 0pt\hsize=300pt
- shifting the $U$-, $V$- and $R$-rows by one step to the east,
\vskip 1pt
- appending the elements of~$U$ to the end of the $R$-row,
\vskip 1pt
- possibly performing the subtraction of~$b$,
\vskip 1pt
- restoring the data in their initial encoding.
}
\hfill}
\vskip 4pt
   The first step is performed by a run from the $W$-column to the easmost
non-blank cell: this requires at most $\delta$+4 steps, as the $W$-column has
4~elements. For appending the elements of~$U$, we have to consider the
travel of~$C$ on the blank until it meets the $U$-row. Then, a copy is delivered
every second step and each element, traveling at speed~1, advancing by one cell 
per time, we have at most $\delta$+$b$+4 steps. To estimate the time needed
by the possible subtraction, we have to decompose this stage into sub-stages.
First, when the eastmost~$R$ is written, a coloration goes back until it can
see~$V3$ on the $V$-row: this takes at most $b$ steps. Then, a $Z$~appears
which moves at speed~1 towards the $W$-row. At the same, time, a signal goes
to the east, at speed~1 too, to the eastmost~$R$, which takes at most~$b$
steps and at the end of this time there is a moving zone of $(ZR0)$'s. 
The second sub-stage is the shrinking of the $(ZR0)$ zone which takes at most
a number of steps equal to its length: at most $2b$. But the westmost $Z$
can then be at~$b$ steps at most from the $W$-column and so, this
second sub-stage requires at most $3b$ steps.
Now, the restoration occurs during the subtraction and it is estimated
by the time needed for signal 1 to go from the cell it appeared to the
$W$-column: at most $3b$. As we have seen, an additional delay of 3~steps
is required by signal~2. 
Accordingly, summing up all these times we have 10$b$+11 steps.

   Now, thanks to the study of Sections~\ref{scenario} and~\ref{rules}, the 
correctness of the rules boils down to checking that there are no contradictory
rules. As the number of rules is over than 1,000 rules, this was
performed by a computer program written for this purpose. In fact, the computer
program helped us to devise the rules at the different stages of the cycle.
Also, checking a finite number of suitable executions was enough to prove
the correctness of the program: indeed, as the working of the algorithm
is linear in the size of the data, and as the structure of a naive discrete  
line is periodic, if the execution works for a particular
choice of general parameters, it works for all of them. It is only needed
to check the particular cases when at least one parameter is small, which
we did for many cases. We have seen that 300 steps of execution are enough
to get convinced of the correctness of the computation performed by the 
automaton.

   Accordingly we have proved:

\begin{thm}
There is a deterministic cellular automaton which simulates the construction of a naive
discrete line given by the equation
\vskip 2pt
\ligne{\hfill
\hbox{$\mu\leq ax-by< \mu+\max\{\vert a\vert,\vert b\vert\}$}, 
\hfill}
\vskip 2pt
\noindent
where we may assume to satisfy, $\vert\mu\vert\leq\max\{\vert a\vert,\vert b\vert\}$.
Moreover, there is such an automaton whose working is linear in the
length of the data and of the segment of the discrete line to be constructed.
\end{thm}

   This latter point raises an interesting question: in a concrete implementation,
we could define the halting of the computation in a different way.

\subsection{Finite executions}

   In fact, for concrete applications, we necessarily have a cellular automaton
whose space is finite. The simplest way is to define the space of the cellular
automaton as a rectangle of $(H$+$2)\times (L$+$2)$ cells. Putting $(0,0)$ as the
coordinates of the lower left-hand side corner, or the south-west one according to
the terminology of the paper, the coordinates of the north-east corner would be
$(H$+$1,L$+$1)$. Of course, the cells have to know when they are at the boundary
of the area. The simplest way is to signalize the limit by a {\bf frame}
surrounding the cells devoted to the computation of the line. The cells of the
frame are an additional state, say \#, and the coordinates of these cells
are of the form $(x,0)$ and $(x,H$+$1)$ with $0\leq x\leq L$+1 for the horizontal 
limits of the rectangle and $(0,y)$ with $(L$+$1,y)$ where $0\leq y\leq H$ for
the vertical limits. There is no rule for the cells of the frame which, by definition
are in a fixed state. For the blank cells which are in contact of the frame, we have
the following conservative rules: 
\hbox{\small\tt B \#{} B B B B} for the northern limit,
\hbox{\small\tt B B \#{} B B B} for the western limit, 
\hbox{\small\tt B B B \#{} B B} for the southern limit
and \hbox{\small\tt B B B B \#{} B} for the eastern limit.

   During the execution, a problem may arise when we have to lift the data by one
step upward. In this case, state~1 should occur when instead of the blank, it
sees \#{} through its northern neighbour. This means that the rules
\hbox{\small\tt 1 B B U B U}, \hbox{\small\tt 1 B B U U U} and 
\hbox{\small\tt 1 B B B B B} have to be replaced by the rules
\hbox{\small\tt 1 \#{} B U B U}, \hbox{\small\tt 1 \#{} B U U U} and 
\hbox{\small\tt 1 \#{} B B B B} respectively. Also, when 1~just vanished, the 
rule \hbox{\small\tt W B X W1 1 WW} is replaced by the rule
\hbox{\small\tt W \#{} X W1 1 WW}. We know that $WW$ disappears and should trigger
the transformation of its northern neighbour by~$X$. Now, $WW$ may 
be changed to~$W3$,
which means that the rule \hbox{\small\tt WW B X W1 B W3} is replaced
by \hbox{\small\tt WW \#{} X W1 B W3}, but the northern neighbour, which is 
now \#, cannot be replaced by~$X$. We have also to replace
the rule \hbox{\small\tt W3 B X B B B} by the rule
\hbox{\small\tt W3 \#{} X B B B}. At this point, all rules which can be
applied are conservative rules, so that the computation stops as,
after the application of these rules, we obtain the same configuration. 
Indeed, if two
consecutive configurations are identical, this situation is repeated endlessly
and so, we can imagine a mechanism which detects the situation, which is always
possible, in principle, if we start from a finite configuration.

\section{{\Large Conclusion}}
\label{conclusion}
We think that there are many possible continuations for this work. As an example,
what was done for the line could be viewed for curves, or for planes in the $2D$-grids
or sub-spaces of $k$-dimensional grids. Now, for lines in the square grid, there
are also possible continuations. We can indicate the following ones.

   First, we could try to improve the scenario described in Section~\ref{scenario}.
It has to be completed for a few particular cases, especially for the rules
covering them. However, from a complexity point of view, there might be some
improvement. In the section, we indicated the scenario as a naive version where 
the different stages are well delimited. We could lower a bit the complexity 
established in Section~\ref{complexity}. Indeed, the starting of the copy of 
the elements of the $U$-row could be already placed when the first element of 
the $U$-row is erased by the $W$-column in construction.
Also later, other stages could be more intricate by starting as early as possible. 
But is this worth the work? We could lower the upper bound of $4\delta$ down to 
to no more than $3\delta$ and perhaps somehow below. But the price to pay would 
be a more difficult proof. Here, as the stages are well delimited, the starting
configuration of a cycle is well characterized, so that it is enough to check
that the execution of a cycle leads from one starting configuration to the next
one. The overlapping of the stages would make it difficult to define the notion 
of a cycle itself. It would be more difficult to check that there is no interaction
between a finishing cycle and the already started next one, as such an interaction
might ruin the computation.

   For what are complexity issues, a more promising improvement could be given by 
the following remark.
Our simulation is based on a representation of the integers in {\it unary}. What
could be done for a {\it binary} representation? Basically, the same scenario
could be performed, with this important difference that adding here would not
be simply appending and that subtracting would not be simply dragging back.
However, an appropriate disposal of the data could make it possible
to perform addition and subtraction: each element represents a bit in a certain
position. Appropriate markings can be managed to do the job as expected.
Now, the number of states would most probably be more important and this would
also increase the number of rules. However, the automaton would still be linear in
time with respect to the size of the data but its programming in cellular automata
would be more difficult than for the automaton of this paper. Now, this time,
the complexity would be much lower.

   Another continuation would be to devise a cellular automaton which would recognize
whether a given pattern in the $2D$-grid is or not a discrete line.

   We hope that this paper opens a new promising avenue giving a new connection
between discrete geometry and cellular automata.

\section*{Acknowledgment}

The second author wishes to express his deep thanks to LORIA and CNRS who made 
the possibility of this collaboration highly effective.
 
\bibliographystyle{eptcs}

\end{document}